\documentclass[12pt, reqno]{amsart}
\usepackage{amsmath}
\usepackage{microtype}
\usepackage[margin=1.04in]{geometry} 
\usepackage{tikz}
\usetikzlibrary{positioning}
\usepackage{newclude}
\usepackage{graphicx}

\usepackage[hidelinks,hyperindex,breaklinks]{hyperref}
\usepackage{xurl}
\usepackage{array}
\usepackage{arydshln}

\author{Shuai Guo}
\address{School of Mathematical Sciences, Peking University,
	No 5. Yiheyuan Road, Beijing, 100871, China}
\email{guoshuai@math.pku.edu.cn}

\author{Qingsheng Zhang}
\address{School of Sciences, Great Bay University,
	Great Bay Institute for Advanced Study,
	Dongguan, 523000, China}
\email{zhangqingsheng@gbu.edu.cn}

\newcommand{\<}{\langle}
\renewcommand{\>}{\rangle}

\newcommand{\sfq}{\mathsf{q}}
\newcommand{\cq}{q}

\newcommand{\cmn}{{\substack{\mu\\ \nu}}}
\newcommand{\ccmn}{{\substack{\mu_0\\ \nu_0}}}

\newcommand{\ve}{\epsilon}

\newcommand{\ii}{{\mathbf i}}

\usepackage{color,amssymb}

\def\beq{\begin{equation}}
\def\eeq{\end{equation}}
\newcommand{\be}{\begin{equation*}}
\newcommand{\ee}{\end{equation*}}

\DeclareMathOperator{\Res}{Res}

\newtheorem{dummy}{}[section]
\newtheorem{lemma}[dummy]{Lemma}
\newtheorem{proposition}[dummy]{Proposition}
\newtheorem{theorem}[dummy]{Theorem}

\newtheorem{conjecture}[dummy]{Conjecture}

\theoremstyle{definition}
\newtheorem{definition}[dummy]{Definition}

\newtheorem{remark}[dummy]{Remark}

\newcommand{\pd}{\partial}

\newcommand{\cyA}{\mathfrak{A}}
\newcommand{\cyB}{\mathfrak{B}}

\newcommand{\cD}{\mathcal{D}}

\newcommand{\cT}{\mathcal{T}}

\newcommand{\I}{\mathcal{I}}

\newcommand{\cC}{\mathcal{C}}

\newtheorem{manualtheoreminner}{Theorem}
\newenvironment{manualtheorem}[1]{%
	\renewcommand\themanualtheoreminner{#1}%
	\manualtheoreminner
}{\endmanualtheoreminner}

\def\cW{{\mathcal W}}
\def\cH{\mathcal{H}}

\begin{document}
\title[Canonical quantum curve]
{Canonical quantum curve I}

\begin{abstract}

We formulate a canonical quantum curve conjecture $\widehat H\Psi_*=0$ for spectral curves of positive genus.
Here $\Psi_*$ is the non-perturbative wave function associated with Eynard--Orantin topological recursion, specialized at the \emph{canonical vanishing point}, a distinguished zero of the closed non-perturbative partition function.
The operator $\widehat H$ is obtained by Weyl quantization of $H(x,y)$, with $\widehat x=x\cdot$ and $\widehat y=\hbar\partial_x$, together with quantum corrections to its coefficients.
These corrections depend only on closed moduli and $\hbar$ and vanish classically.
Whenever a geometric definition is available, we conjecture that the corrected coefficients are suitably normalized \emph{Wilson-loop} expectation values.
When a corresponding refined theory exists, we conjecture a relation between the canonical vanishing point and the \emph{closed Nekrasov--Shatashvili free energy}.

We prove the quantum curve conjecture for Weierstrass curves and the mirror curves of canonical bundles of smooth projective toric weak del Pezzo surfaces, including local $\mathbb P^2$.
{For the local surfaces considered here, we  relate the canonical vanishing point to a derivative of the closed NS free energy.}
For local $\mathbb P^2$, we also identify the corrected coefficient with a suitably normalized, geometrically defined NS Wilson-loop expectation value.
\end{abstract}

\maketitle

\setcounter{section}{-1}
\setcounter{tocdepth}{1}
\tableofcontents

\section{Introduction}

\subsection{History and motivation}
Quantum curves arise in several theories connected with enumerative geometry,
including matrix models~\cite{BEH02}, integrable systems~\cite{KS91}, knot theory and Chern--Simons theory~\cite{Gar04,Guk05}, topological string theory~\cite{ADKMV06}, and supersymmetric gauge theories~\cite{MM10,NS09}.
These developments motivate a common quantization principle: a classical spectral curve \(H(x,y)=0\) arising from the underlying theory should admit a quantization by a differential or difference operator \(\widehat H\) that annihilates the associated wave function $\Psi$:
\[
\widehat H\Psi=0.
\]

In~\cite{GS}, Gukov and Su\l{}kowski proposed a construction of quantum curves via Eynard--Orantin topological recursion~\cite{EO07}, with applications to knot $A$-polynomials and mirror curves of non-compact Calabi--Yau threefolds.

For spectral curves of genus zero, quantum curve equations are known in the
(descendent/open) Gromov--Witten (GW) theories of a point~\cite{Zhou12a},
$\mathbb P^1$~\cite{DBMNPS17,DOSS14,NS14},
$\mathbb C^3$~\cite{ADKMV06,Zhou12b}, and the resolved
conifold~\cite{HY06,KP07,Zhou12b,TN17}, and in stationary orbifold
GW theory of $\mathbb P[r]$~\cite{CG24}.
Other examples include Hurwitz numbers~\cite{BHLM14,DDM17,DLN16,MS15,Zhou12b}
and matrix models~\cite{MS17,Ale18,DN18,WYZadv25};
general genus-zero results appear in~\cite{BE17,BKW24,Wel24,EGMO24}.

For spectral curves of positive genus, the construction of quantum curves generally involves non-perturbative (NP) completions of the wave function~\cite{EM11,BE12}.
Bouchard--\allowbreak Chidambaram--\allowbreak Dauphinee~\cite{BCD18} verified the quantum curve property up to order $\hbar^5$ for the special Weierstrass curve $y^2=4(x^3-1)$. 
For a family of elliptic spectral curves, Iwaki~\cite{Iwa20} proved that the NP wave functions satisfy the second-order scalar equation in the Lax pair for the first Painlev\'e equation.
This result was generalized to hyperelliptic curves in~\cite{EG23,MO22} and further to algebraic spectral curves of arbitrary projection degree under suitable admissibility assumptions in~\cite{EGMO24}. 
These positive-genus constructions establish quantum curve equations, but the coefficients of the resulting scalar operators may have additional poles corresponding to apparent singularities~\cite{EG23,EGMO24,Iwa20,MO22}.

A different approach to quantum curve equations comes from refined topological string theory.
In~\cite{ACDKV12}, Aganagic, Cheng, Dijkgraaf, Krefl, and Vafa proposed that the brane partition function in the Nekrasov--Shatashvili (NS) limit, referred to here as the NS open-string wave function, satisfies a quantum curve equation obtained by canonical quantization of the defining equation of the classical curve.
{The parameters appearing in this operator are expressed in terms of the closed-string flat coordinates through the quantum mirror map.}
They derived such equations when a dual matrix-model description is available.
In~\cite{HKRS15}, Huang, Klemm, Reuter, and Schiereck used quantum periods to compute closed NS free energies for local del Pezzo surfaces.
However, this approach does not itself provide an enumerative definition of the NS open-string wave function.
These works motivate the following mathematical questions: how can the NS open-string wave function be defined directly from enumerative data, independently of the quantum curve equation, and what is the enumerative meaning of the quantum corrections to the curve parameters?

Our approach to the first question is motivated by an independent but related work.
In a joint work with Fu~\cite{FGZblowup}, we find that, for local weak del Pezzo surfaces, the local-relative correspondence~\cite{BFGW21} is equivalent to the NS limit of the vanishing blowup equation for refined topological string theory.
On the one hand, in the framework of refined GW theory~\cite{BS24}, the closed NS limit potential is identified with the relative GW potential.
On the other hand, the NS limit of the vanishing blowup equation takes the form of a vanishing condition for the closed NP partition function of topological recursion, with the vanishing point determined by a derivative of the closed NS potential~\cite[Remark 1.4]{FGZblowup}.
Motivated by this closed-theory relation, we define the \emph{canonical wave function} by specializing the NP wave function at this special point.
This relation also suggests its interpretation as an NS open-string wave function, while the definition itself uses only unrefined data.

We prove the canonical quantum curve conjecture for Weierstrass curves
and the mirror curves of canonical bundles of smooth projective toric
weak del Pezzo surfaces. For local $\mathbb P^2$, we also identify the
scalar coefficient with the NS Wilson loop.

\subsection{Canonical quantum curve}
The NP partition function associated with Eynard--\allowbreak Orantin topological recursion gives a family of wave functions $\Psi_{\mu,\nu}(\epsilon;z;\hbar;w)$ through principal specialization and the normalization specified in \S \ref{sec:wavefunction}.
Here $\epsilon$ denotes the vector of filling fractions, $z$ is a point on the spectral curve, and $\hbar$ is the formal quantization parameter. The vectors $\mu,\nu$ specify the theta characteristic, and $w$ is the corresponding theta parameter. We seek a distinguished specialization determined by these non-perturbative data, independently of the quantum curve equation.

We choose a characteristic $(\mu_0,\nu_0)$ and a formal point $w_*$
satisfying the closed non-perturbative vanishing equation
\[
 Z^{\rm NP}_{\mu_0,\nu_0}(\epsilon,\mathbf0;\hbar;w_*)=0,
\]
with $w_*+\frac{1}{2\pi i\hbar}\partial_{\boldsymbol\epsilon}F_0=O(\hbar)$.
In genus one, the leading-order condition selects the unique odd
characteristic $(\mu_0,\nu_0)=(\tfrac12,\tfrac12)$ modulo integral shifts.
The full vanishing equation then determines a unique solution
$w_*\in\hbar^{-1}\mathbb C[[\hbar^2]]$
(see Lemma~\ref{lem:closed-branch-uniqueness}),
which we call the \emph{canonical vanishing point}.
In higher genus, additional conditions are needed to select a distinguished point;
we will study this in the second paper of this series.

We define the canonical  wave function by taking $(\mu,\nu)=(\mu_0,\nu_0)$ and $w=w_*$ in the wave function:
\[
\Psi_*(\epsilon;z;\hbar)
:=\left.\Psi_{\mu_0,\nu_0}(\epsilon;z;\hbar;w)\right|_{w=w_*}.
\]
\begingroup
\renewcommand{\thedummy}{1}
\begin{conjecture}[Canonical quantum curve conjecture]
	Let $H(x,y)=0$ be a defining equation of a spectral curve with genus $\frak g\geq 1$.
	We conjecture the following canonical quantum curve equation:
	$$
	\widehat{H}\Psi_{*}=0,
	$$
where the operator $\widehat{H}$ is obtained by quantizing $x$ and $y$ according to
$\widehat x=x\cdot$ and $\widehat y=\hbar\,\frac{d}{dx}$,
with  Weyl ordering,
and replacing certain classical parameters of $H$ by quantum-corrected versions.
The corrections depend only on the closed moduli and $\hbar$, and vanish 
as $\hbar\to 0$.

\end{conjecture}
\endgroup

We further conjecture the following geometric explanation of the canonical vanishing point and the quantum corrections to the curve parameters.

\subsubsection*{Nekrasov--Shatashvili limit}
For mirror curves of toric Calabi--Yau threefolds, refined topological
string theory provides a closed free energy depending on two
equivariant parameters~\cite{ACDKV12}. Its NS limit gives the closed
free energy $F^{\rm NS}$.
For a general spectral curve, we expect an analogous two-parameter
refinement whose unrefined specialization recovers the free energies
of topological recursion.

Write $F_g^{\rm NS}$ for the coefficient of $\hbar^{2g-1}$ in $F^{\rm NS}$.
\begingroup
\renewcommand{\thedummy}{2}
\begin{conjecture}\label{conj:NS-vanishing-point}
	The NS potential determines the canonical point by
	\begin{equation}\label{eqn:w_can-NS}
	w_{*,j}=\frac{1}{2\pi i}\pd_{\epsilon_j}F^{\rm NS}
	=\frac{1}{2\pi i}\sum_{g\geq0}\hbar^{2g-1}\pd_{\epsilon_j}F_g^{\rm NS},
	\qquad j=1,\cdots,\mathfrak g.
	\end{equation}
\end{conjecture}
\endgroup
For the local weak del Pezzo surfaces, Conjecture~\ref{conj:NS-vanishing-point} has been established in~\cite{FGZblowup}.

Let $\Psi_*^X$ denote the principal specialization of the NP partition function $Z^{\rm NP}_{\mu_0,\nu_0}$
at the canonical vanishing point (see~\S~\ref{sec:wavefunction} and \S~\ref{sec:canonical-wave}).
For the genus-one cases considered here, it has zero constant term and
a nonzero linear term in the chosen boundary coordinate $X$.
Let $K_1=[X]\Psi_*^X$ denote this linear coefficient. We conjecture the following relation:
	\beq\label{eqn:def_NS}
	\pd_{\hbar}F^{\rm NS}(\epsilon;\hbar)=\log K_1.
	\eeq
When a geometric definition of $F^{\rm NS}(\epsilon;\hbar)$ is unavailable, we define $F^{\rm NS}$ by equation~\eqref{eqn:def_NS} with normalization $[\hbar^0]F^{\rm NS}=0$.

\subsubsection*{Wilson loop}

Let $\widehat H_0$ be the Weyl quantization of $H$ with its classical
coefficients unchanged. For a spectral curve of genus $\mathfrak g\geq1$,
we expect $\widehat H$ to differ from $\widehat H_0$ by $\mathfrak g$
correction terms. When geometric Wilson loops are defined with matching
 normalization, we conjecture that the corrected
coefficients are their expectation values.

For the genus-one curves considered here, the correction is scalar,
so the canonical quantum curve conjecture takes the form
\begin{equation}\label{eqn:canonical_quantum_curve}
 (\widehat H_0+\cW^+)\Psi_*=0,
 \qquad
 \cW^+=\cW(\epsilon;\hbar)-\cW(\epsilon;0).
\end{equation}
Here $\cW$ replaces certain moduli parameter in the classical 
curve equation $H(x,y)=0$,
while $\cW^+$ is its quantum correction.
\footnote{
The ratio of the quadratic and linear coefficients of $\Psi_*^X$
determines $\cW$ through an explicit formula for each model; see
\S~\ref{sec:canonical-wave}.}

\medskip

\begingroup
\renewcommand{\thedummy}{3}
\begin{conjecture}\label{conj:one-loop-geometric-Wilson}
For a smooth weak del Pezzo surface $S$, put $E=-K_S$
and let $\mathbf Q^E$ be its Novikov monomial. Let
$\mathcal W^{\rm NS}_1(\mathbf Q;\hbar)$ be the connected one-point
NS Wilson-loop series~\cite{GuoWangWu} defined in
Section~\ref{sec:wilson-loops}. 
The scalar $\cW$ in the canonical quantum curve satisfies
\begin{equation}\label{eqn:one-loop-geometric-Wilson}
 \cW(\mathbf Q;\hbar)
 =\bigl(\mathbf Q^E\bigr)^{-1/E^2}
       \mathcal W^{\rm NS}_1(\mathbf Q;\hbar)
\end{equation}
as a formal identity in $\hbar$ and the Novikov variables.
\end{conjecture}
\endgroup

\subsection{Main results}
Our main results concern genus one.

\subsubsection*{Weierstrass curve}

For $\operatorname{Im}\tau>0$, let $g_2=g_2(\tau)$ and
$g_3=g_3(\tau)$ be the Weierstrass invariants of the period lattice
$\mathbb Z+\tau\mathbb Z$, normalized by
\[
 g_2(\tau)=60\sum_{(m,n)\in\mathbb Z^2\setminus\{(0,0)\}}
                 \frac{1}{(m+n\tau)^4},
 \qquad
 g_3(\tau)=140\sum_{(m,n)\in\mathbb Z^2\setminus\{(0,0)\}}
                 \frac{1}{(m+n\tau)^6}.
\]
Let $r\ne0$ be the scaling parameter.
We use the closed coordinates $\epsilon$ and $t=-\frac{r^4g_2}{2}$ of
\S \ref{sec:weierstrass}.
\begin{manualtheorem}{I}[{=Theorems~\ref{thm:quantum_weierstrass} and~\ref{thm:weierstrass-NS}}]
\label{thm:intro-weierstrass}
For the Weierstrass curve
\[
 y^2=4x^3-r^4g_2x-r^6g_3,
\]
the canonical wave function satisfies
\[
 \big(\hbar^2\partial_x^2-(4x^3-r^4g_2x-\cW(\epsilon,t;\hbar))\big)
 \Psi_*=0,
\]
where $\cW(\epsilon,t;\hbar)$ replaces the classical coefficient
$r^6g_3$, with $\cW(\epsilon,t;0)=r^6g_3$.
Explicitly, it is given by $\cW=-2\hbar\partial_X\log\tfrac{\Psi^X_*}{X}\big|_{X=0}$.

Moreover, if $F^{\rm NS}(\epsilon,t;\hbar)$ is defined from the canonical
wave function by
\[
\partial_\hbar F^{\rm NS}(\epsilon,t;\hbar)
 =\log\left(\left.\partial_X\Psi_*^X(\epsilon,t,X;\hbar)\right|_{X=0}\right), 
\]
with vanishing $\hbar^0$ coefficient, then
$F_0^{\rm NS}=-F_0$ and
Conjecture~\ref{conj:NS-vanishing-point} holds for this potential.
\end{manualtheorem}

\subsubsection*{Local $\mathbb P^2$}
Let $\cq$ be the mirror parameter and $t$ the corresponding flat
coordinate given by the mirror map, with $Q=e^{-t}$ (see \S~\ref{sec:localp2-TR}).
Let $\sfq=e^{-\hbar}$. In the coordinates $X=e^{-x}$ and $Y=e^{-y}$,
the canonical quantization conventions give
\[
 (\widehat{X^aY^b}f)(X)
 =\sfq^{-ab/2}X^a f(\sfq^{-b}X),
 \qquad a,b\in\mathbb Z.
\]
\begin{manualtheorem}{II}[={Theorems}~\ref{thm:quantum-localp2} and~\ref{thm:localp2-wilson-identification}]
\label{thm:intro-localp2}
For the local $\mathbb P^2$ geometry with the mirror curve
\[
X^{-1}+X^{-1}Y+X^2Y^{-1}+\cq^{-1/3}=0,
\]
the canonical wave function satisfies
\[
 \big(X^{-1}+\sfq^{1/2}X^{-1}\widehat Y
       +\sfq X^2\widehat Y^{-1}+\cW(t;\hbar)\big)\Psi_*=0,
\]
where $\cW$   is
explicitly given by
$\cW(t;\hbar)
 =(\sfq^{-1}-1)\left.\partial_X
   \log\frac{\Psi_*^X}{X}\right|_{X=0}$.
It replaces the classical parameter $\cW(t;0)=\cq^{-1/3}$, so
$\cW^+=\cW(t;\hbar)-\cq^{-1/3}$.

Moreover,  $\cW$ agrees with the normalized geometric NS Wilson loop:
\[
 \cW(t;\hbar)=Q^{-1/3}\mathcal W^{\rm NS}_1(Q;\hbar),
\]
where $\mathcal W^{\rm NS}_1$ is the connected one-point series of
\S~\ref{sec:wilson-loops}. Thus Conjecture~\ref{conj:one-loop-geometric-Wilson}
holds for $S=\mathbb P^2$.
\end{manualtheorem}

\subsubsection*{Local weak del Pezzo surfaces}
The quantum-curve construction for local $\mathbb P^2$ extends to
$\mathcal X=K_S$, where $S$ is a smooth projective toric weak del
Pezzo surface, with $-K_S$ nef and $K_S^2>0$.
Such surfaces arise from the $16$ reflexive polygons by unimodular
star triangulations using every boundary lattice point.

Fix an outer leg and choose coordinates $X=e^{-x},Y=e^{-y}$. 
In the lattice coordinates adapted to this leg, we take the unique interior lattice point of the Newton polygon to be \((0,0)\) and write its boundary lattice points as \((m_k,n_k)\), \(1\leq k\leq\mathfrak m\).
The mirror curve then has the form
\[
 H(X,Y;\cq)=
       \sum_{k=1}^{\mathfrak m}r_kX^{m_k}Y^{n_k}+\gamma=0.
\]
Here $\gamma,r_k$ depend on the complex moduli $\cq$, which are
related to the closed K\"ahler parameters ${\bf Q}$ by the mirror map.
The chosen outer leg determines the boundary point $(X,Y)=(0,-1)$
on the compactified curve.

\begingroup
\begin{manualtheorem}{III}[=Theorem~\ref{thm:one-loop-TR-quantum-curve}]
\label{thm:intro-local-surfaces}
For $S$ and the chosen outer leg above, the canonical wave function
satisfies the quantum curve equation
\[
 \bigg(\sum_{k=1}^{\mathfrak m}
 r_k\sfq^{-m_kn_k/2}X^{m_k}\widehat Y^{\,n_k}
       +\cW({\bf Q};\hbar)\bigg)\Psi_*=0,
\]
where $\cW$ is the scalar defined in \eqref{eqn:wilsonloopdelpezzo}  and
$\cW({\bf Q};0)=\gamma$. Thus $\cW^+=\cW({\bf Q};\hbar)-\gamma$.
\end{manualtheorem}
 
\par
\endgroup

\subsection{Outline of the paper} 
We first recall the NP partition function of topological recursion and introduce the associated wave function in \S \ref{sec:quantum_curve}.
We then formulate the canonical quantum curve conjecture in \S \ref{sec:canonical-quantum-curve} and explain how the closed NS free energy and Wilson-loops are expected to determine the canonical vanishing point and the quantum corrections, respectively.
In \S \ref{sec:weierstrass}, we establish the Weierstrass quantum curve as a warm-up.
In \S \ref{sec:localp2}--\ref{sec:localp2-wilson}, we prove the quantum curve equation for local \(\mathbb P^2\) and establish its relationship with the NS free energy and the NS Wilson-loop expectation value. 
We extend the quantum curve construction to local toric weak del Pezzo surfaces in \S \ref{sec:one-loop}.

\subsection*{Acknowledgments}  
The authors thank Jianjun Fu for collaboration on related work, 
and thank Ce Ji and Xin Wang for helpful discussions.

\subsection*{AI disclosure.}
We used ChatGPT to improve the exposition, assist with extensive
computational tests and checks, and prepare tables.
The mathematical content of this paper was written by the authors,
who take full responsibility for its correctness and for the final manuscript.

$$
$$

\section{Non-perturbative partition functions and wave functions}
\label{sec:quantum_curve}
In this section, we first recall topological recursion and the non-perturbative partition functions. 
We then introduce the associated wave functions of the spectral curve.

\subsection{Topological recursion}
\label{sec:TR}
The input data of the topological recursion are given by $(\Sigma_{\frak g},x,y)$, 
where $\Sigma_{\frak g}$ is a Riemann surface of genus $\frak g$ with a fixed symplectic basis $\{\cyA_i,\cyB_i\}_{i=1}^{\frak g}$ for $H_{1}(\Sigma,\mathbb Z)$,
$x=x(z)$ is a (possibly multi-valued) function on $\Sigma_{\frak g}$ satisfying that $dx$ is meromorphic and has finitely many simple zeros $\{z^\beta\}$,
$y=y(z)$ is a holomorphic function defined near each $z^\beta$ satisfying $dy(z^\beta)\ne 0$.
Here $z$ is a coordinate on $\Sigma_{\frak g}$.

Starting from the input data, the topological recursion gives a sequence of multi-differentials $\{\omega_{g,n}(z_1,\cdots,z_n)\}_{g,n\geq 0}$ on $\Sigma_{\frak g}$ satisfying certain properties~\cite{EO07}.
Firstly,
$$
\omega_{0,1}(z):=y(z)dx(z), \quad \omega_{0,2}(z_1, z_2):=B(z_1, z_2),
$$
where $B(z_1, z_2)$ is the Bergman kernel of the Riemann surface $\Sigma_{\frak g}$ normalized by $A$-cycles.
Let $K_\beta$ be the recursion kernel around the critical point $z^\beta$, defined by
$$
K_\beta(z_0,z)=\frac{\int_{z'=\bar z}^{z}B(z_0,z')}{2(y(z)-y(\bar z))dx(z)},
$$
where $\bar z\in \Sigma$ is the unique local involution of $z\in\Sigma$ around the critical point $z^\beta$ such that $x(\bar{z})=x(z)$. 
Then for $2g-2+n+1>0$, the multi-differentials $\omega_{g,n+1}$ are defined recursively as follows:
\begin{align}
	\omega_{g,n+1}(z_0,z_{[n]})
	:=&\, \sum\limits_{\beta}\mathop{\Res}_{z=z^\beta}K_\beta(z_0,z)\bigg(\omega_{g-1,n+2}(z,\bar z,z_{[n]}) \nonumber\\
	&\, +\sum^{\prime}_{\substack{g_1+g_2=g\\ I\bigsqcup J=[n]}}\omega_{g_1,|I|+1}(z,z_{I})\omega_{g_2,|J|+1}(\bar z,z_{J})\bigg).
	\label{eqn:def-TR}
\end{align}
Here $[n]=\{1,\cdots,n\}$ and for any subset $I\subset [n]$, $z_{I}:=\{z_i\}_{i\in I}$.
The symbol $\sum\limits^{\prime}$ means we exclude $\omega_{0,1}$ in the summation.
For $g\geq 2$,
\beq\label{def:wg0}
F_g=\omega_{g,0}:=\frac{1}{2-2g}\sum_{\beta}\mathop{\Res}_{z=z^\beta}\omega_{g,1}(z)\int_{z'=z^\beta}^{z}\omega_{0,1}(z').
\eeq
The unstable free energies $F_0$ and $F_1$ are fixed separately.
Their period relations are given in~\S~\ref{sec:NP-generating-series},
and their explicit normalizations for the Weierstrass and local $\mathbb P^2$
curves in~\S~\ref{sec:weierstrass} and~\S~\ref{sec:localp2-TR}, respectively.

We follow the treatment of TR descendents in~\cite{GJZ23},
extracting them from the boundary Laurent expansion of the multi-differentials and
packaging them into generating functions.  The associated CohFT and ancestor
potential are described in~\cite{DOSS14,Eyn14,GJZ23}.
Fix a pole $b$ of $dx$ and choose a local coordinate $X$ centered at $b$, so that $X(b)=0$. Write $X_i=X_i(z_i)$ and denote by $\omega_{0,1}^{\mathrm{sing},b}$ the singular part of $\omega_{0,1}=y\,dx$ at $b$ in this coordinate. Then
\be
\omega_{g,n}=\delta_{(g,n)}^{(0,1)}\cdot \omega_{0,1}^{\mathrm{sing},b}(X_1)+\delta_{(g,n)}^{(0,2)}\cdot \frac{dX_1dX_2}{(X_1-X_2)^2}
+
\sum_{k_1,\cdots,k_n\geq 1}\<\alpha_{k_1},\cdots,\alpha_{k_n}\>_{g,n}\frac{dX_1^{k_1}\cdots dX_n^{k_n}}{k_1\cdots k_n}.
\ee
For $n\geq1$, we define
$$
F_{g,n}({\bf p})
=\sum_{k_1,\cdots,k_n\geq 1}\<\alpha_{k_1},\cdots,\alpha_{k_n}\>_{g,n}\frac{p_{k_1}\cdots p_{k_n}}{k_1\cdots k_n}.
$$
We set $F_{g,0}=F_g$. The TR descendent partition function is
$$
Z({\bf p};\hbar)=\exp\left(\sum_{g,n\geq 0}\frac{\hbar^{2g-2+n}}{n!}F_{g,n}({\bf p})\right).
$$

\subsection{Non-perturbative generating series}
\label{sec:NP-generating-series}
We follow the treatment of non-perturbative TR descendent generating series
in~\cite[\S~4]{GJZ23}.
The non-perturbative partition function was introduced in~\cite{Eyn09,EM11}
and developed further in~\cite{BE12}.

For a spectral curve of genus $\frak g\geq 1$,
we denote by $\{du_i\}_{i=1}^{\frak g}$ the normalized holomorphic differentials and by $\tau=(\tau_{ij})_{i,j=1}^{\frak g}$ the period matrix, such that
\begin{equation*}
\oint_{\cyA_j}du_i=\delta_{ij},
\qquad
\tau_{ij}=\oint_{\cyB_j}du_i.
\end{equation*}
For such curves, the multi-differentials may have nonzero $B$-cycle periods.  
For $m\geq0$ and $n+m\geq1$, set
\beq\label{def:NP-multidifferentials}
\omega^{(m)}_{g,n;j_1,\ldots,j_m}(z_1,\ldots,z_n)
:=\oint_{\cyB_{j_1}}\!\cdots\!\oint_{\cyB_{j_m}}\omega_{g,n+m}(z_1,\ldots,z_{n+m}).
\eeq
For the unstable cases, we have
$$
\omega_{0,1;i}^{(1)}=\oint_{\cyB_i}B(z,\mathord{\cdot})=2\pi {\bf i}\,du_i(z),\qquad
\omega_{0,0;i,j}^{(2)}=\oint_{\cyB_i}\oint_{\cyB_j}B=2\pi {\bf i}\,\tau_{ij}.
$$
For $m\geq 1$, expanding $\omega_{g,n}^{(m)}$ at the boundary point $b$ gives
\beq\label{def:NP-descendents}
\omega^{(m)}_{g,n;j_1,\ldots,j_m}
=\sum_{k_1,\ldots,k_n\geq1}
\left\langle\alpha_{k_1},\ldots,\alpha_{k_n}\right\rangle^{(m)}_{g,n;j_1,\ldots,j_m}
\frac{dX_1^{k_1}\cdots dX_n^{k_n}}
{k_1\cdots k_n}.
\eeq
For $m=0$, we set $F^{(0)}_{g,n}=F_{g,n}$.
For $m\geq1$, we introduce the generating series
\be
F^{(m)}_{g,n;j_1,\ldots,j_m}({\bf p})
:=\sum_{k_1,\ldots,k_n\geq1}
\left\langle\alpha_{k_1},\ldots,\alpha_{k_n}\right\rangle^{(m)}_{g,n;j_1,\ldots,j_m}
\frac{p_{k_1}\cdots p_{k_n}}{k_1\cdots k_n}.
\ee
For a sequence of independent variables $w=(w_1,\ldots,w_{\frak g})$, 
letting $\nabla_w=(2\pi i)^{-1}(\pd_{w_1},\ldots,\pd_{w_{\frak g}})$,
we denote
$$
F^{(m)}_{g,n}({\bf p})\cdot\nabla_w^{\otimes m}
=\frac{1}{(2\pi i)^{m}}\sum_{j_1,\cdots,j_m=1}^{\frak g}F^{(m)}_{g,n;j_1,\ldots,j_m}({\bf p})
\pd_{w_{j_1}}\cdots\pd_{w_{j_m}}.
$$

With $\tau$ as above, let $\theta[\substack{\mu\\ \nu}](w|\tau)$ be the theta function 
with characteristic $\{\substack{\mu\\ \nu}\}\in\mathbb R^{\frak g}\times\mathbb R^{\frak g}$.
Precisely,
$$
\theta[\cmn](w|\tau) =\sum_{l\in\mathbb Z^{\frak g}} e^{\pi \ii (l+\mu)^{t}\tau (l+\mu)+2\pi \ii (l+\mu)^{t}(w+\nu)}.
$$
We introduce a shift on variables $\{w_i\}_{i=1}^{\frak g}$ by
\beq\label{def:w-tilde}
w_i^{+}:=w_i+\frac{1}{2\pi {\bf i} \hbar}F^{(1)}_{0,0;i},\qquad i=1,\cdots,\frak g.
\eeq

\begin{definition}\label{def:NP-partition}
	Given the spectral curve $\mathcal C=(\Sigma_{\frak g},x,y)$ with $\frak g\geq 1$, we choose a boundary point $b$ and fix the local coordinate $X$ near the boundary point.
	The corresponding non-perturbative descendent partition function is defined by
	\begin{align}\label{def:Z-NP-theta}
		Z^{\rm NP}_{\mu,\nu}({\bf p};\hbar;w)
		:={}&e^{F_{\rm un}({\bf p};\hbar)}
		\exp\!\left(
		\sum_{\rm stable}
		\frac{\hbar^{2g-2+n+m}}{n!m!}
		F^{(m)}_{g,n}({\bf p})\cdot\nabla_w^{\otimes m}
		\right)
		\theta[\substack{\mu\\ \nu}]\!\left(
		w^{+}+\widehat{\mathbf u}({\bf p})
		\,\middle|\,\tau\right).
	\end{align}
	Here ``stable'' means $g,n,m\geq 0$ and $2g-2+n+m>0$, 
	$F_{\rm un}({\bf p};\hbar)$ is the collection of the (perturbative) unstable terms:
	\beq\label{def:F-un}
	F_{\rm un}({\bf p};\hbar)
	:=\hbar^{-2}F_0+F_1
	+\hbar^{-1}F_{0,1}({\bf p})
	+\frac{1}{2}F_{0,2}({\bf p}),
	\eeq
	and writing $du_i(z)=\sum_{k\geq1}u_{i,k}\frac{dX^{k}}{k}$ near $b$, then $\widehat{\mathbf u}({\bf p})=(\widehat u_1({\bf p}),\ldots,\widehat u_{\frak g}({\bf p}))$ is defined by
	\beq\label{def:u-hat}
	\widehat u_i({\bf p}):=\sum_{k\geq1}u_{i,k}\frac{p_k}{k},\qquad i=1,\cdots,\frak g.
	\eeq
\end{definition}

The non-perturbative partition function can be expressed in a Fourier transformation form by using the filling fraction $$\ve_i:=\frac{1}{2\pi i}\oint_{\cyA_i}ydx,\qquad i=1,\ldots,\frak g$$ introduced in~\cite{EO07}.
Assume a local deformation with coordinates
$(\boldsymbol\epsilon,\mathbf m)$, where $\mathbf m$ denotes the
complementary moduli, and transport the symplectic basis by the
Gauss--Manin connection. At fixed $\mathbf m$, the variation is
\[
 \partial_{\epsilon_i}(y\,dx)=2\pi i\,du_i
 =\oint_{\cyB_i}B(z,\mathord{\cdot}),\qquad
 \partial_{\epsilon_j}F_0=\oint_{\cyB_j}y\,dx,\qquad
 \partial_{\epsilon_i}\partial_{\epsilon_j}F_0=2\pi i\tau_{ij}.
\]
For $(g,n)\ne(0,0),(0,1)$, special geometry~\cite[\S~5.3]{EO07} gives
\begin{equation}\label{assump:ff-var}
 \partial_{\epsilon_i}\omega_{g,n}(-)
 =\oint_{\cyB_i}\omega_{g,n+1}(z,-).
\end{equation}
For the toric families of~\cite{FLZ20}, the period coordinates of
\S~5.7--5.10 and identities (7.7)--(7.9) realize this variation.
We hold $\mathbf m$ fixed and display only the filling-fraction
dependence in $\omega_{g,n}$, $F_g$, $Z$, and their NP analogues.
Thus, with $\partial_\epsilon=(\partial_{\epsilon_i})_i$,
\[
w^+=w+\frac{\partial_\epsilon F_0}{2\pi i\hbar},\qquad
 \widehat{\mathbf u}(\mathbf p)
 =\frac{\partial_\epsilon F_{0,1}(\boldsymbol\epsilon,\mathbf p)}{2\pi i}.
\]
These identities give the following equivalent Fourier representation of the non-perturbative partition function.
\begin{proposition}
	Under the filling-fraction variation assumption, equation~\eqref{def:Z-NP-theta} is equivalent to the following formal Fourier representation in $w$:
	\beq\label{eqn:Z-NP}
	Z^{\rm NP}_{\mu,\nu}(\ve,{\bf p};\hbar;w)
	=\sum_{l\in\mathbb Z^{\frak g}}
	e^{2\pi i(l+\mu)^{t}(w+\nu)}
	Z\bigl(\ve+\hbar(l+\mu), {\bf p};\hbar\bigr).
	\eeq
\end{proposition}
\begin{proof}
	The terms with $2g-2+n+m\leq 0$ are accounted for by
	$F_{\rm un}$, $w^{+}$, $\tau$, and $\widehat{\mathbf u}({\bf p})$, 
	while all the remaining terms satisfy $2g-2+n+m>0$.
	For each $l\in\mathbb Z^{\frak g}$, by the Taylor expansion in the
	filling fractions, the filling-fraction variational formula, and
	$\pd_{\ve}^2F_0=2\pi i\tau$, we have
	\begin{align*}
	&e^{2\pi i(l+\mu)^t(w+\nu)}
	Z(\ve+\hbar(l+\mu),{\bf p};\hbar)\\
	={}&e^{F_{\rm un}(\ve;{\bf p};\hbar)}
	\exp\bigg(
	\sum_{2g-2+n+m>0}
	\frac{\hbar^{2g-2+n+m}}{n!m!}
	F^{(m)}_{g,n}({\bf p})\cdot(l+\mu)^{\otimes m}
	\bigg)\\
	&\quad\cdot
	\exp\Big(\pi i(l+\mu)^t\tau(l+\mu)
	+2\pi i(l+\mu)^t\big(w^{+}+{\widehat{\mathbf u}({\bf p})}+\nu\big)
	\Big).
	\end{align*}
	On the other hand, the last exponential factor is exactly the $l$-th summand in the theta function $\theta[\cmn](w^{+}+{\widehat{\mathbf u}({\bf p})}|\tau)$,
	and \(\nabla_w^{\otimes m}\) acts on this term by multiplication by \((l+\mu)^{\otimes m}\). 
	Applying the operator in equation~\eqref{def:Z-NP-theta} and summing the above identity over
	$l\in\mathbb Z^{\frak g}$, we get~\eqref{eqn:Z-NP}.
\end{proof}

\begin{remark}
	More general non-perturbative sums involve shifts in both the filling fractions and the complementary moduli.
	For toric mirror curves, the shift information is encoded by $r$-fields~\cite{HSW18}.
	In this paper, we mainly consider genus-one curves and hold the complementary moduli fixed, so we work with the characteristic formulation above.
	We leave a systematic discussion of the full shifts and their role in canonical quantum curves to future work.
\end{remark}

\subsection{Wave function associated with the spectral curve} \label{sec:wavefunction}
Fix $u(z)=\int_b^zdu$ and a local coordinate $X$ with $X(b)=0$.
Principal specialization of~\eqref{def:Z-NP-theta} defines
\[
 \Psi^X_{\mu,\nu}(\epsilon;z;\hbar;w)
 :=\left.Z^{\rm NP}_{\mu,\nu}(\epsilon,\mathbf p;\hbar;w)
 \right|_{p_k=X^k,\ k\geq1}.
\]
Write $W^{(m)}_{g,n}(z)=\int_b^z\cdots\int_b^z\omega^{(m)}_{g,n}$ for $2g-2+n+m>0$, with
$W^{(m)}_{g,0}=\omega^{(m)}_{g,0}$ when $n=0$.
In~\eqref{def:Z-NP-theta}, this specialization replaces
$F^{(m)}_{g,n}(\mathbf p)$ by $W^{(m)}_{g,n}(z)$ and
$\widehat{\mathbf u}(\mathbf p)$ by $u(z)$.
The unstable factor~\eqref{def:F-un} uses the regular disk integral
$\int_b^z[ydx]_+$ and the annulus integral of $B-\frac{dX_1dX_2}{(X_1-X_2)^2}$, 
and this is where the choice of $X$ enters.

To define a wave function independent on the local coordinate,
we modify unstable terms in $\Psi^{X}_{\mu,\nu}$.
For $(0,1)$-term, we just take
	$$
	\int_{b}^{z} [ydx]_{+}\quad\longrightarrow\quad \int^{z} ydx.
	$$
For the $(0,2)$-term, we first recall the prime form~\cite{Fay73} on the Riemann surface.
{Here we follow the notation of
\cite[\S 2.1]{GJZ23}.
Let ${\bf c}$ be a nonsingular odd half-integer characteristic, with $dh_{\bf c}(b)\ne0$ for the
differential defined below.}
Let 
$$
dh_{\bf c}(z)=\sum_{i=1}^{\frak g}\pd_i\theta[{\bf c}](0|\tau)du_i(z),
$$
and define the prime form~\cite{Fay73} (it does not depend on the choice of odd characteristic ${\bf c}$)
$$
E(z_1,z_2)=\frac{\theta[{\bf c}](u(z_1)-u(z_2)|\tau)}{\sqrt{d h_{{\bf c}}(z_1)d h_{{\bf c}}(z_2)}},
$$
then we have the following equality
\be
\exp\left(\frac{1}{2}\int_{b}^{z}\int_{b}^{z}\left(B(z_1,z_2)-\frac{dX_1dX_2}{(X_1-X_2)^2}\right)\right)
=\frac{X}{E(z,b)\sqrt{dX(b)}\sqrt{dX(z)}}.
\ee
We see $\Psi^{X}_{\mu,\nu}$ depends on the local coordinate $X$ via $1/\sqrt{-dX^{-1}}$.
We take
$$
\sqrt{\frac{dh_{\bf c}}{dX}(b)} \cdot \frac{1}{\sqrt{-dX^{-1}}}\quad\longrightarrow\quad \frac{1}{\sqrt{dx}}.
$$
We make the following definition.

\begin{definition} \label{normalizedpsi}
Let $Z_{\rm cl}(\epsilon;\hbar)=Z(\epsilon,\mathbf0;\hbar)$ be the closed perturbative partition function.
	For spectral curve $(\Sigma_{\frak g},x,y)$ with $\frak g\geq 1$, we define the normalized wave function with characteristic $[\cmn]$ by
	$$
	\Psi_{\mu,\nu}(\ve,z;\hbar;w)=Z_{\rm cl}(\epsilon;\hbar)\cdot e^{\frac{1}{\hbar}\int_{b}^{z}ydx+\frac{1}{2}\log(\frac{dh_{\bf c}}{dx})}
	\cdot \Psi^{\rm st}_{\mu,\nu}(\ve,z;\hbar;w),
	$$
	where
	\be
	\Psi^{\rm st}_{\mu,\nu}(\ve,z;\hbar;w)
	=\exp\left({\sum_{\substack{2g-2+n+m>0 \\ m+n>0}} \hbar^{2g-2+n+m}\frac{W^{(m)}_{g,n}(z)}{n!}\cdot \frac{\nabla_{w}^{\otimes m}}{m!}}\right)
	\frac{\theta[\substack{\mu \\ \nu}](w^{+}+u(z)|\tau)}{\theta[{\bf c}](u(z)|\tau)}.
	\ee
\end{definition}
Clearly, the normalized wave function is independent of the local coordinate $X$. 
It is initially defined near the boundary point \(b\).
Moreover, at zeros of the theta denominator, the normalized wave function is interpreted using the prime-form identity
\[
\frac{\sqrt{dh_{\bf c}(z)/dx(z)}}
{\theta[{\bf c}](u(z)|\tau)}
=\frac{1}{E(z,b)\sqrt{dx(z)}\sqrt{dh_{\bf c}(b)}}.
\]
Since \(E(z,b)\) is holomorphic, with a simple zero at \(z=b\) and no other zeros, the zeros of the theta denominator away from \(b\) introduce no additional singularities.
\begin{remark}
	An annulus normalization based on $x^{-1}$ may diverge at the boundary:
	if $x=X^{-r}$ with $r>1$, then
	$dX/d(x^{-1})=(rX^{r-1})^{-1}$ has a pole at $X=0$.
	The prime-form normalization above avoids this coordinate singularity.
\end{remark}

Clearly, we have the following relation between $\Psi^{X}_{\mu,\nu}$ and $\Psi^{\rm st}_{\mu,\nu}$:
\beq\label{eqn:Psi_lambda_st}
\Psi^{X}_{\mu,\nu}=X\cdot \sqrt{\frac{dh_{\bf c}}{dX}(b)}\, \sqrt{\frac{dh_{\bf c}}{dX}(z)} \cdot Z_{\rm cl}(\epsilon;\hbar)\cdot e^{\hbar^{-1}F_{0,1}(X)}\cdot \Psi^{\rm st}_{\mu,\nu}.
\eeq
Since $u(b)=0$ and ${\bf c}$ is odd,
\[
 \theta[{\bf c}](u(z)\mid\tau)
 =\frac{dh_{\bf c}}{dX}(b)\,X+O(X^2),
 \qquad \frac{dh_{\bf c}}{dX}(b)\ne0.
\]
The stable integrals are regular at $b$, and $\nabla_w$ acts only on
the numerator theta function. Thus $\Psi^{\rm st}_{\mu,\nu}$ has at
most a simple pole at $b$, coefficientwise in $\hbar$.

{At a special value of \(w\), the principal specialization may vanish identically in \(z\), and the normalization above does not remove this vanishing. We allow this possibility in the present paper. Additional conditions are needed to obtain a nonzero canonical wave function.}

$$
$$

\section{Canonical specialization and the quantum curve conjecture}
\label{sec:canonical-quantum-curve}

The canonical specialization is defined by evaluating the wave function at a chosen vanishing point of the closed NP partition function. We describe this construction for spectral curves of positive genus, then formulate the quantum curve conjecture and discuss the proposed NS and Wilson-loop interpretations.

We fix a defining equation in one of the two forms
$$
H(x,y)=0 \qquad\text{or}\qquad H(e^{-x},e^{-y})=0.
$$
In the first case, $H$ is a polynomial in its two arguments; this
includes hyperelliptic models $y^2=P(x)$, with $P$ a polynomial, and
in particular Weierstrass curves. In the second case, $H$ is a Laurent
polynomial, that is, a finite sum of monomials with integer exponents.
This includes the mirror curves of toric Calabi--Yau threefolds:
$X=e^{-x},Y=e^{-y}$ are coordinates on $(\mathbb C^*)^2$, with
$x=-\log X$ and $y=-\log Y$ locally. The coefficients of $H$
may depend on the closed moduli.

\subsection{Quantization conventions}\label{sec:quantization-conventions}

We take $\widehat x=x\cdot$ and $\widehat y=\hbar\partial_x$,
and use Weyl ordering in both cases.
For a polynomial monomial, this is given by
$$
\widehat{x^a y^b}
=\frac{1}{2^b}\sum_{k=0}^{b}\binom{b}{k}
\widehat y^{\,k}\widehat x^{\,a}\widehat y^{\,b-k},
\qquad a,b\in\mathbb Z_{\geq0}.
$$
For the exponential model, set $\sfq=e^{-\hbar}$,
$\widehat X=e^{-\widehat x}=X\cdot$, and
$\widehat Y=e^{-\widehat y}$.
Then $(\widehat Yf)(X)=f(\sfq^{-1}X)$ and
$\widehat YX=\sfq^{-1}X\widehat Y$, and Weyl ordering gives
$$
\widehat{X^aY^b}
=X^{a/2}\widehat Y^{\,b}X^{a/2}
=\sfq^{-ab/2}X^a\widehat Y^{\,b},
\qquad a,b\in\mathbb Z.
$$
We denote by $\widehat H_0$ the Weyl
quantization of $H$ with the classical coefficients unchanged.
We also allow the moduli-dependent coefficients $\gamma_i(\cq)$
to be replaced by scalar formal series $\cW_i(\cq;\hbar)$ in the quantization,
where $\cq$ denotes the moduli parameter for the closed string. These scalars are
independent of the curve coordinate and satisfy
$\cW_i(\cq;0)=\gamma_i(\cq)$.
When a compatible geometric theory is available, we expect them to
be normalized NS Wilson-loop expectation values.

After the coefficient replacements, we denote the resulting
operator by $\widehat H$ and seek a distinguished specialization of the wave function that it annihilates.

\subsection{The canonical vanishing point}
We consider the NP descendent partition function
$Z^{\rm NP}_{\mu,\nu}(\epsilon,\mathbf p;\hbar;w)$
defined in Definition~\ref{def:NP-partition}.
We choose a characteristic $(\mu_0,\nu_0)$ and a formal point $w_*$ satisfying the closed non-perturbative vanishing equation
\[
Z^{\rm NP}_{\mu_0,\nu_0}(\epsilon,\mathbf 0;\hbar;w_*)=0,
\]
subject to the leading-order condition
\[
w_*^+:=w_*+\frac{\partial_{\boldsymbol\epsilon}F_0}{2\pi i\hbar}
=O(\hbar).
\]
Taking the coefficient of $\hbar^0$ after dividing by the nonzero closed perturbative factor gives
\[
\theta\big[\substack{\mu_0\\ \nu_0}\big](0\mid\tau)=0.
\]
The higher-order terms of the full vanishing equation impose further
conditions on $w_*^+$.

With this characteristic fixed, we write
\[
 Z^{\rm NP}_{\rm cl}(\epsilon;\hbar;w)
 :=Z^{\rm NP}_{\mu_0,\nu_0}(\epsilon,\mathbf0;\hbar;w)
\]
for the closed non-perturbative partition function.
 
For a fixed smooth genus-one curve, the leading
condition selects $(\mu_0,\nu_0)=(\tfrac12,\tfrac12)$ modulo integral
shifts, and the full equation has a unique formal solution
satisfying $w_*^+=O(\hbar)$; see Lemma~\ref{lem:closed-branch-uniqueness}.
In higher genus, additional conditions are needed to determine the specialization. For the construction below, we fix a solution of the closed vanishing equation.
 
We now specialize the closed vanishing equation to genus one.
We take $(\mu_0,\nu_0)=(\tfrac12,\tfrac12)$, the odd characteristic.
The complementary moduli are held fixed, and no complementary shift is
made.
Dividing by the nonzero closed perturbative factor in
Definition~\ref{def:NP-partition}, the closed vanishing equation becomes
\begin{equation}\label{eq:closed-vanishing-theta}
 \exp\left(\sum_{\substack{g\geq0,\,m\geq1\\2g-2+m>0}}
 \frac{\hbar^{2g-2+m}}{m!\, (2\pi i)^m}F^{(m)}_{g,0}({\bf0})
 \partial_{w^{+}}^m\right)
 \theta\!\left[\substack{\frac12\\ \frac12}\right]
 (w^{+}\mid\tau)=0.
\end{equation}
\begin{lemma}\label{lem:closed-branch-uniqueness}
For a smooth genus-one curve, the closed
vanishing equation has a unique formal solution satisfying
$w_*^+=O(\hbar)$.
\end{lemma}
\begin{proof}
The expression on the left of~\eqref{eq:closed-vanishing-theta},  has leading term
$\theta[\substack{1/2\\1/2}](w^{+}\mid\tau)$ at $\hbar=0$.
This theta function has a simple zero at $w^{+}=0$. The formal
implicit function theorem gives a unique solution
$w^{+}\in\hbar\mathbb C[[\hbar]]$.
Every term in the exponent has $\hbar$-degree congruent to $m$ modulo
$2$. Together with the oddness of the theta function, this shows that
the left-hand side changes sign under
$(\hbar,w^{+})\mapsto(-\hbar,-w^{+})$.
Uniqueness therefore implies that the solution is odd in $\hbar$.
\end{proof}

Let
\(
 \eta(\tau)={\bf q}^{\frac{1}{24}}\prod_{n\geq1}(1-{\bf q}^{n})
\)
be the Dedekind eta-function, where ${\bf q}=e^{2\pi i \tau}$. Using the expansion
\[
 \theta\!\left[\substack{\frac12\\ \frac12}\right](z|\tau)
 =-2\pi\eta(\tau)^3 z
 \exp\bigg(\sum_{k\geq1}\frac{B_{2k}E_{2k}(\tau)}{2k(2k)!}(2\pi iz)^{2k}\bigg),
\]
where $B_{2k}$ are the Bernoulli numbers and $E_{2k}$ are the normalized
Eisenstein series, we obtain the leading contribution
$$
w_{*}
=-\frac{\pd_\epsilon F_0}{2\pi i\hbar}
-\frac{\hbar}{2\pi i}\left(\pd_\epsilon F_1+\frac{E_2(\tau)}{24}\pd_\epsilon^3F_0\right)+O(\hbar^3).
$$
This follows directly from the coefficient of $\hbar$ in
\eqref{eq:closed-vanishing-theta}, after setting
$z=w^{+}$.

\subsection{The canonical wave function and canonical quantization}
\label{sec:canonical-wave}

\begin{definition}
For the chosen characteristic $[\ccmn]$ and closed vanishing point
$w_*$, define
\[
 \Psi^a_*(\epsilon,z;\hbar)
 :=\Psi^a_{\mu_0,\nu_0}(\epsilon,z;\hbar;w_*),
 \qquad a\in\{\varnothing,X,{\rm st}\}.
\]
Here, for $a=\varnothing$, $\Psi^{a}$ denotes the normalized wave function in Definition \ref{normalizedpsi}.
In particular, $\Psi^X_*$ is the principal specialization of the NP
partition function $Z^{\rm NP}$ at $w_*$.
\end{definition}

The quantum curve conjecture asks for an operator $\widehat H$
of the form described above such that
$$
\widehat H\Psi_{*}(\epsilon,z;\hbar)=0.
$$
For the genus-one models considered here, the conjecture takes a
scalar form: only the constant term $\gamma_1(\cq)$ is corrected.
We have
\[
 \widehat H=\widehat H_0+\cW^+,
 \qquad
 \cW^+=\cW(\cq;\hbar)-\cW(\cq;0),
\]
where $\cW(\cq;0)=\gamma_1(\cq)$ and $\widehat H_0$ is
the Weyl quantization with unchanged classical coefficients.
 
\begin{conjecture}\label{conj:canonical-quantum-curve}
For the genus-one spectral curves considered in this paper, let
$\Psi_*$ be the canonical wave function. There exists a
scalar formal series $\cW(\cq;\hbar)$, independent of $z$ and
satisfying $\cW(\cq;0)=\gamma_1(\cq)$, such that
\[
 (\widehat H_0+\cW^+)\Psi_*=0.
\]
\end{conjecture}

At $X=0$, principal specialization sets all open variables to zero,
so
\[
 \Psi^X_{\mu,\nu}(\epsilon,0;\hbar;w)
 =Z^{\rm NP}_{\mu,\nu}(\epsilon,\mathbf0;\hbar;w).
\]
The closed vanishing condition therefore gives $\Psi_*^X(0)=0$.
The boundary expansion~\eqref{eqn:Psi_lambda_st} shows that $\Psi^{\rm st}_*$ is
regular at $b$.

For the genus-one models considered below, the first boundary coefficient is nonzero.
In the fixed coordinate $X$, we may therefore write
\[
 \Psi_*^X(X)=K_1X+\frac{K_2}{2}X^2+O(X^3),\qquad K_1\ne0.
\]
Moreover, the scalar $\cW$ is expressed in
terms of the boundary ratio $K_2/(2K_1)$.
The conversion from $K_2/(2K_1)$ may include model-dependent factors
and an additive term; see~\eqref{eqn:weierstrass-wilson-loop},
\eqref{eqn:localp2-Wilson-coefficient},
and~\eqref{eqn:one-loop-TR-Wilson-coefficient}.

When an NS Wilson loop is independently defined, we conjecture that
its expectation value, with the model's prescribed normalization,
equals the scalar $\cW$ in Conjecture~\ref{conj:canonical-quantum-curve}.
The geometric definition of NS Wilson loop and the local $\mathbb P^2$ comparison
are given in Section~\ref{sec:localp2-wilson}.

\subsection{The vanishing point and the closed NS potential}\label{sec:canonical-NS}
Conjecture~\ref{conj:NS-vanishing-point} relates the closed vanishing
point $w_*$ to the filling-fraction derivatives of the closed NS potential.
We specify the NS normalization and
discuss the vanishing-point identity~\eqref{eqn:w_can-NS} here.

For a compatible closed refined free energy $F^{\rm ref}(\ve;\eta,\hbar)$ with equivariant parameters $\eta,\hbar$, we define its NS limit by
$$
	F^{\rm NS}(\ve;\hbar)
	:=\lim_{\eta\to0}\eta\,F^{\rm ref}(\ve;\eta,\hbar)
	=\sum_{g\geq 0}\hbar^{2g-1}F^{\rm NS}_{g}(\ve).
$$
We normalize the classical term by $F_0^{\rm NS}=-F_0$.
For smooth quasi-projective Calabi--Yau threefolds $X$,
Brini--Schuler~\cite{BS24} construct a refinement of GW theory by using equivariant stable maps to the fivefold $X\times\mathbb C^2$.
A refined theory compatible with our non-perturbative construction
is not known for a general spectral curve.
When no refined theory is specified, we conjecture that $\log K_1$
is an even Laurent expansion beginning with $F_0/\hbar^2$, and define
$F^{\rm NS}$ by~\eqref{eqn:def_NS}, with vanishing $\hbar^0$ coefficient.
An $\hbar$- and moduli-independent rescaling of $X$ changes only
the additive constant in $F_1^{\rm NS}$.

For the local del Pezzo theories of~\cite{FGZblowup}, the inverse
local--relative correspondence and the identification of relative
and NS potentials in~\cite[Corollary~4.2]{BS24} give,   after matching the
full closed potentials, filling direction, coupling, and periods,
\[
 \sum_{l\in\mathbb Z}(-1)^l
 \exp\!\left(\left(l+\tfrac12\right)\partial_\epsilon F^{\rm NS}\right)
 {Z_{\rm cl}}\!\left(\epsilon+\hbar\left(l+\tfrac12\right);\hbar\right)=0.
\]
This is~\eqref{eqn:Z-NP} at $\mu=\nu=\tfrac12$ and
$2\pi iw=\partial_\epsilon F^{\rm NS}$, with the common factor $i$
removed. Since $F_0^{\rm NS}=-F_0$, this point satisfies $w^+=O(\hbar)$.
The uniqueness in Lemma~\ref{lem:closed-branch-uniqueness}
now identifies this point with $w_*$, completing the proof of
Conjecture~\ref{conj:NS-vanishing-point} in this setting.

$$
$$

\section{The Weierstrass quantum curve}\label{sec:weierstrass}
As a warm-up, we first consider the Weierstrass curve, the simplest
genus-one example of the construction in
Section~\ref{sec:canonical-quantum-curve}. Its spectral data are
$$
\mathcal C=(\Sigma=\mathbb C/(\mathbb Z+\tau\mathbb Z),\quad
x(z)=r^2\wp(z;\tau),\quad 
y(z)=r^3\wp'(z;\tau)).
$$
We prove its canonical quantum curve equation and relate the closed
vanishing point to the NS potential defined by~\eqref{eqn:def_NS}.

\subsection{Topological recursion of Weierstrass curve and its properties}
We first introduce some notations.
For $k\geq 1$, let $G_{2k}(\tau)$ denote the weight-$2k$ Eisenstein series
$$
G_{2k}(\tau)=\sum_{(m,n)\in\mathbb Z^2\setminus\{(0,0)\}}\frac{1}{(m+n\tau)^{2k}}.
$$
We denote $g_2(\tau)=60G_4(\tau)$ and $g_3(\tau)=140G_6(\tau)$.
Then it is well-known that $x=x(z)$ and $y=y(z)$ satisfy the following algebraic relation
\beq\label{eqn:wp-relation}
y^2=4x^3-r^4g_2(\tau)x-r^6g_3(\tau).
\eeq

Start from the spectral curve data $\cC$, we have immediately 
$$
\omega_{0,1}(z)=y(z)dx(z),\qquad
\omega_{0,2}(z_1,z_2)=B(z_1,z_2).
$$
Here, the Bergman kernel of the genus-one curve is given by 
$$
B(z_1,z_2)=\big(\wp(z_1-z_2;\tau)+G_2(\tau)\big)dz_1dz_2.
$$
Furthermore, the differential $dx(z)$ has zeros $\frac{1}{2}$, $\frac{\tau}{2}$, and $\frac{1+\tau}{2}$, and the local involution $\bar z$ is given by $-z$ modulo the periods $\mathbb Z+\tau\mathbb Z$.
By the topological recursion procedure~\eqref{eqn:def-TR} and \eqref{def:wg0}, one obtains stable differentials 
$$
\omega_{g,n}(z_1,\cdots,z_n),\qquad 2g-2+n>0.
$$
For remaining two unstable terms, let 
\be
\epsilon=\frac{1}{2\pi i}\oint_{\cyA} ydx =-\frac{r^5}{10}D(g_2),\qquad
t=-\frac{r^4g_2}{2},
\ee
where $D=\frac{d}{d\tau}$. Following Iwaki~\cite{Iwa20}, we define
\be
F_0(\epsilon,t)=(2\pi i \tau)\cdot \frac{\epsilon^2}{2}+\frac{t}{5}\cdot (4\pi i r \epsilon- r^6 g_3),
\ee
and
\be
F_1(\epsilon,t)=-\frac{1}{12}\log(r^6\cdot (g_2^3-27g_3^2)).
\ee
We note here that our $\epsilon$ is denoted by $\nu$ in~\cite{Iwa20}, and we have $r=\frac{1}{\omega_A},\tau=\frac{\omega_{B}}{\omega_A}$, where $\omega_{A,B}$ are periods used by Iwaki.

The multi-differentials defined by the topological recursion satisfy the following homogeneity conditions.
Fixing the moduli parameter $\tau$ and rescaling $r$ by $r\mapsto ar$, we have
$\epsilon\mapsto a^5\epsilon$, $t\mapsto a^4 t$, and for $g,n\geq 0$ and $(g,n)\notin\{(0,0),(1,0)\}$,
\begin{equation*}
	\omega_{g,n}\longmapsto a^{-5(2g-2+n)}\omega_{g,n}.
\end{equation*}
The two unstable free energies satisfy
$$
F_0(a^5\epsilon,a^4t)=a^{10}F_0(\epsilon,t),\qquad
F_1(a^5\epsilon,a^4t)=F_1(\epsilon,t)-\frac12\log a.
$$

The function $x$ has a double pole at $z=0$. 
We choose the local coordinate $X\sim \frac{z}{r}$ near $z=0$ by $x=X^{-2}$, 
and define $Z(\epsilon,t,\mathbf p;\hbar)$ and $Z^{\rm NP}_{\mu,\nu}(\epsilon,t,\mathbf p;\hbar;w)$ as in \S \ref{sec:TR}.
By~\cite{GJZ23}, we know that $Z^{\rm NP}_{\mu,\nu}$ is a tau-function of KdV hierarchy.
On the other hand, the calibrated descendent theory~\cite{GZ25} associated with the Weierstrass curve is described in~\cite[\S 5.4]{GZ25b}. While $r$ is normalized to be $1$ in~\cite{GZ25b}, generic $r$ can be recovered by the homogeneity condition.
Denote by $\mathcal D(\epsilon,t,\mathbf t;\hbar)$ the calibrated descendent potential, where ${\bf t}=\{t_k^i\}_{i=0,1,2; k\geq 0}$ (see~\cite{GZ25b} for more details).
We have 
$\mathcal D(\epsilon,t,\hbar\mathbf t(\mathbf p);\hbar)=Z(\epsilon,t,\mathbf p;\hbar)$, where ${\bf t(p)}$ is given by $t_k^0=t_k^2=0$ and $t_k^1=\frac{(2k-1)!!}{2^{k+1/2}}p_{2k+1}$ for $k\geq 0$.
Moreover, $\cD$ depends on $\epsilon,t$ through $t_0^0+\epsilon$ and $t_0^1+t/\sqrt2$.
Since $t_0^1=p_1/\sqrt2$, we see $\hbar\partial_tZ=\partial_{p_1}Z$.
By equation~\eqref{eqn:Z-NP}, we obtain
\begin{equation}\label{eqn:ddt-ddp}
	\hbar\partial_t Z^{\rm NP}_{\mu,\nu}
	(\epsilon,t,\mathbf p;\hbar;w)
	=\partial_{p_1}Z^{\rm NP}_{\mu,\nu}
	(\epsilon,t,\mathbf p;\hbar;w).
\end{equation}

\subsection{Weierstrass canonical quantum curve}
\label{sec:weierstrass-canonical-curve}

For the Weierstrass curve, $u(z)=z$ and
$\mathbf c=\tfrac12+\tfrac12\tau$ is the unique odd characteristic.
In our normalization (obtained by dividing $\Psi_{\mu,\nu}$ in \S \ref{sec:quantum_curve} by a $z$-independent factor $\sqrt{\theta'[{\bf c}](0|\tau)}$ ), the wave function 
factorizes as
\[
\Psi_{\mu,\nu}(\epsilon,t,z;\hbar;w)
={Z_{\rm cl}}(\epsilon,t;\hbar)\,
\frac{\exp(\frac{1}{\hbar}\int^z y\,dx)}{\sqrt{x'(z)}}\,
\Psi^{\rm st}_{\mu,\nu}(\epsilon,t,z;\hbar;w).
\]
Here ${Z_{\rm cl}}(\epsilon,t;\hbar)=Z(\epsilon,t,\mathbf0;\hbar)$ is the closed
perturbative partition function, and $\Psi^{\rm st}_{\mu,\nu}$ is the
stable theta factor defined there, with the shifted argument $w^+$
of~\eqref{def:w-tilde}.

	Before proving the canonical quantum curve conjecture, 
	we first explain the choice of characteristic, closed zero, and scalar correction by taking certain local expansion.
	For $w^+=O(\hbar)$ and $\cW^+=O(\hbar^2)$, consider
	\begin{equation}\label{eqn:qc-Psi}
		\left[\hbar^2\partial_x^2-(4x^3-r^4g_2x-r^6g_3)+\cW^+\right]
		\Psi_{\mu,\nu}.
	\end{equation}
	Dividing by ${Z_{\rm cl}}(\epsilon,t;\hbar)e^{\hbar^{-1}\int^z ydx}/\sqrt{x'(z)}$, this expression becomes
	\begin{equation}\label{eqn:qc-Psi-st}
		\left[\left(\hbar\partial_x+y-\frac{\hbar x''(z)}{2x'(z)^2}\right)^2
		-(4x^3-r^4g_2x-r^6g_3)+\cW^+\right]\Psi^{\rm st}_{\mu,\nu}.
	\end{equation}
	Its classical term vanishes by~\eqref{eqn:wp-relation}.
	Since $\Psi^{\rm st}_{\mu,\nu}=\theta[\cmn](z|\tau)/\theta[{\bf c}](z|\tau)
	+O(\hbar)$, the coefficient of $\hbar$ is
	$2y\partial_x\bigl(\theta[\cmn](z|\tau)/\theta[{\bf c}](z|\tau)\bigr)$.
	Its vanishing requires $\mu+\tau\nu={\bf c}=\tfrac12+\tfrac{\tau}{2}$.
	
	For this characteristic, $\theta[{\bf c}](z|\tau)
	=\theta'[{\bf c}](0|\tau)z+O(z^3)$. Write
	\[
	\Psi^{\rm st}_{\bf c}=\frac{a_{-1}}z+a_0+a_1z+O(z^2),\qquad
	a_{-1}=\frac{Z^{\rm NP}_{\rm cl}(\epsilon,t;\hbar;w)}
	{\theta'[{\bf c}](0|\tau){Z_{\rm cl}}(\epsilon,t;\hbar)}.
	\]
	Then~\eqref{eqn:qc-Psi-st} has expansion
	\[
	-\frac{2r\hbar a_{-1}}{z^2}+\frac{\cW^+a_{-1}}z
	+2r\hbar a_1+\cW^+a_0+O(z).
	\] Thus its vanishing forces $a_{-1}=0$.
	Under the condition $w^+=O(\hbar)$, this is equivalent to $w=w_*$
	by Lemma~\ref{lem:closed-branch-uniqueness}.
	
	At the point $w=w_*$, $a_0=1+O(\hbar)$, so the constant term gives
	\[
	\cW^+=-2r\hbar\frac{a_1}{a_0}
	=-\left.2r\hbar\partial_z\log\Psi^{\rm st}_*\right|_{z=0}.
	\]
	Since $z=rX+O(X^5)$, we have $\omega_{0,1}=\omega_{0,1}^{\rm sing,b}-\frac{r^6g_3}{2}dX+dO(X^3)$.
	Equation~\eqref{eqn:Psi_lambda_st} then gives
	\begin{equation}\label{eqn:weierstrass-wilson-loop}
		\cW(\epsilon,t;\hbar)=r^6g_3+\cW^+ =-2\hbar\partial_X\log\tfrac{\Psi^X_*}{X}\big|_{X=0}.
	\end{equation}

We now prove that these necessary choices are sufficient.
\begin{theorem}\label{thm:quantum_weierstrass}
	We have the following canonical quantum curve equation:
	$$
	\bigg[\hbar^2\frac{\pd^2}{\pd x^2}-\Big(4x^3-r^4g_2x-\cW(\epsilon,t;\hbar)\Big)\bigg] \Psi_{*}(\epsilon,t,z;\hbar) =0,
	$$
	where $\cW(\epsilon,t;\hbar)=-2\hbar\partial_X\log\tfrac{\Psi^X_*}{X}\big|_{X=0}$.
\end{theorem}
\begin{proof}
	According to~\cite{Iwa20}, the wave function $\Psi_{\mu,\nu}$ satisfies the following equation
	~\footnote{Comparing with Iwaki's notation, we have $r=\frac{1}{\omega_{A}}$, $\tau=\frac{\omega_{B}}{\omega_A}$, and we identify Iwaki's $\rho$ (resp. $\nu$) with our $\hbar (w+\nu)$ (resp. $\epsilon$).
		Then the equation can be deduced by acting $e^{\hbar\mu\pd_\epsilon}$ on equation (1.12) in~\cite{Iwa20}.}:
	\beq\label{eqn:diff-Psi}
	\bigg[\hbar^2\frac{\pd^2}{\pd x^2}
	-\frac{\hbar^2}{x-\mathcal H_2}\bigg(\frac{\pd}{\pd x}-\mathcal H_3\bigg)
	-\Big(4x^3-r^4g_2x-2\mathcal H_1\Big)\bigg] \Psi_{\mu,\nu}(\epsilon,t;x;\hbar;w) =0,
	\eeq
	where for $k=1,2,3$, $\mathcal H_{k}$ is given by
	$\mathcal H_{k}=-\hbar^2{\pd_{t}^k \log Z^{\rm NP}_{\mu,\nu} (\epsilon,t,\mathbf0;\hbar;w)}$.
	We take $\mu=\nu=\frac{1}{2}$ and $w=w_{*}$ in equation~\eqref{eqn:diff-Psi}.
	Recall that by definition,
	$$
	Z^{\rm NP}_{\rm cl}(\epsilon,t;\hbar;w_{*})=0.
	$$
	To trace the dependence of $w_{*}$ on the variable $t$, we write $w_{*}(t)$.
	By using~\eqref{eqn:ddt-ddp} and the independence of $p_2$ of $Z^{\rm NP}$, 
	at $(\epsilon,t,\mathbf p=\mathbf0;\hbar;w_*)$ we have
	\[
	\hbar^i\partial^i_tZ^{\rm NP}_{\rm cl}
	=\partial_{p_1}^iZ^{\rm NP}
	=\left.\partial^i_X\Psi^X_*\right|_{X=0}
	=K_i,\qquad i=1,2.
	\]
	Therefore, we have expansion
	$$
	Z^{\rm NP}_{\rm cl}(\epsilon,t+s;\hbar;w_{*}(t))=\frac{K_1}{\hbar}s+\frac{K_2}{2\hbar^2}s^2+O(s^3).
	$$
	The nonvanishing of \(K_1\) follows from \(a_0=1+O(\hbar)\) and \eqref{eqn:Psi_lambda_st}.
	By taking derivative of $\log(Z^{\rm NP}_{\rm cl}(\epsilon,t+s;\hbar;w_{*}(t)))$ with respect to $s$, we get
	$$
	\mathcal H_1=-\frac{\hbar^2}{s}-\frac{\hbar}{2}\frac{K_2}{K_1}+O(s).
	\qquad
	\mathcal H_2=\frac{\hbar^2}{s^2}+O(1),
	\qquad
	\mathcal H_3=-\frac{2\hbar^2}{s^3}+O(1).
	$$
	This gives
	$$
	-\frac{\hbar^2}{x-\mathcal H_2}\bigg(\frac{\pd}{\pd x}-\mathcal H_3\bigg)+2\cH_1
	=-\hbar\frac{K_2}{K_1}+O(s).
	$$
	Taking $s\to 0$, we get
	\be
	\bigg[\hbar^2\frac{\pd^2}{\pd x^2}-\Big(4x^3-r^4g_2x+\hbar\frac{K_2}{K_1}\Big)\bigg] \Psi_{*}(\epsilon,t;x;\hbar) =0.
	\ee
	The Theorem then follows from 
	\(
	\hbar\frac{K_2}{K_1}
	=2\hbar\partial_X\log\frac{\Psi^X_*}{X}\big|_{X=0}
	=-\cW(\epsilon,t;\hbar).
	\)
\end{proof}

\subsection{Nekrasov--Shatashvili limit}\label{sec:weierstrass-NS}
For the Weierstrass curve, we define $F^{\rm NS}$ by
$\partial_\hbar F^{\rm NS}=\log K_1$, as in~\eqref{eqn:def_NS}, and prove
Conjecture~\ref{conj:NS-vanishing-point} for this potential.

We first do some preparation.
By~\cite[Theorem~4.3]{Iwa20}, we have known that $Z^{\rm NP}_{\mu,\nu}(\epsilon,t;\hbar;w)$ is a formal tau-function of the first Painlev\'e equation.
We now determine its full logarithmic differential with respect to $(\epsilon,t,w)$, in the form of the extended tau-function differential studied in~\cite[\S 3.1]{LR17}.
Precisely, we have the following lemma:

\begin{lemma}
	Let $\mathcal H_{k}=-\hbar^2{\pd_{t}^k \log Z^{\rm NP}_{\mu,\nu} (\epsilon,t;\hbar;w)}$, $k=1,2,3$.
	We have
	\begin{equation}\label{eqn:weierstrass-closed-full-variation}
		d\log Z^{\rm NP}_{\mu,\nu}(\epsilon,t;\hbar;w)
		=\frac{\mathcal H_1}{5\hbar^2}\,dt
		-\frac{4t}{5\hbar^2}\,d\mathcal H_1
		+\frac{3}{5}\mathcal H_3\,d\mathcal H_2
		-\frac{2}{5}\mathcal H_2\,d\mathcal H_3
		-\frac{2\pi i\epsilon}{\hbar}\,dw.
	\end{equation}
	Here, $\mu,\nu$, and $\hbar$ are viewed as constants.
\end{lemma}
\begin{proof}
	Proving \eqref{eqn:weierstrass-closed-full-variation} amounts to determining the partial derivatives of \(\log Z^{\mathrm{NP}}_{\mu,\nu}(\epsilon,t;\hbar;w)\) with respect to \(t\), \(\epsilon\), and \(w\). 
	Iwaki's results provide the necessary tools for computing the derivatives with respect to \(t\).
	In fact, according to~\cite[Theorem~4.3]{Iwa20}, we have
	$$
	\hbar^2\pd_t\mathcal H_3=6\mathcal H_2^2+t,\qquad
	-\mathcal H_1=\tfrac{\hbar^2}{2}\mathcal H_3^2-2\mathcal H_2^3-t\mathcal H_2.
	$$
	Let $\Omega$ denote the right-hand side of~\eqref{eqn:weierstrass-closed-full-variation}.
	Then by using these two equations, we have
	\[
	d\log Z^{\rm NP}_{\mu,\nu}-\Omega
	=R_\epsilon\,d\epsilon+R_w\,dw,
	\]
	for some functions $R_{\epsilon}$ and $R_{w}$ satisfying $\pd_tR_\epsilon=\pd_tR_w=0$.
	
	We next prove that $R_{\epsilon}$ and $R_{w}$ do not depend on $w$ and $\epsilon$ by using the periodicity.
	Firstly, according to the theta representation~\eqref{def:Z-NP-theta}, for $\star=\epsilon,w$, $R_{\star}$ has an expansion:
	\[
	R_\star=\sum_{n\geq-2}\hbar^nR_{\star,n}(\epsilon,t;w^+).
	\]
	By using $Z^{\rm NP}_{\mu,\nu}(\epsilon,t;\hbar;w+1)=e^{2\pi i\mu}Z^{\rm NP}_{\mu,\nu}(\epsilon,t;\hbar;w)$, 
	we see that $d\log Z^{\rm NP}_{\mu,\nu}$ and the $\mathcal H_k$ are $1$-periodic in $w$.
	The same holds for $\Omega$, and hence for $R_\star$.
	By comparing coefficients, we have
	$$
	R_{\star,n}(\epsilon,t;w^++1)=R_{\star,n}(\epsilon,t;w^+).
	$$
	On the other hand, by using $\pd_tR_\star=0$ and $\pd_tw^+=r/\hbar$, we have
	\[
	\pd_{w^+}R_{\star,n}=-\tfrac{1}{r}\left.\pd_tR_{\star,n-1}\right|_{w^+}.
	\]
	Set $R_{\star,-3}=0$.
	Suppose that $n\geq-2$ is the smallest integer such that $\pd_{w^+}R_{\star,n}\not\equiv0$.
	Then $R_{\star,n-1}$ is independent of $w^+$, so the right-hand side of the preceding identity is constant with respect to $w^+$.
	Since $R_{\star,n}$ is $1$-periodic, this constant derivative must vanish, which contradicts the choice of $n$.
	This proves that $R_{\epsilon}$ and $R_{w}$ do not depend on $w$. 
	Similarly, by using 
	\(
	Z^{\rm NP}_{\mu,\nu}(\epsilon+\hbar,t;\hbar;w)
	=e^{-2\pi i(w+\nu)}Z^{\rm NP}_{\mu,\nu}(\epsilon,t;\hbar;w),
	\)
	one can prove that $R_{\epsilon}$ and $R_{w}$ are also independent of $\epsilon$.
	
	Finally, by using the homogeneity condition
	\(
	Z^{\rm NP}_{\mu,\nu}(a^5\epsilon,a^4t;a^5\hbar;w)
	=a^{-1/2}Z^{\rm NP}_{\mu,\nu}(\epsilon,t;\hbar;w)
	\)
	and by a similar argument, we have $R_\epsilon(a^5\hbar)=a^{-5}R_\epsilon(\hbar)$ and $R_w(a^5\hbar)=R_w(\hbar)$ (since $R_{\epsilon}d\epsilon$ and $R_{w}dw$ are unchanged).
	This implies
	\[
	R_\epsilon=\frac{c_\epsilon}{\hbar},\qquad R_w=c_w,
	\]
	By expanding \(d\log Z^{\mathrm{NP}}_{\mu,\nu}-\Omega\) in powers of \(\hbar\) and comparing the coefficients of \(\hbar^{-1}\) and \(\hbar^0\), we obtain \(c_\epsilon=c_w=0\). 
	The lemma is proved.
\end{proof}

\begin{theorem}\label{thm:weierstrass-NS}
	Let $F^{\rm NS}=F^{\rm NS}(\epsilon,t;\hbar)$ be the formal Laurent series defined by~\eqref{eqn:def_NS} and normalized to have zero constant term in \(\hbar\).
	Then
	\[
	w_{*}=\frac{1}{2\pi i}\pd_\epsilon F^{\rm NS},
	\qquad
	\cW=2\hbar\pd_tF^{\rm NS}.
	\]
	Recall here that $w_{*}$ is defined by $Z^{\rm NP}_{\rm cl}(\epsilon,t;\hbar;w_{*})=0$ and $\cW=-2\hbar\partial_X\log\tfrac{\Psi^X_*}{X}\big|_{X=0}$.
\end{theorem}

\begin{proof}
	We first prove the following identity:
	\begin{equation}\label{eqn:weierstrass-NS-jet-variation}\textstyle
		d\log K_1
		=-\frac{1}{10\hbar}\frac{K_2}{K_1}\,dt
		+\frac{2t}{5\hbar}\,d\big(\frac{K_2}{K_1}\big)
		-\frac{2\pi i\epsilon}{\hbar}\,dw_{*}.
	\end{equation}
	In fact, by the expansion
	$Z^{\rm NP}_{\rm cl}\bigl(\epsilon,t+s;\hbar;w_{*}\bigr)=\frac{K_1}{\hbar}s+\frac{K_2}{2\hbar^2}s^2+O(s^3)$,
	we have
	\[\textstyle
	d\log Z^{\rm NP}_{\rm cl}=d\log K_1+O(s),\qquad {\text{and}}\qquad
	\mathcal H_1=-\frac{\hbar^2}{s}
	-\frac{\hbar}{2}\frac{K_2}{K_1}+O(s).
	\]
	By the Painlev\'e equation, we have further 
	\[\textstyle
	\mathcal H_1=-\frac{\hbar^2}{s}
	-\frac{\hbar}{2}\frac{K_2}{K_1}
	-\frac{t}{30\hbar^2}s^3
	-\frac{1}{24\hbar^2}s^4+O(s^5).
	\]
	By differentiating with respect to \(s\), we have
	\[\textstyle
	\mathcal H_2=\frac{\hbar^2}{s^2}-\frac{t}{10\hbar^2}s^2-\frac{1}{6\hbar^2}s^3+O(s^4),\qquad
	\mathcal H_3=-\frac{2\hbar^2}{s^3}-\frac{t}{5\hbar^2}s-\frac{1}{2\hbar^2}s^2+O(s^3).
	\]
	Substituting these expansions into~\eqref{eqn:weierstrass-closed-full-variation} and then taking $s\to 0$ (singular terms cancel with each other in computations), we get~\eqref{eqn:weierstrass-NS-jet-variation}.
	
	By using equation~\eqref{eqn:weierstrass-NS-jet-variation} and $dd\log K_1=0$, we have
	\[\textstyle
	\pd_\epsilon\big(\frac{K_2}{K_1}\big)
	=-4\pi i\pd_t w_{*}.
	\]
	Further by using the homogeneity properties:
	$$\textstyle
	\big(\frac{2t}{5}\pd_{t}+\frac{\epsilon}{2}\pd_{\epsilon}+\frac{\hbar}{2}\pd_{\hbar}\big)w_{*}=0,
	\qquad
	\big(\frac{2t}{5}\pd_{t}+\frac{\epsilon}{2}\pd_{\epsilon}+\frac{\hbar}{2}\pd_{\hbar}\big)\frac{K_2}{K_1}=\frac{1}{10}\frac{K_2}{K_1},
	$$
	we obtain
	$$\textstyle
	\pd_t\log K_1
	=-\frac{1}{10\hbar}\frac{K_2}{K_1}+\frac{2t}{5\hbar}\pd_t\big(\frac{K_2}{K_1}\big)+\frac{\epsilon}{2\hbar}\pd_\epsilon\big(\frac{K_2}{K_1}\big)
	=-\pd_\hbar\big(\frac{1}{2}\frac{K_2}{K_1}\big),
	$$
	and
	$$\textstyle
	\pd_\epsilon\log K_1
	=-\frac{2\pi i}{5\hbar}\left(4t\pd_t w_{*}+5\epsilon\pd_\epsilon w_{*}\right)
	=2\pi i\pd_\hbar w_{*}.
	$$
	By using $\pd_\hbar F^{\rm NS}=\log K_1$, we have
	\[\textstyle
	\pd_\hbar\big(\pd_tF^{\rm NS}+\frac{1}{2}\frac{K_2}{K_1}\big)=0,
	\qquad
	\pd_\hbar\left(\pd_\epsilon F^{\rm NS}-2\pi i w_{*}\right)=0.
	\]
	Since $w_{*}$ and $K_2/K_1$ are odd in $\hbar$,
	with the convention that $F^{\rm NS}$ is a Laurent series in $\hbar$ with vanishing constant term,
	we obtain the desired equations.
\end{proof}

$$
$$

\section{The local $\mathbb P^2$ quantum curve}\label{sec:localp2}

We prove the canonical quantum curve equation for the mirror curve of local \(\mathbb P^2\). 
After specifying the spectral curve data, we reduce the quantum curve equation to a relation among the Schur coefficients of the perturbative TR partition function.
We prove this relation in the next section by using the topological vertex.

\subsection{Geometric setup}
Let $\mathcal X$ be a toric Calabi--Yau threefold and $L$ an outer
Aganagic--Vafa brane with framing $f$. We use the geometric
definitions and conventions of~\cite{FLZ20}: $N^{\mathcal X}_{g,\beta}$
denotes a closed GW invariant, and
$N^{\mathcal X,L}_{g,\beta;\mu_1,\ldots,\mu_n}$ an open invariant
with positive boundary winding numbers $\mu_1,\ldots,\mu_n$.
For $g,n\geq 0$, we define the correlators by
\begin{equation}\label{eqn:corr-GW}
	\<\alpha_{\mu_1},\cdots,\alpha_{\mu_n}\>_{g,n}^{\mathcal X,L}:=
	\bigg(\prod_{i=1}^n\mu_i\bigg)\cdot\sum_{\beta}N_{g,\beta;\mu_1,\ldots,\mu_n}^{\mathcal X,L}Q^\beta.
\end{equation}

We consider the local $\mathbb P^2$ case.
Let $\mathcal X=K_{\mathbb P^2}=\operatorname{Tot}\bigl(\mathcal O_{\mathbb P^2}(-3)\bigr)$ be the total space of the vector bundle $\mathcal O_{\mathbb P^2}(-3)\to \mathbb P^2$,
and $T=(\mathbb C^*)^3$ acts diagonally on the homogeneous coordinates of $\mathbb P^2$.
We use the convention that $-\lambda_i$ is the equivariant first Chern class of the $i$-th standard character.
We choose the outer brane $L$ on the noncompact toric leg given by the fiber over $p_2=[0:1:0]$.
The torus action on $T_{p_2}\mathcal X$ has weights
$-3\lambda_2,\lambda_2-\lambda_3,\lambda_2-\lambda_1$,
corresponding respectively to the fiber direction and
the two directions in $\mathbb P^2$ towards $p_3$ and $p_1$.
With this ordering, the framing-zero specialization is $\lambda_2=\lambda_3$ (see \cite[\S 3.13]{FLZ20}).
We take the equivariant parameters as
\(
(\lambda_1,\lambda_2,\lambda_3)=(-\frac23,\frac13,\frac13).
\)

\subsection{Mirror curve}\label{sec:localp2-TR}
The mirror curve of local $\mathbb P^2$ model is given by
\beq\label{eqn:localp2-mirror-curve}
X^{-1}+X^{-1}Y+X^2Y^{-1}+q^{-1/3}=0,
\eeq
and the spectral curve data of the topological recursion is given by
$$
\mathcal C^{K_{\mathbb P^2}}=\big(\mathbb C/(\mathbb Z+3\tau\mathbb Z),\quad x=-\log(X),\quad y(z)=-\log(Y)\big).
$$
The variables $Q$, ${\cq}$, and $\tau$ are related to each other in the following way.
Let $I({\bf \cq},z)$ be the $I$-function for the GW theory of local $\mathbb P^2$, precisely,
\[
I(\cq,z)=\sum_{k=0}^2I_k(\cq){\rm H}^k z^{1-k}
=z e^{{\rm H}\log\cq/z}\sum_{d\geq 0}\cq^d\frac{\prod_{k=0}^{3d-1}(-3{\rm H}-kz)}{\prod_{k=1}^{d}({\rm H}+kz)^3}.
\]
In particular, we have $I_0(\cq)=1$. The variables $\cq$ and $Q$ are related by the mirror map
$$
\log(Q)=I_1(\cq)=\log\cq+\sum_{d\geq 1}\frac{(-1)^d(3d)!}{d(d!)^3}\cq^d.
$$
Introduce $I_{11}=\cq\frac{d I_1(\cq)}{d\cq}$ and $I_{12}=\cq\frac{d I_2(\cq)}{d\cq}$, then 
\begin{equation}\label{eqn:localp2-modular-parameter}
\tau=\frac{1}{2}+\frac{1}{2\pi i}\frac{I_{12}(\cq)}{I_{11}(\cq)}.
\end{equation}
We refer readers to~\cite{BFGW21,FRZZ19} for more discussions.
In what follows, we will use notation
$$
t=-\log(Q)=-I_1(\cq).
$$

Let $-3\phi=\cq^{-1/3}$ and $\kappa=\frac{\sqrt{-1}}{2^{{4}/{3}}\pi}\frac{3\phi}{I_{11}}$,
then the spectral curve has the following parameterization:
$$
X= -4^{1/3}\kappa^2 \wp(z;3\tau) +\frac{3}{4}\phi^2,
\qquad
Y= \kappa^3 \wp'(z;3\tau)-\frac{1-3\phi X}{2}.
$$
From the spectral curve data, one has immediately 
$$
\omega_{0,1}(z)=ydx=\log(Y)\frac{dX}{X},\qquad
\omega_{0,2}=\big(\wp(z_1-z_2;3\tau)+G_2(3\tau)\big)dz_1dz_2.
$$
The topological recursion procedure~\eqref{eqn:def-TR} and~\eqref{def:wg0} determines $\omega_{g,n}$ for $2g-2+n>0$. 
By the genus-zero and genus-one mirror theorem~\cite{FLZ20}, we choose the normalization
\[
F_0=-\frac{t^3}{18}+\frac{\pi i t^2}{6}-\sum_{d\geq1}N^{K_{\mathbb P^2}}_{0,d}Q^d,\qquad
F_1=\frac{t}{12}+\sum_{d\geq1}N^{K_{\mathbb P^2}}_{1,d}Q^d.
\]

At the boundary point $(X,Y)=(0,-1)$, we define $\langle-\rangle^{\rm TR}_{g,n}$ as introduced in \S \ref{sec:TR}, where we have $\omega_{0,1}^{\mathrm{sing},b}=\pi i\,\frac{dX}{X}$. 
The open and closed mirror theorem~\cite{BKMP,EO15,FLZ20} reads
\begin{equation*}
	\langle\alpha_{\mu_1},\ldots,\alpha_{\mu_n}\rangle^{\rm TR}_{g,n}
	=(-1)^{g-1+n}Q^{-|\mu|/3}
	\langle\alpha_{\mu_1},\ldots,\alpha_{\mu_n}\rangle^{\mathcal X,L}_{g,n},
\end{equation*}
for $(g,n)\ne (0,0), (1,0)$. Here $|\mu|=\mu_1+\cdots+\mu_n$.

\subsection{Canonical vanishing point}
We first consider the filling fraction $\epsilon=\frac{1}{2\pi i}\int_{\cyA}\omega_{0,1}(z)$.
Fixing $X$ as an independent variable with respect to the moduli variable $Q$ (equivalently, ${\cq},\tau$), we have
$$
\frac{\pd \omega_{0,1}(z)}{\pd \log\cq}
=-\frac{\phi \cdot dX}{2Y+1-3\phi X}
=-\frac{2\pi i}{3}I_{11}dz.
$$
By using $d\log Q=I_{11}d\log \cq$, we see that the filling fraction satisfies $\pd_t\epsilon=\frac13$.
Therefore, up to a constant with respect to $t$, we define
$$\epsilon=\frac{t}{3}.$$
We denote by $Z(\epsilon,{\bf p};\hbar)$ and $Z^{\rm NP}_{\mu,\nu}(\epsilon,{\bf p};\hbar;w)$
the perturbative and non-perturbative partition functions, respectively.
We note here that the non-perturbative invariants have no geometric explanation.

For the closed non-perturbative partition function $Z^{\rm NP}_{\mu,\nu}(\epsilon,{\bf 0};\hbar;w)$, we have by definition
$$
Z^{\rm NP}_{\mu,\nu}(\epsilon,{\bf 0};\hbar;w)=e^{\frac{1}{\hbar^2}F_0+F_1}
e^{\sum_{2g-2+m>0}\frac{\hbar^{2g-2+m}}{m!(2\pi i)^m}F_g^{(m)}{\pd_w^m}}\theta[\cmn](w^{+}|3\tau).
$$
Here $F_g^{(m)}=\pd_\epsilon^mF_g$ and $\pd_\epsilon=-3Q\pd_Q$.
Take $\mu=\nu=\frac{1}{2}$.
\begin{proposition}[\cite{FGZblowup}]\label{prop:w-NS}
	The vanishing point $w=w_{*}$ of ${Z^{\rm NP}_{\rm cl}(\epsilon;\hbar;w)}$ with $w_{*}^{+}=O(\hbar)$ is unique and satisfies
	$$
	w_{*}^{+}=\frac{3}{2\pi i}\sum_{g\geq1}\hbar^{2g-1}\pd_tF_g^{\mathbb P^2/E}(Q).
	$$
	Here, $F_g^{\mathbb P^2/E}(Q)$ is the generating series of relative GW invariants of pair $(\mathbb P^2,E)$ where $E\in |-K_{\mathbb P^2}|$ (see~\cite{BFGW21} or ~\cite{FGZblowup}).
	According to~\cite{BS24}, it is the NS limit of the refined GW potential of local $\mathbb P^2$.
\end{proposition}
\begin{remark}
This Proposition was proved in~\cite{FGZblowup} for local surfaces.
Since that work uses a different normalization, we give the correspondence with the quantities $\cT_{g,m}$ and $z$ used there.
For local $\mathbb P^2$, the mirror theorem gives 
$$6\pi i\tau=\cT_{0,2}(-Q)  \quad \text{and}  \quad
F_g^{(m)}(Q)=(-1)^m\cT_{g,m}(-Q) \ \text{ for } \ 2g-2+m>0 .$$
Set $w^{+}=-\frac{z}{2\pi i}$, then the closed vanishing equation becomes
$$
e^{\sum_{2g-2+m>0}\frac{\hbar^{2g-2+m}}{m!}\cT_{g,m}(-Q){\pd_z^m}}\theta[\cmn]\big(\tfrac{z}{2\pi i}\big|\tfrac{\cT_{0,2}(-Q)}{2\pi i}\big)=0.
$$
By~\cite{FGZblowup}, its unique formal solution with $z_{*}=O(\hbar)$ is exactly $-2\pi i w_{*}^{+}$.
\end{remark}

\subsection{Schur expansion of generating series}
\label{sec:localp2-NP-expansion}
Define the Schur coefficients
$Z_\lambda(\epsilon;\hbar)$ of the TR partition function by
\begin{equation}\label{eqn:localp2-TR-TV-partition}
 Z(\epsilon,\mathbf p;\hbar)
 =\sum_\lambda Z_\lambda(\epsilon;\hbar)s_\lambda(\mathbf p).
\end{equation}
Here $s_\lambda$ is the Schur function in the power sums $\mathbf p$.
The coefficients satisfy the following relation.
\begin{proposition}\label{prop:localp2-TR-row-hook}
For $n\geq1$,
\begin{equation}\label{eqn:localp2-TR-row-hook}
 (1-\sfq^{-n})Z_{(n+1)}-(\sfq-1)Z_{(n,1)}
 -\sfq^{n-3/2}Z_{(n-2)}=0.
\end{equation}
Here $Z_\lambda=Z_\lambda(\epsilon;\hbar)$ and $\sfq=e^{-\hbar}$.
\end{proposition}
We defer the proof of this proposition to Section~\ref{sec:localp2-2},
where we derive the relation using the topological vertex.

\medskip

Write the Schur expansion of the NP partition function as
\begin{equation}\label{eqn:localp2-NP-Schur}
 Z^{\rm NP}_{\mu,\nu}(\epsilon,\mathbf p;\hbar;w)
 =\sum_\lambda Z^{\rm NP}_{\lambda;\mu,\nu}(\epsilon;\hbar;w)
 s_\lambda(\mathbf p).
\end{equation}
Since $\epsilon=t/3$, the Fourier representation~\eqref{eqn:Z-NP} gives
\begin{equation}\label{eqn:localp2-total-coefficients}
 Z^{\rm NP}_{\lambda;\mu,\nu}(\epsilon;\hbar;w)
 =\sum_{l\in\mathbb Z}e^{2\pi i(l+\mu)(w+\nu)}
 Z_\lambda(\epsilon+\hbar(l+\mu);\hbar).
\end{equation}
We write $Z^{\rm NP}_\lambda$ for these coefficients when the
parameters are fixed.

Take $\mu=\nu=\frac12$, and we define
$$
Z^{\rm NP}_{\lambda,*}:=Z^{\rm NP}_{\lambda;\frac12,\frac12}(\epsilon;\hbar;w_*).
$$
Then the closed vanishing equation is equivalent to 
$Z^{\rm NP}_{\varnothing,*}=0$.
Moreover, we have expansion
\[
{Z^{\rm NP}_{(1),*}}\big/{Z_\varnothing(\epsilon;\hbar)}
=\theta'\!\left[\substack{\frac12\\ \frac12}\right](0\mid3\tau)
\frac{dz}{dX}\Big|_b+O(\hbar).
\]
Since the odd theta function has a simple zero at the origin, we have $Z^{\rm NP}_{(1),*}\ne0$.

\subsection{The canonical quantum curve}
\label{sec:localp2-quantum-curve}
Since $s_\lambda(X,X^2,\ldots)$ is $X^n$ for $\lambda=(n)$ and vanishes otherwise, 
by taking the principal specialization $p_k=X^k$, we have
\[
\Psi_{\mu,\nu}^{X}(X)=\sum_{n\geq0}Z^{\rm NP}_{(n);\mu,\nu}(\epsilon;\hbar;w)X^n.
\]
Moreover, by using $\omega_{0,1}^{\mathrm{sing},b}=\pi i\,dX/X$ and $dX/dx=-X$, 
up to a nonzero constant factor, we have $ \Psi_{\mu,\nu}(X)=X^{\pi i/\hbar-1/2}\Psi_{\mu,\nu}^{X}(X)$.
Taking $\mu=\nu=\frac12$ and $w=w_{*}$, we write
\[
\Psi^X_{*}(X):=\Psi^X_{\frac12,\frac12}(X)\big|_{w=w_{*}},\qquad
\Psi_{*}(X):=\Psi_{\frac12,\frac12}(X)\big|_{w=w_{*}}.
\]
Then we have $\Psi^X_{*}(0)=0$ and $\partial_X\Psi^X_{*}(0)\ne0$.

For $x=-\log X$ and $y=-\log Y$, the quantization
$\widehat y=\hbar\partial_x$ gives
$$
\widehat X=X\cdot,\qquad
\widehat Y=e^{-\hbar\partial_x}=\sfq^{-X\partial_X}.
$$ 
{We quantize~\eqref{eqn:localp2-mirror-curve} using the symmetric ordering}
$\widehat{X^aY^b}=X^{a/2}\widehat Y^{\,b}X^{a/2}
=\sfq^{-ab/2}X^a\widehat Y^{\,b}$.

\begin{theorem}\label{thm:quantum-localp2}
The canonical wave function $\Psi_{*}$ satisfies
\begin{equation}\label{eqn:localp2-quantum-curve}
 \left(X^{-1}+\sfq^{1/2}X^{-1}\widehat Y
 +\sfq X^2\widehat Y^{-1}-3\phi+\cW^{+}\right)\Psi_{*}
 =0,
\end{equation}
where $\cW^{+}=\cW(t;\hbar)-\cW(t;0)$ with
\begin{equation}\label{eqn:localp2-Wilson-coefficient}
\cW=(\sfq^{-1}-1)\frac{Z^{\rm NP}_{(2),*}}{Z^{\rm NP}_{(1),*}}
=(\sfq^{-1}-1)\partial_X\log\frac{\Psi^X_{*}(X)}{X}\bigg|_{X=0}.
\end{equation}
\end{theorem}
\begin{proof}
Taking the $\hbar$-expansion, the leading disk contribution gives
\[
\partial_X\log\tfrac{\Psi^X_{*}(X)}{X}\Big|_{X=0}
=\tfrac{1}{\hbar}\partial_XF_{0,1}(0)+O(1)
=-\tfrac{3\phi}{\hbar}+O(1),
\]
where the last equality follows from $Y=-1+3\phi X+O(X^3)$.
Since $\sfq^{-1}-1=\hbar+O(\hbar^2)$, we obtain
$\cW(t;0)=-3\phi$.
Therefore, equation~\eqref{eqn:localp2-quantum-curve} can be rewritten as
\beq\label{eqn:localp2-quantum-curve-2}
\left(X^{-1}+\sfq^{1/2}X^{-1}\widehat Y
+\sfq X^2\widehat Y^{-1}+\cW\right)\Psi_{*}=0.
\eeq

Dividing the left-hand side of~\eqref{eqn:localp2-quantum-curve-2}
by the common factor
$X^{\pi i/\hbar-1/2}$
and expanding in $X$, we see that the equation is equivalent to
\[
 (1-\sfq^{-n})Z^{\rm NP}_{(n+1),*}+\cW Z^{\rm NP}_{(n),*}
 -\sfq^{n-3/2}Z^{\rm NP}_{(n-2),*}=0,
 \qquad n\geq0.
\]
Applying Proposition~\ref{prop:localp2-TR-row-hook} with $\epsilon$ replaced
by $\epsilon+\hbar(l+\mu)$ and summing as
in~\eqref{eqn:localp2-total-coefficients}, we obtain
\begin{equation}\label{eqn:localp2-NP-row-hook}
 (1-\sfq^{-n})Z^{\rm NP}_{(n+1)}-(\sfq-1)Z^{\rm NP}_{(n,1)}
 -\sfq^{n-3/2}Z^{\rm NP}_{(n-2)}=0,\qquad n\geq1.
\end{equation}
In particular, $Z^{\rm NP}_{(1,1)}=\sfq^{-1}Z^{\rm NP}_{(2)}$.

	It is known that the total partition function is a tau-function of the KP hierarchy~\cite{ADKMV06,DZ12,WYZvertex24,WYZ25}, and hence so is $Z^{\rm NP}_{\mu,\nu}$ (see Remark~\ref{rem:localp2-WYZ-total} in the next section).
	See also~\cite{ABDKS24} for a proof from topological recursion.
	The Schur coefficients of a KP tau-function satisfy the homogeneous Pl\"ucker relations.
	For \(n\geq2\), the relation corresponding to \((n,2)=(n-1,0\mid1,0)\) in Frobenius notation is
\begin{equation}\label{eqn:localp2-TR-rank-two}
 Z^{\rm NP}_\varnothing Z^{\rm NP}_{(n,2)}
 =Z^{\rm NP}_{(n,1)}Z^{\rm NP}_{(1)}
 -Z^{\rm NP}_{(n)}Z^{\rm NP}_{(1,1)}.
\end{equation}
This polynomial identity remains valid when $Z^{\rm NP}_\varnothing=0$.

Combining~\eqref{eqn:localp2-NP-row-hook} with~\eqref{eqn:localp2-TR-rank-two}, we obtain for $n\geq 2$,
\begin{equation*}
 Z^{\rm NP}_{(1)}\bigl[(1-\sfq^{-n})Z^{\rm NP}_{(n+1)}
 -\sfq^{n-3/2}Z^{\rm NP}_{(n-2)}\bigr]
 -(1-\sfq^{-1})Z^{\rm NP}_{(2)}Z^{\rm NP}_{(n)}
 =(\sfq-1)Z^{\rm NP}_\varnothing Z^{\rm NP}_{(n,2)}.
\end{equation*}
Setting $\mu=\nu=\frac12$ and $w=w_*$, the right-hand side vanishes since $Z^{\rm NP}_{\varnothing,*}=0$.
Dividing by $Z^{\rm NP}_{(1),*}$ gives the required identity for $n\geq 2$.
The remaining two cases follow from the closed vanishing equation and the definition of $\cW$.
This proves~\eqref{eqn:localp2-quantum-curve-2}, and hence \eqref{eqn:localp2-quantum-curve}.
\end{proof}

$$
$$

\section{Topological vertex identities}\label{sec:localp2-2}
The open and closed GW invariants of smooth toric Calabi--Yau threefolds, 
with open boundary conditions specified by framed Aganagic--Vafa branes, 
can be computed in all genera by gluing topological vertices~\cite{AKMV,LLLZ09,MOOP11,DZ12}.
We prove Proposition~\ref{prop:localp2-TR-row-hook} by combining the
vertex symmetry with the gluing formula. The resulting relation among
the coefficients \(V_\lambda\) of the glued state gives the required relation for the
TR coefficients $Z_\lambda$ through their comparison formula.

\subsection{Topological vertex}
\label{sec:localp2-vertices}
The topological vertex encodes the all-genus open GW theory of $\mathbb C^3$ with three framed Aganagic--Vafa branes.
Its algebraic construction is given by the formal relative GW theory~\cite{LLLZ09}.
We recall its realization in the half-infinite wedge space.
We refer to~\cite[Appendix~A]{Oko01} for more details for the infinite-wedge formalism..

Let $G^{(0)}$ be the charge-zero half-infinite wedge space of the vector space $\bigoplus_{n\in\mathbb Z}\mathbb C z^{n+\frac{1}{2}}$ with basis:
\[
|\mu\rangle
=z^{\mu_1-1/2}\wedge z^{\mu_2-3/2}\wedge\cdots,
\]
where $\mu=(\mu_1,\cdots,\mu_{l})$ is a partition with $\mu_1\geq \cdots\geq \mu_l>0$ (we set $\mu_{k}=0$ for $k>l$). 
The vacuum vector is $|0\rangle=|\varnothing\rangle=z^{-1/2}\wedge z^{-3/2}\wedge\cdots$.
We denote by $\{\<\mu|\}$ the corresponding dual basis, i.e., $\<\mu|\nu\>=\delta_{\mu,\nu}$.
On tensor products, we use the notation
\[
|\mu^{1},\cdots,\mu^{m}\rangle
=|\mu^{1}\rangle\otimes\cdots\otimes|\mu^{m}\rangle,
\qquad
\langle\mu^{m},\cdots,\mu^{1}|
=\langle\mu^{m}|\otimes\cdots\otimes\langle\mu^{1}|.
\]

Let $E_{a,b}$ be the matrix given by $E_{a,b}z^c=\delta_{b,c}z^a$. It acts on the wedge space by the Leibniz rule. 
For $k\ne0$, set
\[
 \alpha_k=\sum_{j\in\mathbb Z+1/2}E_{j-k,j}.
\]
The Schur and skew Schur functions in the variables ${\bf p}=(p_1,p_2,\ldots)$ are characterized by
\[
 s_\lambda({\bf p})
 =\left\langle0\left|
 e^{\sum_{k\geq1}\frac{p_k\alpha_k}{k}}
 \right|\lambda\right\rangle,
 \qquad
 s_{\lambda/\eta}({\bf p})
 =\left\langle\eta\left|
 e^{\sum_{k\geq1}\frac{p_k\alpha_k}{k}}
 \right|\lambda\right\rangle.
\]
For a partition $\mu$, write $\mu^t$ for its transpose and set
\[
|\mu|=\sum_{i}\mu_i,\qquad
\kappa_\mu=\sum_{i\geq1}\mu_i(\mu_i-2i+1).
\]
Let $\rho=(-\frac12,-\frac32,\ldots)$ and write $\mu+\rho=(\mu_1-\frac12,\mu_2-\frac32,\dots)$. 
The topological vertex in canonical framing is given by
\begin{equation}\label{eqn:localp2-vertex}
	C_{\lambda\mu\nu}(\sfq)
	=\sfq^{\kappa_\lambda/2}s_\nu(\sfq^\rho) \sum_\eta s_{\lambda^t/\eta}(\sfq^{\nu+\rho})s_{\mu/\eta}(\sfq^{\nu^t+\rho}).
\end{equation}
Here $s_{\lambda/\eta}(\sfq^{\nu+\rho})$ denotes the specialization
\(
p_k=\sum_{j\geq1}\sfq^{k(\nu_j-j+1/2)},
\)
where the sums are understood as rational functions of $\sfq^{1/2}$.
In particular, $C_{\varnothing\varnothing\varnothing}=1$.
We define the corresponding vertex state by
\[
 |C\rangle=\sum_{\lambda,\mu,\nu}
 C_{\lambda\mu\nu}|\lambda,\mu,\nu\rangle.
\]
\begin{remark}

The total state of the topological vertex can be expressed as a Bogoliubov transform $e^{A}|0,0,0\>$, where $A=(A_{k,l}^{i,j})_{i,j=1,2,3;k,l\geq 0}$ is explicitly known~\cite{ADKMV06,WYZvertex24}.
	The state $|C\rangle$ is the component of this transform having charge zero in each tensor factor.
\end{remark}

The vertex state $|C\rangle$ satisfies the following symmetry identities~\cite{ADKMV06,DP25} which we call the Ward identities following~\cite{ADKMV06}.
Firstly, following~\cite{OP06}, for each $r\in\mathbb Z$, define
\[\textstyle
\mathcal E_r(u)
=\sum_{j\in\mathbb Z+1/2}
e^{u(j-r/2)}:E_{j-r,j}:
+\frac{\delta_{r,0}}{e^{u/2}-e^{-u/2}}\operatorname{id},
\]
where $:E_{a,b}:$ is the normal ordering given by
\(
:E_{a,b}:
=E_{a,b}-\delta_{a,b}\mathbf1_{\{b<0\}}\operatorname{id}.
\)
We set
\begin{equation}\label{eqn:localp2-current}
	\mathcal E_{r,k}
	:=\mathcal E_r(k\hbar),
	\qquad (r,k)\in\mathbb Z^2\setminus\{(0,0)\}.
\end{equation}
For $r\ne0$, the operator $\mathcal E_r(u)$ is regular at
$u=0$, and $\mathcal E_{r,0}=\alpha_r$.
Then, for $(a,b)\in\mathbb Z^2\setminus \{(0,0)\}$,
\begin{equation}\label{eqn:localp2-vertex-symmetry}
	\big(\mathcal E_{a,b}^{(1)}+(-1)^b\mathcal E_{-b,a-b}^{(2)}
	+(-1)^a\mathcal E_{b-a,-a}^{(3)}\big)|C\rangle=0.
\end{equation}
The superscript indicates the tensor factor.
\begin{remark}
	The state and operator are denoted by $\mathfrak T$ and $\mathcal W$ in~\cite{DP25}. 
	The identifications with our notation are
	\[
	 \frac{\mathfrak T}{\mathfrak T_\varnothing}
	 \longleftrightarrow\sum_{\lambda,\mu,\nu}(-1)^{|\lambda|+|\mu|+|\nu|}
	 C_{\lambda\mu\nu}|\lambda,\mu,\nu\rangle ,\qquad
	 \mathcal W_{rw_{\ell^-}+kn_{\ell^-},\ell^-}
	 \longleftrightarrow\frac{(-1)^k}{\sqrt{-1}}\mathcal E^{(\ell)}_{r,-k},
	\]
	where $\mathfrak T_\varnothing$ is the degree $(0,0,0)$ part of $\mathfrak T$, and $w_{\ell^-}$ and $n_{\ell^-}$ denote the framing vector and the normal vector of the outgoing leg $\ell^-$, respectively, with $\det(w_{\ell^-},n_{\ell^-})=1$.
	The first identification uses the vertex comparison~\cite{LLLZ09,MOOP11}, and the second is the operator correspondence in~\cite[Remark~13.1]{DP25}.
	Via these identifications, equation~\eqref{eqn:localp2-vertex-symmetry} is just their Theorem~15.1
\end{remark}

\subsection{Gluing the topological vertices}
Given a smooth toric Calabi--Yau threefold $\mathcal X$, it can be expressed by a trivalent diagram, and the generating series can be read off from this diagram~\cite{AKMV,LLLZ09,MOOP11,DZ12}.
At each vertex $v$, place a copy $|C_v\rangle$ of the vertex state $|C\rangle$, 
with its three tensor factors corresponding to the three legs in counterclockwise order.
For an internal edge $e$, let $Q_e$ be the corresponding Novikov parameter and let $f_e\in\mathbb Z$ be the framing integer, the pairing of its two ordered legs is given by
\begin{equation}\label{eqn:framed-gluing-pair}
	\langle P_e|
	=\sum_\mu
	(-1)^{(f_e+1)|\mu|}Q_e^{|\mu|}\sfq^{f_e\kappa_\mu/2}
	\langle\mu^t|\otimes\langle\mu|.
\end{equation}
The gluing construction gives
\begin{equation}\label{eqn:glued-state}
|V\rangle
=\bigg(\bigotimes_{e\in {\rm internal\ edges}}\langle P_e|\bigg)
\bigg(\bigotimes_{v\in {\rm vertex}}|C_v\rangle\bigg).
\end{equation}
Here $|V\>$ denotes the charge-zero part of the state associated with the diagram.

\begin{figure}[htbp]
	\centering
\begin{tikzpicture}[
        font=\fontsize{9}{11}\selectfont,
        line width=0.5pt,
        vertex/.style={
          circle,draw,fill=gray!5,
          minimum size=8mm,inner sep=1pt
        },
        pairing/.style={
          rectangle,draw,fill=gray!8,rounded corners=1pt,
          minimum width=10mm,minimum height=4.5mm,inner sep=1pt
        }
        ]
        \node[vertex] (c1) at (3,0) {$|C_1\rangle$};
        \node[vertex] (c2) at (3,3) {$|C_2\rangle$};
        \node[vertex] (c3) at (0,0) {$|C_3\rangle$};

        \node[pairing] (b12) at (3,1.5) {$\langle P_{12}|$};
        \node[pairing] (b23) at (1.5,1.5) {$\langle P_{23}|$};
        \node[pairing] (b31) at (1.5,0) {$\langle P_{31}|$};

        \draw (c1) --
          node[pos=.53,right=2pt] {$\eta_1$} (b12);
        \draw (b12) --
          node[pos=.47,right=2pt] {$\eta_1^t$} (c2);

        \draw (c2) --
          node[above left=1pt] {$\eta_2$} (b23);
        \draw (b23) --
          node[above left=1pt] {$\eta_2^t$} (c3);

        \draw (c3) --
          node[pos=.53,below=2pt] {$\eta_3$} (b31);
        \draw (b31) --
          node[pos=.47,below=2pt] {$\eta_3^t$} (c1);

        \draw (c1) -- (3.5,-.5) node[below right] {$\lambda$};
        \draw (c2) -- (3.7,4.4) node[above] {$\mu$};
        \draw (c3) -- (-1,-.5) node[left] {$\nu$};
      \end{tikzpicture}
	\caption{Gluing vertices for local $\mathbb P^2$; cf.~\cite[Figure~2]{AKMV}.}
	\label{fig:localp2-gluing}
\end{figure}
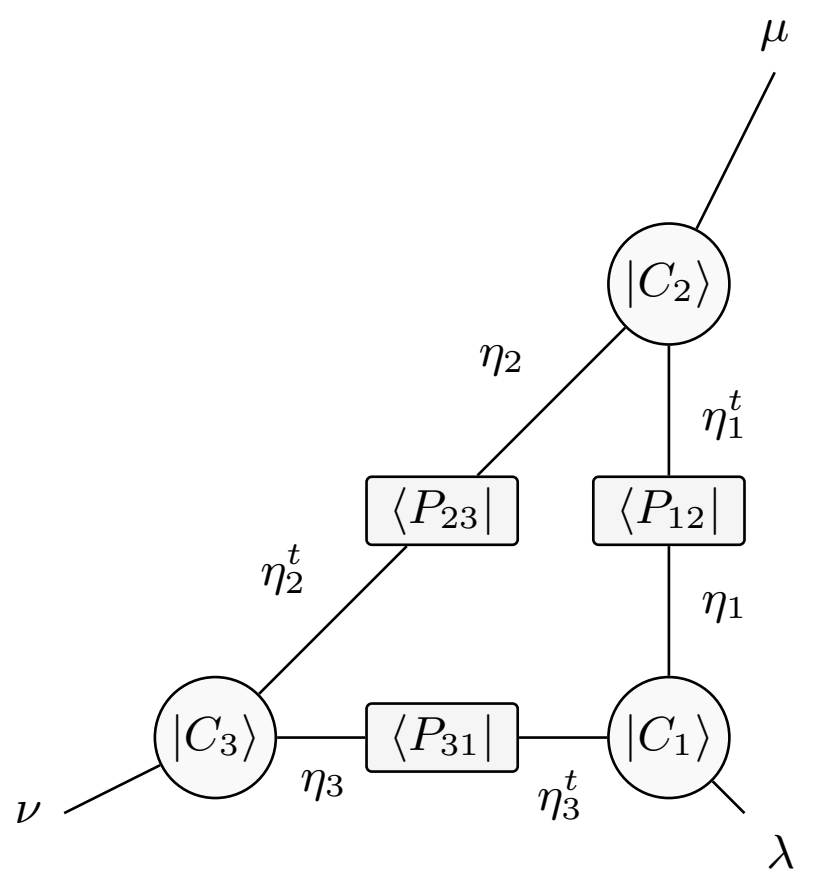

For $K_{\mathbb P^2}$, the diagram and the gluing rule are shown in Figure~\ref{fig:localp2-gluing}.
Set three vertices $|C_i\rangle$ in counterclockwise order,
and their tensor factors are the external, outgoing, and incoming legs. 
The latter two join $i$ to $i+1$ and $i-1$, respectively.
Each internal edge has $Q_e=Q$ and $f_e=2$ in incoming--outgoing order,
so~\eqref{eqn:framed-gluing-pair} specializes to
\[
 \langle P|=\sum_\mu(-Q)^{|\mu|}\sfq^{\kappa_\mu}
 \langle\mu^t|\otimes\langle\mu|.
\]
Let $\langle P_{ij}|$ pair the incoming leg of $C_j$ with the outgoing leg of $C_i$. 
By contracting $P_{12},P_{23},P_{31}$, we get $|V\rangle=\sum_{\lambda,\mu,\nu}V_{\lambda\mu\nu}|\lambda,\mu,\nu\rangle$.
If $\eta_i$ labels the edge from $i$ to $i+1$, its coefficients are
\begin{equation}\label{eqn:localp2-gluing}
 V_{\lambda\mu\nu}=V_{\lambda\mu\nu}(Q;\sfq)
 =\sum_{\eta_1,\eta_2,\eta_3}(-Q)^{\sum_i|\eta_i|}\sfq^{\sum_i\kappa_{\eta_i}}
 C_{\lambda,\eta_1,\eta_3^t}C_{\mu,\eta_2,\eta_1^t}C_{\nu,\eta_3,\eta_2^t}.
\end{equation}
Cyclically relabelling the internal partitions in
\eqref{eqn:localp2-gluing} gives
\(
V_{\lambda\mu\nu}
=V_{\mu\nu\lambda}
=V_{\nu\lambda\mu}.
\)

We have the following Ward identity for state $|V\>$.
\begin{proposition}
	For $(a,b)\in\mathbb Z^{2}\setminus \{(0,0)\}$, 
	\begin{equation}\label{eqn:localp2-three-leg-general-symmetry}
		\big(
		\mathcal E_{a,b}^{(1)}
		+(-1)^aQ^b\mathcal E_{a-3b,a-2b}^{(2)}
		+(-1)^bQ^{a-b}\mathcal E_{3b-2a,b-a}^{(3)}
		\big)|V\rangle=0.
	\end{equation}
\end{proposition}

\begin{proof}
	We then derive this symmetry property from the symmetry property for the vertex state $|C\>$.
	We first prove that the edge pairing $\<P|$ satisfies
	\begin{equation}\label{eqn:localp2-edge-transfer}
		\langle P|(\mathcal E_{r,k}\otimes1)
		=\langle P|(1\otimes[-Q^{-r}\mathcal E_{-r,-2r-k}]).
	\end{equation}
	Indeed, matrix transposition sends $\mathcal E_{r,k}$ to
	$\mathcal E_{-r,k}$, while partition transposition sends it to
	$(-1)^{r+1}\mathcal E_{r,-k}$.
	For a nonzero particle move $j\mapsto j-r$, the changes in
	$|\lambda|$ and $\kappa_\lambda$ are $-r$ and $-2r(j-r/2)$.
	Conjugation by the inverse diagonal edge weight therefore gives
	$(-Q)^r\mathcal E_{r,k-2r}$. 
	Combining these operations gives \eqref{eqn:localp2-edge-transfer}.
	For $(a,b)\in\mathbb Z^2\setminus\{(0,0)\}$, apply
	\eqref{eqn:localp2-vertex-symmetry} to the three vertices in
	\eqref{eqn:localp2-gluing}, with parameters
	$(a,b)$, $(a-3b,a-2b)$, and $(3b-2a,b-a)$ and multipliers
	$1$, $(-1)^aQ^b$, and $(-1)^bQ^{a-b}$, respectively.
	By~\eqref{eqn:localp2-edge-transfer}, the currents on the outgoing
	internal legs are transferred to the incoming internal legs at the
	next vertices as follows:
	\begin{align*}
		(-1)^b\mathcal E_{-b,a-b}
		&\longmapsto-(-1)^bQ^b\mathcal E_{b,3b-a},\\
		Q^b\mathcal E_{2b-a,-b}
		&\longmapsto-Q^{a-b}\mathcal E_{a-2b,2a-3b},\\
		(-1)^aQ^{a-b}\mathcal E_{a-b,2b-a}
		&\longmapsto-(-1)^a\mathcal E_{b-a,-a}.
	\end{align*}
	These cancel the incoming currents at vertices $2$, $3$, and $1$, respectively. 
	The remaining external currents therefore give the identity~\eqref{eqn:localp2-three-leg-general-symmetry}.
\end{proof}

	\begin{remark}\label{rem:localp2-WYZ-total}
		Following the proposal of Aganagic--Dijkgraaf--Klemm--Mari\~no--Vafa
		in~\cite[\S~4.7]{ADKMV06}, we sum over all fermion number fluxes around the loop.
		Let $N\in\mathbb Z$ denote the flux and let $\Xi$ record it with weight $\Xi^{-N}$.
		For local $\mathbb P^2$ with one outer brane, the total partition function is given by~\cite[Theorem~4.2]{WYZ25}
		$$
		Z^{\rm total}(Q,{\bf p};\sfq;\Xi)
		=\sum_{N\in\mathbb Z}\sum_\lambda
		\Xi^{-N}(-Q)^{3N^2/2}\sfq^{\frac{3N(4N^2-1)}8-N|\lambda|}
		\cdot V_{\lambda\varnothing\varnothing}(Q\sfq^{3N};\sfq)s_\lambda({\bf p}).
		$$
		The $N=0$ term recovers the ordinary vertex partition function.

		To compare this sum with the NP partition function, recall that
		$\epsilon=t/3$, $Q=e^{-t}$, and $\sfq=e^{-\hbar}$.
		Thus the shift $\epsilon\mapsto\epsilon+N\hbar$ induces $Q\mapsto Q\sfq^{3N}$.
		The closed normalizations in Section~\ref{sec:localp2-TR} give
		\[
		\frac{Z_\varnothing(\epsilon;\hbar)}{V_{\varnothing\varnothing\varnothing}(Q;\sfq)}
		=\exp\!\left(-\frac{t^3}{18\hbar^2}
		+\frac{\pi i t^2}{6\hbar^2}+\frac{t}{12}\right).
		\]
		Using the Schur-coefficient comparison~\eqref{eqn:localp2-Schur-comparison} below,
		we can therefore identify the charge sum term by term with the Fourier
		representation~\eqref{eqn:Z-NP}. This yields
		$$
		Z^{\rm NP}_{0,0}(\epsilon,{\bf p};\hbar;w)
		=\frac{Z_\varnothing(\epsilon;\hbar)}{V_{\varnothing\varnothing\varnothing}(Q;\sfq)}
		\cdot Z^{\rm total} \!\left(Q,\{(-Q^{-1/3})^kp_k\}_{k\geq1};\sfq;\Xi\right),
		$$
		where
		\(
		\Xi=e^{-2\pi iw+\frac{t^2}{2\hbar}-\frac{\pi it}{\hbar}+\frac{\hbar}{8}}.
		\)

		The total partition function is a tau-function of the KP hierarchy with times $p_k/k$~\cite{ADKMV06,DZ12,WYZvertex24,WYZ25}.
		The prefactor in the comparison is independent of $\mathbf p$, and the
		rescaling $p_k\mapsto(-Q^{-1/3})^kp_k$ preserves the KP equations.
		Hence $Z^{\rm NP}_{0,0}$ is also a KP tau-function.
		The general $\mu,\nu$ case follows from
		\[
		Z^{\rm NP}_{\mu,\nu}(\epsilon,{\bf p};\hbar;w)
		=e^{2\pi i\mu(w+\nu)}Z^{\rm NP}_{0,0}(\epsilon+\mu\hbar,{\bf p};\hbar;w+\nu).
		\]
		The Schur coefficients of $Z^{\rm NP}_{\mu,\nu}$ therefore
		satisfy the Pl\"ucker relations used above.
		A parallel argument gives an analogous identification for more general toric geometries,
		including the local weak del Pezzo surfaces considered in Section~\ref{sec:one-loop}.
	\end{remark}

\subsection{Proof of Proposition~\ref{prop:localp2-TR-row-hook}}
For the chosen outer leg, write
$V_\lambda=V_{\lambda\varnothing\varnothing}$.
By cyclic symmetry,
$V_\lambda=V_{\varnothing\lambda\varnothing}
=V_{\varnothing\varnothing\lambda}$.
We denote $|V^{(1)}\>=\sum_{\lambda}V_{\lambda}|\lambda\>$.
The corresponding partition function is
\begin{equation*}
	Z^{\rm TV}(Q,{\bf p};\sfq)
	=\big\<0\big|e^{\sum_{k\geq1}\frac{p_k\alpha_k}k}\big|V^{(1)}\big\rangle
	=\sum_\lambda V_\lambda(Q;\sfq)s_\lambda({\bf p}).
\end{equation*}
The connected vertex correlators are defined by
\begin{equation*}
 \langle\alpha_{\mu_1},\ldots,\alpha_{\mu_n}\rangle^{\rm TV}_{g,n}
 :=[\hbar^{2g-2+n}]\prod_{i=1}^n\left(\mu_i\partial_{p_{\mu_i}}\right)
 \log Z^{\rm TV}(Q,{\bf p};\sfq)\bigg|_{{\bf p}=0}.
\end{equation*}
For $n\geq1$, we have the following relations between correlators~\cite{AKMV,LLLZ09,MOOP11,FLZ20}
$$
 \langle\alpha_{\mu_1},\ldots,\alpha_{\mu_n}\rangle^{\mathcal X,L}_{g,n}
 =(-1)^{g-1+n+|\mu|}\langle\alpha_{\mu_1},\ldots,\alpha_{\mu_n}\rangle^{\rm TV}_{g,n},
$$
For $n=0$, i.e., the closed correlators, we have $V_\varnothing(0;\sfq)=1$, and
\begin{equation*}
 \<\, \>^{\rm TV}_{g,0}=(-1)^{g-1}\sum_{d\geq1}N^{\mathcal X}_{g,d}Q^d.
\end{equation*}
Combining these identities with the mirror theorem in
Section~\ref{sec:localp2-TR}, we obtain
\begin{equation}\label{eqn:localp2-Schur-comparison}
 \frac{Z_\lambda(\epsilon;\hbar)}{Z_\varnothing(\epsilon;\hbar)}
 =(-Q^{-1/3})^{|\lambda|}
 \frac{V_\lambda(Q;\sfq)}{V_\varnothing(Q;\sfq)}.
\end{equation}

\begin{proof}[Proof of Proposition~\ref{prop:localp2-TR-row-hook}]
	We first prove that, for $n\geq1$,
	\begin{equation}\label{eqn:localp2-row-hook}
		(1-\sfq^{-n})V_{(n+1)}-(\sfq-1)V_{(n,1)}
		+Q\sfq^{n-3/2}V_{(n-2)}=0.
	\end{equation}
	Here, $V_{(0)}=V_{\emptyset}$ and $V_{(k)}=0$ for $k<0$.
	In this proof, $|V\rangle$ denotes the three-leg state, and we write
	\(
	|V^{(1)}\rangle=\sum_\lambda V_\lambda|\lambda\rangle
	\)
	for the one-leg state.
	By taking $(a,b)=(1,0)$ in equation~\eqref{eqn:localp2-three-leg-general-symmetry}, we have
	\begin{equation*}
		\big(\mathcal E_{1,0}^{(1)}-\mathcal E_{1,1}^{(2)}+Q\mathcal E_{-2,-1}^{(3)}\big)|V\rangle=0.
	\end{equation*}
	We deduce a similar equation for the one-leg case.
	We prove that
	\begin{equation}\label{eqn:localp2-one-leg-symmetry}
		(\mathcal E_{1,0}-\mathcal E_{1,1}+Q\mathcal E_{-2,-1})
		|V^{(1)}\rangle=0.
	\end{equation}
	In fact, by using $\langle0|\mathcal E_{1,0}=\langle0|\mathcal E_{1,1}=\langle(1)|$ and 
	$\langle0|\mathcal E_{-2,-1}=0$,
	we have for any partition $\lambda$,
	$$
	\<\lambda|\mathcal E_{1,0} |V^{(1)}\>
	=\langle0,0,\lambda|\mathcal E_{1,0}^{(1)}|V\rangle 
	=\langle0,0,\lambda|(\mathcal E_{1,1}^{(2)}-Q\mathcal E_{-2,-1}^{(3)})|V\rangle 
	=V_{\lambda, (1),\varnothing}.
	$$
	Similarly,
	$$
	-\<\lambda|\mathcal E_{1,1} |V^{(1)}\>
	=-\langle 0,\lambda, 0|\mathcal E_{1,1}^{(2)}|V\rangle 
	=-V_{(1),\lambda,\varnothing},
	$$
	$$
	Q\<\lambda| \mathcal E_{-2,-1} |V^{(1)}\>
	=Q\langle \lambda, 0,0|\mathcal E_{-2,-1}^{(3)}|V\rangle
	=V_{\varnothing,(1),\lambda}-V_{(1),\varnothing,\lambda}.
	$$
	Adding these three equations and using cyclic symmetry, we obtain equation~\eqref{eqn:localp2-one-leg-symmetry}.
	
	We now pair \eqref{eqn:localp2-one-leg-symmetry} with $\langle(n)|$, for $n\geq1$.
	We have firstly,
	\begin{align*}
		\langle(n)|\mathcal E_{1,0}=\langle(n+1)|+\langle(n,1)|,\qquad
		\langle(n)|\mathcal E_{1,1}=\sfq^{-n}\langle(n+1)|+\sfq\langle(n,1)|,
	\end{align*}
	and 
	$$\langle(n)|\mathcal E_{-2,-1}=\sfq^{n-3/2}\langle(n-2)|.$$
	Substituting these matrix elements gives \eqref{eqn:localp2-row-hook}.
	Multiplying~\eqref{eqn:localp2-row-hook} by
	\[
	 \frac{Z_\varnothing(\epsilon;\hbar)}{V_\varnothing(Q;\sfq)}
	 (-Q^{-1/3})^{n+1}
	\]
	and using~\eqref{eqn:localp2-Schur-comparison}, together with
	$Q(-Q^{-1/3})^3=-1$, gives
	\eqref{eqn:localp2-TR-row-hook}.
\end{proof}

$$
$$

\section{Wilson loops and quantum corrections}\label{sec:localp2-wilson}
 
In this section, we recall the geometric definition of NS Wilson loops for local surfaces and prove the relation \(\mathcal W=Q^{-1/3}\mathcal W^{\mathrm{NS}}_1(Q;\hbar)\) for local \(\mathbb P^2\). 
This establishes Conjecture~\ref{conj:one-loop-geometric-Wilson} in this case and gives a geometric interpretation of the scalar quantum correction.

\subsection{Wilson loops for local surfaces}
\label{sec:wilson-loops}

Guo--Wang--Wu~\cite{GuoWangWu} realize Wilson loops for local weak Fano
surfaces as generating series of GW invariants with point
insertions on a compactification of the local surface.
We recall the NS version of this construction.

For a smooth projective surface $S$, let
$\pi:\overline Y=\mathbb P_S(\mathcal O_S\oplus K_S)\to S$
compactify its canonical bundle, with zero section
$\iota_0$, infinity divisor $D_\infty$, and fibre class $f$.

\begin{definition}
For $n\geq1$, the connected $n$-point NS Wilson-loop series~\cite{GuoWangWu} is
\begin{equation*}
	\begin{aligned}
		\mathcal W^{\mathrm{NS}}_n(Q;\hbar)
		&:=\hbar^{n-1}\sum_{\substack{g\geq0\\ \beta\ \mathrm{effective}}}Q^\beta
		\int_{[\overline M_{g,n}(\overline Y,\iota_{0*}\beta+nf)]^{\mathrm{vir}}}
		(-1)^g\lambda_g\Lambda_g^\vee(\hbar)\,
		\prod_{i=1}^n\operatorname{ev}_i^*\!\left([D_\infty]\pi^*[\mathrm{pt}]_S\right).
	\end{aligned}
\end{equation*}
Here $Q^\beta$ records the effective integral curve class $\beta$;
$\overline M_{g,n}$ parametrizes connected stable maps with $n$ labelled markings,
with virtual class $[\,\cdot\,]^{\mathrm{vir}}$ and evaluation map
$\operatorname{ev}_i$ at the $i$th marking. 
For the rank-$g$ Hodge bundle $\mathbb E$, set $\lambda_j=c_j(\mathbb E)$ and
$\Lambda_g^\vee(u)=\sum_{j=0}^g(-1)^j\lambda_j u^{g-j}$, with $\lambda_0=1$.
The equivariant action fixes $S$ and has weight $-\hbar$ on $K_S$;
$[D_\infty]$ is the equivariant divisor class and
$[\mathrm{pt}]_S$ is the point class of $S$.
\end{definition}

For the smooth projective toric weak del Pezzo surfaces $S$ of
Section~\ref{sec:one-loop}, put $E=-K_S$ and let $\mathbf Q^E$ be its
Novikov monomial. 
Conjecture~\ref{conj:one-loop-geometric-Wilson}
identifies the full scalar in the quantum correction as
$\cW=(\mathbf Q^E)^{-1/E^2}\mathcal W^{\rm NS}_1$.
The scalar quantum curve of Section~\ref{sec:canonical-wave} therefore
has the conjectural geometric form
\begin{equation*}
 \widehat H_0\Psi_*
 =-(\mathbf Q^E)^{-1/E^2}
 \bigl(\mathcal W^{\rm NS}_1(\mathbf Q;\hbar)
       -\mathcal W^{\rm NS}_1(\mathbf Q;0)\bigr)\Psi_*.
\end{equation*}
The subtraction is taken at fixed Novikov variables, and the identity
is formal in $\hbar$ and those variables after adjoining the displayed
fractional monomial.

For $S=\mathbb P^2$, let $\cW$ be the scalar defined
in~\eqref{eqn:localp2-Wilson-coefficient}, and put $E=-K_S$.
Then $\mathbf Q^E=Q^3$ and $E^2=9$.
We have the following theorem.
\begin{theorem}\label{thm:localp2-wilson-identification}
	
	The following equation holds:
	\begin{equation}\label{eqn:localp2-wilson-identification}
		\mathcal W=Q^{-1/3}\mathcal W^{\mathrm{NS}}_1(Q;\hbar).
	\end{equation}
\end{theorem}

We first express Wilson loops in terms of blowup invariants in
Section~\ref{sec:localp2-wilson-blowup} and derive a variation formula for
$\mathcal W$ in Section~\ref{sec:localp2-wilson-variation}.
We then use these formulas and the flop formula for $K_{\mathbb F_1}$
to prove the theorem in Section~\ref{sec:localp2-wilson-proof}.

\subsection{Wilson loops and blowup invariants}\label{sec:localp2-wilson-blowup}
We compare Wilson loop invariants with GW invariants of blowups of
$\mathbb P^2$.
We use the case $n=1$ of this comparison in
Section~\ref{sec:localp2-wilson-proof}.

For $n\geq1$ distinct points $p_1,\ldots,p_n$, put
$S_n=\operatorname{Bl}_{p_1,\ldots,p_n}\mathbb P^2$, with $H$ the pullback
of a line and $e_i$ the exceptional classes. Set $S_0=\mathbb P^2$.
For $n=1$, $S_1=\mathbb F_1$ has negative section
$e_1$ and ruling fiber $f_1=H-e_1$, so
\begin{equation}\label{eq:app-wilson-F1-class}
	dH-e_1=(d-1){e_1}+df_1.
\end{equation}

Let $\overline Y_n=\mathbb P_{S_n}(\mathcal O_{S_n}\oplus K_{S_n})$,
with zero section $\iota_n$ of normal bundle $K_{S_n}$.
Let the $p_i$ lie on a smooth cubic $E\subset\mathbb P^2$.
For $d\geq1$ with $m=3d-n>0$, the marked NS local--relative formula
of Guo--Wang--Wu~\cite{GuoWangWu}, in our normalization, is
\begin{equation}\label{eq:app-wilson-marked-input}
	[Q^d]\mathcal W_n^{\rm NS}(Q;\hbar)
	=\frac{(-1)^{m-1}\hbar^{n-1}}m
	\sum_{g\geq0}\hbar^{2g}
	\int_{[\overline M_{g,(m,1^n)}(\mathbb P^2/E,dH)]^{\rm vir}}
	\lambda_g\prod_{i=1}^n\operatorname{ev}_i^*[p_i]_E .
\end{equation}
The moduli space parametrizes connected relative stable maps of genus
$g$ and degree $dH$. The order-$m$ contact is free; the $n$ labelled
simple contacts are fixed at $p_i$ by the point insertions $[p_i]_E$.
The fibre weight in $\mathcal W_n^{\rm NS}$ is $-\hbar$.

With fibre weight $-\hbar$, define the closed NS coefficient by
\[
F^{\rm NS}_{d,n}(\hbar)=\hbar^{-1}\sum_{g\geq0}
\int_{[\overline M_{g,0}(K_{S_n},dH-\sum_{i=1}^n e_i)]^{\rm vir}}
(-1)^g\lambda_g\Lambda_g^\vee(\hbar).
\]

\begin{proposition}\label{prop:app-wilson-NS-blowup}
	Let $p_1,\ldots,p_n$ be distinct points on a smooth cubic
	$E\subset\mathbb P^2$, with $n\geq1$, such that
	$S_n=\operatorname{Bl}_{p_1,\ldots,p_n}\mathbb P^2$ is del Pezzo.
	For an integer $d\geq1$, assume that
	$dH-\sum_{i=1}^n e_i$ is effective and $m=3d-n>0$. Then
	\begin{equation}\label{eq:app-wilson-NS-blowup}
		[Q^d]\mathcal W_n^{\rm NS}(Q;\hbar)
		=\bigl(2\sinh(\hbar/2)\bigr)^n F^{\rm NS}_{d,n}(\hbar).
	\end{equation}
\end{proposition}
\begin{proof}
	The strict transform $E_n$ of $E$ is smooth anticanonical, with
	$E_n^2=9-n>0$ and $E_n\cdot(dH-\sum_{i=1}^n e_i)=m>0$.
	Thus~\cite[Corollary~4.2]{BS24} gives
	\[
	F^{\rm NS}_{d,n}(\hbar)
	=\frac{(-1)^{m-1}}{m\hbar}\sum_{g\geq0}\hbar^{2g}
	\int_{[\overline M_{g,(m)}(S_n/E_n,dH-\sum_{i=1}^n e_i)]^{\rm vir}}\lambda_g ,
	\]
	with one free contact of order $m$.
	
	Deform $\mathbb P^2$ to the normal cone of $E$ and blow up the disjoint
	sections through $p_i$. The central fibre consists of $\mathbb P^2$ and
	$\mathbb P_E(\mathcal O_E(-E)\oplus\mathcal O_E)$ blown up at the $p_i$
	on its infinity section. The vanishing and dimension arguments
	of~\cite[Lemmas~6.1--6.4 and the proof of Theorem~6.5]{GRZ24}
	apply separately at these distinct points. Only one degree-$m$ fibre
	joining the two boundary divisors and one degree-one tail over each
	$p_i$ survive; the exceptional coefficients $-1$ force the tail degrees.
	All attach once, so the remaining vertex gives the marked integral
	in~\eqref{eq:app-wilson-marked-input}.
	The ordinary fibre contributes $1/m$, cancelling the gluing multiplicity
	$m$, and has no higher-genus term by $\lambda_g^2=0$.
	
	Each tail has normal bundle $\mathcal O_{\mathbb P^1}(-1)$ and series
	$u/(2\sin(u/2))$ with insertion $(-1)^g\lambda_g$
	\cite[Theorem~5.1]{BP05}. Substituting $u=i\hbar$ removes the genus sign
	and gives $\hbar/(2\sinh(\hbar/2))$ per tail, with no factorial since
	the points are labelled. Comparing with~\eqref{eq:app-wilson-marked-input},
	\[
	F^{\rm NS}_{d,n}(\hbar)
	=\hbar^{-n}\left(\frac{\hbar}{2\sinh(\hbar/2)}\right)^n
	[Q^d]\mathcal W_n^{\rm NS}(Q;\hbar),
	\]
	which is~\eqref{eq:app-wilson-NS-blowup}.
\end{proof}

For $n=1$, adding the pure-fibre term gives
\[
\mathcal W_1^{\rm NS}(Q;\hbar)
=1+2\sinh(\hbar/2)\sum_{d\geq1}Q^d
F^{\rm NS,K_{\mathbb F_1}}_{(d-1){e_1}+df_1}(\hbar).
\]
The constant $1$ is the genus-zero pure-fibre contribution;
its higher-genus terms vanish by $\lambda_g^2=0$.

\begin{remark}
Let $f$ be the fibre class of $\overline Y_0\to\mathbb P^2$.
Write $\mathcal W_n^{\rm un}(Q;u)$ for the ordinary (unrefined) connected
$n$-point GW series on $\overline Y_0$: its $Q^du^{2g}$ coefficient,
for $d,g\geq0$, is the invariant in class $\iota_{0*}(dH)+nf$ with
$n$ labelled non-equivariant point insertions represented by points on
the infinity section. For $d\geq1$,
Guo--Wang--Wu~\cite{GuoWangWu} give
\begin{equation*}
	[Q^d]\mathcal W_n^{\rm un}(Q;u)
	=\left(\frac{2\sin(u/2)}{u}\right)^{2n}
	\sum_{g\geq0}u^{2g}
	\int_{[\overline M_{g,0}(\overline Y_n,\iota_{n*}(dH-\sum_{i=1}^n e_i))]^{\rm vir}}1 .
\end{equation*}
For $n=1$, the integral is the local invariant of $K_{\mathbb F_1}$.
\end{remark}

\subsection{Variation of the scalar quantum correction}\label{sec:localp2-wilson-variation}
We establish the following formula for $\cW$ on the TR side.
\begin{proposition}
	The term $\cW$ defined by equation~\eqref{eqn:localp2-Wilson-coefficient} satisfies the following equation:
	\begin{equation}\label{eqn:localp2-wilson-mixed-identity}
		\partial_\epsilon\mathcal W
		=2\pi i\frac{Z^{\rm NP}_{(1),*}}{\left.\partial_wZ^{\rm NP}_{\varnothing}\right|_{w=w_*}}.
	\end{equation}
\end{proposition}
\begin{proof}
	We first establish the relationship between the right-hand side of~\eqref{eqn:localp2-wilson-mixed-identity} and the wave function. We consider 
	$$
	\sum_{k\geq1}\partial_{p_k}Z^{\rm NP}_{\frac12,\frac12}(\epsilon,{\bf 0};\hbar;w_{*}) d(X^k)
	=\sum_{\substack{g\geq0\\m\geq1}}\frac{\hbar^{2g-1+m}}{m!(2\pi i)^m}\omega^{(m)}_{g,1}(z)\left(\partial_w^mZ^{\rm NP}_{\varnothing}\right)_*,
	$$
	where we have used $Z^{\rm NP}_{\varnothing,*}=0$. 
	This equation holds near the boundary point, and we extend the definition of the differential on the left-hand side to the whole Riemann surface by the one on the right-hand side.
	Then by using $\oint_{\cyA}\omega^{(m)}_{g,1}=\delta_{g,0}\delta_{m,1}2\pi i$, we have at $w=w_{*}$,
	\beq\label{eqn:dw-dp}
	\left.\partial_wZ^{\rm NP}_{\varnothing}\right|_{w=w_*} = \oint_{\cyA} \sum_{k\geq1}\partial_{p_k}Z^{\rm NP}_{\frac12,\frac12}({\epsilon},{\bf 0};\hbar;w_{*}) d(X^k).
	\eeq
	Recalling that $\left.\pd_{p_k}s_{\lambda}({\bf p})\right|_{{\bf p}=0}=\frac1k\sum_{b=0}^{k-1}(-1)^b\delta_{\lambda,(k-b,1^b)}$,
	we see
	$$
	\sum_{k\geq1}\partial_{p_k}Z^{\rm NP}_{\frac12,\frac12}({\epsilon},{\bf 0};\hbar;w_{*}) d(X^k)
	=\sum_{a,b\geq0}(-1)^bZ^{\rm NP}_{(a+1,1^b),*}X^{a+b}\,dX.
	$$
	Moreover, by the homogeneous Pl\"ucker relations, for $a,b\geq0$, 
	\[
	Z^{\rm NP}_{(1),*} Z^{\rm NP}_{(a+1,1^b),*}
	={Z^{\rm NP}_{(a+1),*}Z^{\rm NP}_{(1^{b+1}),*}}.
	\]
	To deal with the partition $1^{b+1}$, we use the identities $Z(\epsilon,-{\bf p};-\hbar)=Z(\epsilon,{\bf p};\hbar)$ and $s_{\lambda}(-{\bf p})=(-1)^{|\lambda|}s_{\lambda^{t}}({\bf p})$, together with the Fourier representation of $Z^{\rm NP}_{\mu,\nu}$, to obtain
	\[
	Z^{\rm NP}_{\lambda,*}(\epsilon;-\hbar)
	=-(-1)^{|\lambda|}Z^{\rm NP}_{\lambda^t,*}(\epsilon;\hbar).
	\]
	In particular, $Z^{\rm NP}_{(1),*}$ is even in $\hbar$.
	Putting all these computations together, we get
	$$
	\frac{\left.\partial_wZ^{\rm NP}_{\varnothing}\right|_{w=w_*}}{Z^{\rm NP}_{(1),*}}=\oint_{\cyA}\psi_+(X;\hbar)\psi_-(X;\hbar)\,\frac{dX}{X},
	$$
	where $\psi_{\pm}(X;\hbar)$ are normalized wave functions defined by
	\[
	\psi_\pm(X;\hbar)
	=\frac{X^{\pm\pi i/\hbar-1/2}\Psi_*^X(X;\pm\hbar)}
	{Z^{\rm NP}_{(1),*}(\epsilon;\hbar)}.
	\]
	
	We next connect this $\cyA$-period with $\pd_{\epsilon}\cW$.
	We first prove that $\psi_{\pm}(X;\hbar)$ satisfy the following equation:
	\beq\label{eqn:psi-pm-relation}
	\sfq X^2\psi_-(X;\hbar)\psi_+(\sfq X;\hbar)
	-\sfq^{-1/2}X^{-1}\psi_-(\sfq X;\hbar)\psi_+(X;\hbar)=1.
	\eeq
	This can be deduced from the quantum curve equation as follows.
	Note first that, as discussed above, $\mathcal W(-\hbar)=\mathcal W(\hbar)$ (use $Z^{\rm NP}_{(1,1),*}(\epsilon;-\hbar)=-Z^{\rm NP}_{(2),*}(\epsilon;\hbar)$ and $Z^{\rm NP}_{(1,1),*}=\sfq^{-1}Z^{\rm NP}_{(2),*}$).
	Hence the quantum curve equation gives
	\begin{align*}
		0={}&(X^{-1}+\mathcal W)\psi_+(X;\hbar)
		+\sfq^{1/2}X^{-1}\psi_+(\sfq^{-1}X;\hbar)
		+\sfq X^2\psi_+(\sfq X;\hbar),\\
		0={}&(X^{-1}+\mathcal W)\psi_-(X;\hbar)
		+\sfq^{-1/2}X^{-1}\psi_-(\sfq X;\hbar)
		+\sfq^{-1}X^2\psi_-(\sfq^{-1}X;\hbar).
	\end{align*}
	Multiplying these equations by the opposite waves and subtracting
	shows that the expression on the left-hand side of ~\eqref{eqn:psi-pm-relation}
	is invariant under $X\mapsto \sfq^{-1}X$.
	After cancellation of the fractional powers, this expression has
	a power-series expansion in $X$, so shift invariance makes it constant, which equals the leading term $1$ of the expression.
	This identity, established near $b$, is continued coefficientwise in the formal expansion with respect to $\hbar$ along the same paths as the full TR wave functions.
	
	Now we consider $(\partial_\epsilon\mathcal W)\oint_{\mathcal A}\psi_+(X;\hbar)\psi_-(X;\hbar)\,\frac{dX}{X}$.
	Since $\epsilon$ is independent of $X$, the quantum curve equation gives
	$(\partial_\epsilon\mathcal W)\psi_+(X;\hbar)=-\widehat H\pd_{\epsilon}\psi_+(X;\hbar)$.
	The two wave equations imply
		\[
		(\partial_\epsilon\mathcal W)\psi_+(X;\hbar)\psi_-(X;\hbar)
		=\varphi(\sfq^{-1}X)-\varphi(X),
		\]
	where
	$$
	\varphi(X)=\sfq X^2\psi_-(X;\hbar)\partial_\epsilon\psi_+(\sfq X;\hbar)
	-\sfq^{-1/2}X^{-1}\psi_-(\sfq X;\hbar)\partial_\epsilon\psi_+(X;\hbar).
	$$
	Now we fix a regular $\cyA$-cycle based at $z_0$ and set $X_0=X(z_0)$.
	Integrating along $\cyA$, we obtain
	$$
	(\partial_\epsilon\mathcal W)\oint_{\mathcal A}\psi_+(X;\hbar)\psi_-(X;\hbar)\,\frac{dX}{X}
	=-\int_{\sfq^{-1}X_{0}}^{X_{0}} (\varphi_{\cyA}(X)-\varphi(X))\frac{dX}{X},
	$$
	where $\varphi_{\cyA}(X)$ is its analytic continuation around the $\cyA$-cycle.
	By Definition~\ref{normalizedpsi}, the stable integrals and theta quotient are $\cyA$-periodic.
	Thus the disk integral determines the following multiplier, up to a common locally constant square-root sign that cancels in the bilinear expressions:
	\[
	\psi_{\pm,\cyA}(X;\hbar)
	=e^{\pm\frac{2\pi i}{\hbar}(\epsilon+n\log({X}/{X_0}))}\psi_\pm(X;\hbar),
	\]
	where $n=(2\pi i)^{-1}\oint_{\cyA}\frac{dY}{Y}\in\mathbb Z$ and the difference is determined by
	$$
	\bigg(\int^z\omega_{0,1}\bigg)_{\cyA}-\int^z\omega_{0,1}
	=2\pi i\epsilon+\int_{z_0}^z\big((\omega_{0,1})_{\cyA}-\omega_{0,1}\big)
	=2\pi i\left(\epsilon+n\log\frac{X}{X_0}\right).
	$$
	Therefore, by direct computation and by using~\eqref{eqn:psi-pm-relation}, we get
	\[
	\varphi_{\cyA}(X)-\varphi(X)=\frac{2\pi i}{\hbar}
	\Big[\sfq X^2\psi_-(X;\hbar)\psi_+(\sfq X;\hbar)
	-\sfq^{-1/2}X^{-1}\psi_-(\sfq X;\hbar)\psi_+(X;\hbar)\Big]
	=\frac{2\pi i}{\hbar}.
	\]
	It follows that
	\[
	(\partial_\epsilon\mathcal W)\oint_{\mathcal A}\psi_+(X;\hbar)\psi_-(X;\hbar)\,\frac{dX}{X}
	=-\frac{2\pi i}{\hbar}
	\int_{\sfq^{-1}X_0}^{X_0}\frac{dX}{X}=2\pi i.
	\]
	Combining this with the first period formula proves
	\eqref{eqn:localp2-wilson-mixed-identity}.
\end{proof}

\subsection{Proof of the Wilson-loop identification}\label{sec:localp2-wilson-proof}

\begin{proof}[Proof of Theorem~\ref{thm:localp2-wilson-identification}]
	 Section~\ref{sec:localp2-wilson-blowup} expresses the geometric Wilson loop
		through the closed NS invariants of $K_{\mathbb F_1}$. It remains to
		identify $Q^{-1/3}\mathcal W^{\rm NS}_1$ with the scalar $\cW$ of
		Section~\ref{sec:localp2-quantum-curve}.
	
	Let $Q_{e_1}$ and $Q_H$ be the Novikov variables corresponding to $e_1$ and $H$, respectively.
	{Proposition~\ref{prop:app-wilson-NS-blowup} gives}
	\begin{equation}\label{eqn:localp2-wilson-geometric-comparison}
		\mathcal W^{\rm NS}_1(Q;\hbar)=1+(\sfq^{-1/2}-\sfq^{1/2})[Q_{e_1}^{-1}] \bar F^{\rm NS,\mathbb F_1}(Q_{e_1},Q_H;\hbar)|_{Q_H=Q}.
	\end{equation}
	Here, $\bar F^{\rm NS,\mathbb F_1}(Q_{e_1},Q_H;\hbar)$ is the positive Novikov-degree part of the closed geometric NS potential of $\mathbb F_1$, and $[Q_{e_1}^{-1}]$ stands for taking the coefficient of $Q_{e_1}^{-1}$.
	Choose a smooth anticanonical curve $E\subset\mathbb F_1$ of class $3H-e_1$. By~\cite[Corollary~4.2]{BS24},
	\[\textstyle
	\bar F^{\rm NS,\mathbb F_1}(Q_{e_1},Q_H;\hbar)=\sum_{g\geq0}\hbar^{2g-1}\bar F_g^{\mathbb F_1/E}(Q_{e_1},Q_H),
	\]
	where $\bar F_g^{\mathbb F_1/E}$ is the positive Novikov-degree part of the normalized maximal-contact relative potential in the conventions of~\cite{BFGW21}.
	
	For $K_{\mathbb F_1}$, let $Z^{\mathbb F_1}_{\varnothing}$ be the perturbative closed TR partition function.
	By the closed mirror theorem~\cite{FLZ20} and the vertex/GW correspondence~\cite{LLLZ09,MOOP11}, it is related to the closed vertex partition function $V^{\mathbb F_1}_{\varnothing}$ by
	$$
	\frac{Z^{\mathbb F_1}_{\varnothing}(Q_{e_1},Q_H;\hbar)}
	{V^{\mathbb F_1}_{\varnothing}(Q_{e_1},Q_H;\sfq)}
	=\exp\left(\frac{(\log(Q_H^3/Q_{e_1}))^3}{384\hbar^2}
	-\frac{\log(Q_H^3/Q_{e_1})}{48}\right).
	$$
	We use the classical and genus-one normalization of~\cite{FGZblowup}, with $E^2=8$ and $\chi(\mathbb F_1)=4$, and omit constant-map contributions.
	Here we express the dependence on the closed moduli in terms of the Novikov variables \(Q_{e_1}\) and \(Q_H\).
	We take $\mu=\nu=\frac12$.
	Since $\partial_\epsilon=-(Q_{e_1}\partial_{Q_{e_1}}+3Q_H\partial_{Q_H})$, 
	the Fourier representation~\eqref{eqn:Z-NP} gives
	\[
	Z^{{\rm NP},\mathbb F_1}_{\varnothing}(Q_{e_1},Q_H;\hbar;w)
	=\sum_{l\in\mathbb Z}e^{2\pi i(l+\frac12)(w+\frac12)}
	Z^{\mathbb F_1}_{\varnothing}(Q_{e_1}\sfq^{l+\frac12},Q_H\sfq^{3l+\frac32};\hbar).
	\]
	The blowup equation is equivalent to the vanishing of
	$Z^{{\rm NP},\mathbb F_1}_{\varnothing}$ (see~\cite{FGZblowup}).
	This determines a unique formal solution $w^{\mathbb F_1}_{*}$ with
	$(w^{\mathbb F_1}_{*})^+=O(\hbar)$, and by~\cite{FGZblowup},
	\begin{equation}\label{eqn:localp2-wilson-F1-root}
		[Q_{e_1}^{-1}]w^{\mathbb F_1}_{*}
		=-\frac{1}{2\pi i}[Q_{e_1}^{-1}]
		\left(Q_{e_1}\partial_{Q_{e_1}}+3Q_H\partial_{Q_H}\right)
		\bar F^{\rm NS,\mathbb F_1}(Q_{e_1},Q_H;\hbar).
	\end{equation}
	Furthermore, setting $D=Q\,d/dQ$ and applying $1-3D$
	to~\eqref{eqn:localp2-wilson-geometric-comparison}, we obtain by~\eqref{eqn:localp2-wilson-F1-root}
	\begin{equation}\label{eqn:D-W-NS}
		(1-3D)\mathcal W^{\rm NS}_1
		=1-2\pi i(\sfq^{1/2}-\sfq^{-1/2})\left.[Q_{e_1}^{-1}]w^{\mathbb F_1}_{*}\right|_{Q_H=Q}.
	\end{equation}
	
	We next consider the threefold obtained by flopping the exceptional curve of class $e_1$ in $K_{\mathbb F_1}$
	(the graph is shown in Section~\ref{sec:one-loop}; after the flop it has a triangular loop and one attached vertex).
	Write $\check e_1$ for the class of the flopped curve and denote the corresponding Novikov variable by $Q_{\check e_1}$.
	On the flopped side, we define
	$\check Z^{\mathbb F_1}_{\varnothing}$ using the normalization of local $\mathbb P^2$:
	$$
	\frac{\check Z^{\mathbb F_1}_{\varnothing}(Q_{\check e_1},Q_H;\hbar)}
	{\check V^{\mathbb F_1}_{\varnothing}(Q_{\check e_1},Q_H;\sfq)}
	=\exp\left(\frac{(\log Q_H)^3}{18\hbar^2}
	+\frac{\pi i(\log Q_H)^2}{6\hbar^2}-\frac{\log Q_H}{12}\right).
	$$
	We consider the closed vanishing equation:
	\[
	\check Z^{{\rm NP},\mathbb F_1}_{\varnothing}(Q_{\check e_1},Q_H;\hbar;w)
	=\sum_{l\in\mathbb Z}e^{2\pi i(l+\frac12)(w+\frac12)}
	\check Z^{\mathbb F_1}_{\varnothing}(Q_{\check e_1}\sfq^{-l-\frac12},Q_H\sfq^{3l+\frac32};\hbar)=0.
	\]
	Write $\check w^{\mathbb F_1}_{*}$ for its simple zero satisfying
	$\check w^{\mathbb F_1}_{*}(0,Q_H;\hbar)=w_*(t;\hbar)$, where $Q_H=e^{-t}$.
	Since the flop sends $e_1$ to $-\check e_1$, we have $Q_{\check e_1}=Q_{e_1}^{-1}$.
	Let $V^{\mathbb F_1}_{\varnothing}$ and $\check V^{\mathbb F_1}_{\varnothing}$
	be the corresponding ordinary closed vertex coefficients.
	By the normalized flop formula~\cite[Theorem~4.4(ii)]{KM08},
	\[
	\frac{V^{\mathbb F_1}_{\varnothing}(Q_{e_1},Q_H;\sfq)}{\Pi_1(Q_{e_1})}
	=\frac{\check V^{\mathbb F_1}_{\varnothing}(Q_{\check e_1},Q_H;\sfq)}{\Pi_1(Q_{\check e_1})},
	\qquad \Pi_1(Q)=\prod_{j\geq1}(1-Q\sfq^j)^j.
	\]
	Substituting this into the closed vanishing equations, we obtain
	\begin{equation}\label{eqn:localp2-wilson-flop-comparison}
		[Q_{\check e_1}]\check w^{\mathbb F_1}_{*}
		=-\frac{1}{2\pi i(\sfq^{1/2}-\sfq^{-1/2})}
		+[Q_{e_1}^{-1}]w^{\mathbb F_1}_{*}.
	\end{equation}
	The first term on the right is
	$-\frac{1}{2\pi i}[Q]\log\frac{\Pi_1(Q\sfq^{-1/2})}{\Pi_1(Q\sfq^{1/2})}$, which is determined by
	\[
	\log\frac{\Pi_1(Q\sfq^{-1/2})}{\Pi_1(Q\sfq^{1/2})}
	=\sum_{d\geq1}\frac{Q^d}{d(\sfq^{d/2}-\sfq^{-d/2})}.
	\]
	
	Using the vertex $C_{\lambda\mu\nu}$ and the gluing rule recalled
	in Section~\ref{sec:localp2-2}, the attached vertex gives
	\[
	\check V^{\mathbb F_1}_{\varnothing}(Q_{\check e_1},Q;\sfq)
	=\sum_\lambda(-Q_{\check e_1})^{|\lambda|}
	C_{\lambda^t,\varnothing,\varnothing}(\sfq)V_\lambda(Q;\sfq).
	\]
	Combining this with~\eqref{eqn:localp2-Schur-comparison} and the Fourier representation,
	the closed vanishing equation becomes
	\[
	\sum_\lambda(Q_{\check e_1}Q^{1/3})^{|\lambda|}
	C_{\lambda^t,\varnothing,\varnothing}(\sfq)Z^{\rm NP}_{\lambda}(\epsilon;\hbar;w)=0.
	\]
	Here $Q_{\check e_1}Q^{1/3}$ is fixed along the filling fraction direction.
	Since $C_{(1),\varnothing,\varnothing}=1/(\sfq^{1/2}-\sfq^{-1/2})$, this gives
	\[
	Z^{\rm NP}_{\varnothing}(\epsilon;\hbar;w)
	+\frac{Q_{\check e_1}Q^{1/3}}{\sfq^{1/2}-\sfq^{-1/2}}Z^{\rm NP}_{(1)}(\epsilon;\hbar;w)
	+O(Q_{\check e_1}^2)=0.
	\]
	By solving the closed vanishing equation and then expanding at the simple zero, we get
	\begin{equation}\label{eqn:localp2-wilson-cap-variation}
		[Q_{\check e_1}]\check w^{\mathbb F_1}_{*}(Q_{\check e_1},Q)
		=-\frac{Q^{1/3}}{\sfq^{1/2}-\sfq^{-1/2}}
		\frac{Z^{\rm NP}_{(1),*}}{\left.\partial_wZ^{\rm NP}_{\varnothing}\right|_{w=w_*}}.
	\end{equation}

	Finally, by the definition~\eqref{eqn:localp2-Wilson-coefficient} of $\cW$ and equations~\eqref{eqn:localp2-wilson-cap-variation} and~\eqref{eqn:localp2-wilson-mixed-identity}, we get
	\[
	(1-3D)(Q^{1/3}\cW)
	=-2\pi i(\sfq^{1/2}-\sfq^{-1/2})[Q_{\check e_1}]\check w^{\mathbb F_1}_{*}(Q_{\check e_1},Q).
	\]
	Further by using~\eqref{eqn:D-W-NS} and~\eqref{eqn:localp2-wilson-flop-comparison}, we obtain
	\[
	(1-3D)\bigl[Q^{1/3}\mathcal W-\mathcal W^{\rm NS}_1\bigr]=0.
	\]
	By noticing that both $Q^{1/3}\mathcal W$ and $\mathcal W^{\rm NS}_1$ have non-negative integral powers of $Q$,
	and by using $(1-3D)Q^d=(1-3d)Q^d$ and $1-3d\ne0$ for every integer $d\geq0$, this gives $Q^{1/3}\mathcal W-\mathcal W^{\rm NS}_1=0$.  The theorem is proved.
\end{proof}

$$
$$

\section{Quantum curves for local toric weak del Pezzo surfaces}
 \label{sec:one-loop}
In this section, we extend the argument of Section~\ref{sec:localp2} to the smooth local surfaces $K_S$ associated with star triangulations of convex lattice polygons with a unique interior lattice point.

\subsection{Toric diagram}
\label{sec:one-loop-vertices}
Let $P$ be a convex lattice polygon with unique interior lattice point $0$.
Such a polygon is \emph{reflexive}: every edge has lattice distance one from $0$.
There are $16$ such polygons up to lattice isomorphism.

We denote by $v_1,\ldots,v_{\mathfrak m}$ its boundary lattice points in counterclockwise order.
Then Pick's formula gives
\(
 \det(v_{e},v_{e+1})=1.
\)
Here and in what follows, we identify $v_{{\mathfrak m}+e}$ with $v_e$.
Consider the star triangulation with triangles $(0,v_e,v_{e+1})$.
The cones $\operatorname{Cone}(v_e,v_{e+1})$ and their faces define a smooth toric surface $S$.
The surface $S$ is projective and weak del Pezzo: $-K_S$ is nef and big. 
Putting this into three dimensional space by $(x,y)\mapsto (x,y,1)$ and consider the cone including $0$ and the polygon gives the fan of $\mathcal X=K_S$. 
See, e.g., ~\cite{Ful93} and~\cite{CLS11} for details.

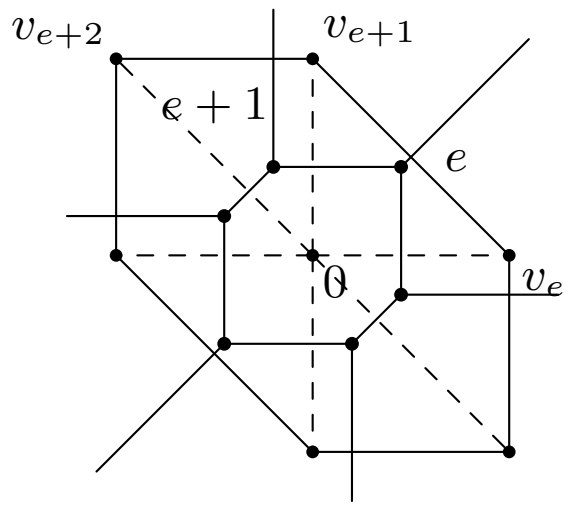
\begin{figure}[htbp]
\centering
\begin{tikzpicture}[x=1.30cm,y=1.30cm,font=\scriptsize,line cap=round,line join=round,
 web/.style={line width=.4pt}]
\draw[line width=.4pt] (1,0)--(0,1)--(-1,1)--(-1,0)--(0,-1)--(1,-1)--cycle;
\foreach \x/\y in {1/0,0/1,-1/1,-1/0,0/-1,1/-1}
  \draw[dashed,line width=.4pt] (0,0)--(\x,\y);
\draw[web] (.45,.45)--(-.20,.45)--(-.45,.20)--(-.45,-.45)--(.20,-.45)--(.45,-.20)--cycle;
\draw[web] (.45,.45)--(1.10,1.10);
\draw[web] (-.20,.45)--(-.20,1.25);
\draw[web] (-.45,.20)--(-1.25,.20);
\draw[web] (-.45,-.45)--(-1.10,-1.10);
\draw[web] (.20,-.45)--(.20,-1.25);
\draw[web] (.45,-.20)--(1.25,-.20);
\foreach \x/\y in {0/0,1/0,0/1,-1/1,-1/0,0/-1,1/-1}
  \fill (\x,\y) circle[radius=1.2pt];
\foreach \x/\y in {.45/.45,-.20/.45,-.45/.20,-.45/-.45,.20/-.45,.45/-.20}
  \fill (\x,\y) circle[radius=1.3pt];
\node[anchor=north west,inner sep=1pt] at (.02,-.025) {$0$};
\node[anchor=north west,inner sep=1pt] at (1.03,-.04) {$v_e$};
\node[anchor=south west,inner sep=1pt] at (.02,1.03) {$v_{e+1}$};
\node[anchor=south east,inner sep=1pt] at (-1.02,1.02) {$v_{e+2}$};
\node[inner sep=1pt] at (.73,.48) {$e$};
\node[inner sep=1pt] at (-.50,.76) {$e+1$};
\end{tikzpicture}
\caption{The star triangulation and its dual graph.}
\label{fig:one-loop-general}
\end{figure}

It is more convenient for us to use the dual graph.
As shown in Fig~\ref{fig:one-loop-general},
the triangle $(0,v_e,v_{e+1})$ is dual to vertex $e$, 
the edge $[v_e,v_{e+1}]$ is dual to its outer leg,
and the edge $[0,v_{e+1}]$ in the polygon is dual to the edge joining vertices $e$ and $e+1$ in the dual graph.
We see that the dual graph consists of one cycle with ${\mathfrak m}$ legs.

\begin{figure}[!tbp]
  \centering
  \begingroup
  \setlength{\tabcolsep}{2pt}
  \renewcommand{\arraystretch}{1.05}
  \scriptsize
  \tikzset{reflexive atlas/.style={x=.50cm,y=.50cm,
    line cap=round,line join=round,line width=.4pt,
    baseline=(current bounding box.center)}}
  \resizebox{\linewidth}{!}{
  \begin{tabular}{|c|*{8}{>{\centering\arraybackslash}m{1.70cm}|}}
  \hline
\raisebox{-.5\height}{\strut $S$} & {\raisebox{-.5\height}{\shortstack{\strut $\mathbb P^2$}}} & {\raisebox{-.5\height}{\shortstack{\strut $\mathbb F_0$}}} & {\raisebox{-.5\height}{\shortstack{\strut $\mathbb F_1$}}} & {\raisebox{-.5\height}{\shortstack{\strut $\mathbb F_2$}}} & {\raisebox{-.5\height}{\shortstack{\strut $dP_2$}}} & {\raisebox{-.5\height}{\shortstack{\strut weak $dP_{2}$\\[-1pt]\strut $(A_{1})$}}} & {\raisebox{-.5\height}{\shortstack{\strut $dP_3$}}} & {\raisebox{-.5\height}{\shortstack{\strut weak $dP_{3}$\\[-1pt]\strut $(2A_{1})$}}} \\[2pt]
  \hline
polygon &
\begin{tikzpicture}[reflexive atlas]
  \path[use as bounding box] (-1.7,-2.1) rectangle (1.7,2.1);
  \begin{scope}[shift={(-0,-0)},scale=1]
  \foreach \x in {-1,0,1}
    \foreach \y in {-1,0,1}
      \fill[black!30] (\x,\y) circle[radius=.3pt];
  \foreach \x/\y in {1/0,-1/1,0/-1}
    \draw[dashed] (0,0)--(\x,\y);
  \draw (1,0)--(-1,1)--(0,-1)--cycle;
  \foreach \x/\y in {1/0,-1/1,0/-1}
    \fill (\x,\y) circle[radius=.85pt];
  \fill (0,0) circle[radius=1.1pt];
  \end{scope}
\end{tikzpicture}
&
\begin{tikzpicture}[reflexive atlas]
  \path[use as bounding box] (-1.7,-2.1) rectangle (1.7,2.1);
  \begin{scope}[shift={(-0,-0)},scale=1]
  \foreach \x in {-1,0,1}
    \foreach \y in {-1,0,1}
      \fill[black!30] (\x,\y) circle[radius=.3pt];
  \foreach \x/\y in {1/0,0/1,-1/0,0/-1}
    \draw[dashed] (0,0)--(\x,\y);
  \draw (1,0)--(0,1)--(-1,0)--(0,-1)--cycle;
  \foreach \x/\y in {1/0,0/1,-1/0,0/-1}
    \fill (\x,\y) circle[radius=.85pt];
  \fill (0,0) circle[radius=1.1pt];
  \end{scope}
\end{tikzpicture}
&
\begin{tikzpicture}[reflexive atlas]
  \path[use as bounding box] (-1.7,-2.1) rectangle (1.7,2.1);
  \begin{scope}[shift={(-0,-0)},scale=1]
  \foreach \x in {-1,0,1}
    \foreach \y in {-1,0,1}
      \fill[black!30] (\x,\y) circle[radius=.3pt];
  \foreach \x/\y in {1/0,0/1,-1/1,0/-1}
    \draw[dashed] (0,0)--(\x,\y);
  \draw (1,0)--(0,1)--(-1,1)--(0,-1)--cycle;
  \foreach \x/\y in {1/0,0/1,-1/1,0/-1}
    \fill (\x,\y) circle[radius=.85pt];
  \fill (0,0) circle[radius=1.1pt];
  \end{scope}
\end{tikzpicture}
&
\begin{tikzpicture}[reflexive atlas]
  \path[use as bounding box] (-1.7,-2.1) rectangle (1.7,2.1);
  \begin{scope}[shift={(-0,-0.5)},scale=1]
  \foreach \x in {-1,0,1}
    \foreach \y in {-1,0,1,2}
      \fill[black!30] (\x,\y) circle[radius=.3pt];
  \foreach \x/\y in {1/0,0/1,-1/2,0/-1}
    \draw[dashed] (0,0)--(\x,\y);
  \draw (1,0)--(0,1)--(-1,2)--(0,-1)--cycle;
  \foreach \x/\y in {1/0,0/1,-1/2,0/-1}
    \fill (\x,\y) circle[radius=.85pt];
  \fill (0,0) circle[radius=1.1pt];
  \end{scope}
\end{tikzpicture}
&
\begin{tikzpicture}[reflexive atlas]
  \path[use as bounding box] (-1.7,-2.1) rectangle (1.7,2.1);
  \begin{scope}[shift={(-0,-0)},scale=1]
  \foreach \x in {-1,0,1}
    \foreach \y in {-1,0,1}
      \fill[black!30] (\x,\y) circle[radius=.3pt];
  \foreach \x/\y in {1/0,0/1,-1/1,-1/0,0/-1}
    \draw[dashed] (0,0)--(\x,\y);
  \draw (1,0)--(0,1)--(-1,1)--(-1,0)--(0,-1)--cycle;
  \foreach \x/\y in {1/0,0/1,-1/1,-1/0,0/-1}
    \fill (\x,\y) circle[radius=.85pt];
  \fill (0,0) circle[radius=1.1pt];
  \end{scope}
\end{tikzpicture}
&
\begin{tikzpicture}[reflexive atlas]
  \path[use as bounding box] (-1.7,-2.1) rectangle (1.7,2.1);
  \begin{scope}[shift={(-0,-0)},scale=1]
  \foreach \x in {-1,0,1}
    \foreach \y in {-1,0,1}
      \fill[black!30] (\x,\y) circle[radius=.3pt];
  \foreach \x/\y in {1/1,0/1,-1/1,-1/0,0/-1}
    \draw[dashed] (0,0)--(\x,\y);
  \draw (1,1)--(0,1)--(-1,1)--(-1,0)--(0,-1)--cycle;
  \foreach \x/\y in {1/1,0/1,-1/1,-1/0,0/-1}
    \fill (\x,\y) circle[radius=.85pt];
  \fill (0,0) circle[radius=1.1pt];
  \end{scope}
\end{tikzpicture}
&
\begin{tikzpicture}[reflexive atlas]
  \path[use as bounding box] (-1.7,-2.1) rectangle (1.7,2.1);
  \begin{scope}[shift={(-0,-0)},scale=1]
  \foreach \x in {-1,0,1}
    \foreach \y in {-1,0,1}
      \fill[black!30] (\x,\y) circle[radius=.3pt];
  \foreach \x/\y in {1/0,0/1,-1/1,-1/0,0/-1,1/-1}
    \draw[dashed] (0,0)--(\x,\y);
  \draw (1,0)--(0,1)--(-1,1)--(-1,0)--(0,-1)--(1,-1)--cycle;
  \foreach \x/\y in {1/0,0/1,-1/1,-1/0,0/-1,1/-1}
    \fill (\x,\y) circle[radius=.85pt];
  \fill (0,0) circle[radius=1.1pt];
  \end{scope}
\end{tikzpicture}
&
\begin{tikzpicture}[reflexive atlas]
  \path[use as bounding box] (-1.7,-2.1) rectangle (1.7,2.1);
  \begin{scope}[shift={(-0,-0)},scale=1]
  \foreach \x in {-1,0,1}
    \foreach \y in {-1,0,1}
      \fill[black!30] (\x,\y) circle[radius=.3pt];
  \foreach \x/\y in {1/1,0/1,-1/1,-1/0,-1/-1,0/-1}
    \draw[dashed] (0,0)--(\x,\y);
  \draw (1,1)--(0,1)--(-1,1)--(-1,0)--(-1,-1)--(0,-1)--cycle;
  \foreach \x/\y in {1/1,0/1,-1/1,-1/0,-1/-1,0/-1}
    \fill (\x,\y) circle[radius=.85pt];
  \fill (0,0) circle[radius=1.1pt];
  \end{scope}
\end{tikzpicture}
\\[1pt]
  \hdashline[2pt/2pt]
\shortstack{dual\\graph} &
\begin{tikzpicture}[reflexive atlas]
  \path[use as bounding box] (-1.7,-2.4) rectangle (1.7,2.4);
  \begin{scope}[shift={(-0.5,0)},scale=1]
  \draw (1,0.5)--(0,-0.5)--(1,-0.5)--cycle;
  \draw (1,0.5)--(1.15,0.8);
  \draw (0,-0.5)--(-0.3,-0.65);
  \draw (1,-0.5)--(1.3,-0.8);
  \foreach \x/\y in {1/0.5,0/-0.5,1/-0.5}
    \fill (\x,\y) circle[radius=.65pt];
  \end{scope}
\end{tikzpicture}
&
\begin{tikzpicture}[reflexive atlas]
  \path[use as bounding box] (-1.7,-2.4) rectangle (1.7,2.4);
  \begin{scope}[shift={(-0.5,0)},scale=1]
  \draw (1,0.5)--(0,0.5)--(0,-0.5)--(1,-0.5)--cycle;
  \draw (1,0.5)--(1.3,0.8);
  \draw (0,0.5)--(-0.3,0.8);
  \draw (0,-0.5)--(-0.3,-0.8);
  \draw (1,-0.5)--(1.3,-0.8);
  \foreach \x/\y in {1/0.5,0/0.5,0/-0.5,1/-0.5}
    \fill (\x,\y) circle[radius=.65pt];
  \end{scope}
\end{tikzpicture}
&
\begin{tikzpicture}[reflexive atlas]
  \path[use as bounding box] (-1.7,-2.4) rectangle (1.7,2.4);
  \begin{scope}[shift={(0.15,0)},scale=1]
  \draw (1,0.5)--(0,0.5)--(-1,-0.5)--(1,-0.5)--cycle;
  \draw (1,0.5)--(1.3,0.8);
  \draw (0,0.5)--(0,0.8);
  \draw (-1,-0.5)--(-1.6,-0.8);
  \draw (1,-0.5)--(1.3,-0.8);
  \foreach \x/\y in {1/0.5,0/0.5,-1/-0.5,1/-0.5}
    \fill (\x,\y) circle[radius=.65pt];
  \end{scope}
\end{tikzpicture}
&
\begin{tikzpicture}[reflexive atlas]
  \path[use as bounding box] (-1.7,-2.4) rectangle (1.7,2.4);
  \begin{scope}[shift={(0.571429,0)},scale=0.714286]
  \draw (1,0.5)--(0,0.5)--(-2,-0.5)--(1,-0.5)--cycle;
  \draw (1,0.5)--(1.3,0.8);
  \draw (0,0.5)--(0.3,0.8);
  \draw (-2,-0.5)--(-2.9,-0.8);
  \draw (1,-0.5)--(1.3,-0.8);
  \foreach \x/\y in {1/0.5,0/0.5,-2/-0.5,1/-0.5}
    \fill (\x,\y) circle[radius=.65pt];
  \end{scope}
\end{tikzpicture}
&
\begin{tikzpicture}[reflexive atlas]
  \path[use as bounding box] (-1.7,-2.4) rectangle (1.7,2.4);
  \begin{scope}[shift={(0,0)},scale=1]
  \draw (0.5,0.5)--(0,0.5)--(-0.5,0)--(-0.5,-0.5)--(0.5,-0.5)--cycle;
  \draw (0.5,0.5)--(0.8,0.8);
  \draw (0,0.5)--(0,0.8);
  \draw (-0.5,0)--(-0.8,0);
  \draw (-0.5,-0.5)--(-0.8,-0.8);
  \draw (0.5,-0.5)--(0.8,-0.8);
  \foreach \x/\y in {0.5/0.5,0/0.5,-0.5/0,-0.5/-0.5,0.5/-0.5}
    \fill (\x,\y) circle[radius=.65pt];
  \end{scope}
\end{tikzpicture}
&
\begin{tikzpicture}[reflexive atlas]
  \path[use as bounding box] (-1.7,-2.4) rectangle (1.7,2.4);
  \begin{scope}[shift={(-1.043478,0)},scale=0.978261]
  \draw (0.666667,0.666667)--(0.333333,0.666667)--(-0.166667,0.166667)--(-0.166667,-0.666667)--(2,-0.666667)--cycle;
  \draw (0.666667,0.666667)--(0.666667,0.966667);
  \draw (0.333333,0.666667)--(0.333333,0.966667);
  \draw (-0.166667,0.166667)--(-0.466667,0.166667);
  \draw (-0.166667,-0.666667)--(-0.466667,-0.966667);
  \draw (2,-0.666667)--(2.6,-0.966667);
  \foreach \x/\y in {0.666667/0.666667,0.333333/0.666667,-0.166667/0.166667,-0.166667/-0.666667,2/-0.666667}
    \fill (\x,\y) circle[radius=.65pt];
  \end{scope}
\end{tikzpicture}
&
\begin{tikzpicture}[reflexive atlas]
  \path[use as bounding box] (-1.7,-2.4) rectangle (1.7,2.4);
  \begin{scope}[shift={(0,0)},scale=1]
  \draw (0.5,0.5)--(-0.2,0.5)--(-0.5,0.2)--(-0.5,-0.5)--(0.2,-0.5)--(0.5,-0.2)--cycle;
  \draw (0.5,0.5)--(0.8,0.8);
  \draw (-0.2,0.5)--(-0.2,0.8);
  \draw (-0.5,0.2)--(-0.8,0.2);
  \draw (-0.5,-0.5)--(-0.8,-0.8);
  \draw (0.2,-0.5)--(0.2,-0.8);
  \draw (0.5,-0.2)--(0.8,-0.2);
  \foreach \x/\y in {0.5/0.5,-0.2/0.5,-0.5/0.2,-0.5/-0.5,0.2/-0.5,0.5/-0.2}
    \fill (\x,\y) circle[radius=.65pt];
  \end{scope}
\end{tikzpicture}
&
\begin{tikzpicture}[reflexive atlas]
  \path[use as bounding box] (-1.7,-2.4) rectangle (1.7,2.4);
  \begin{scope}[shift={(-1.043478,0)},scale=0.978261]
  \draw (0.666667,0.666667)--(0.333333,0.666667)--(-0.166667,0.166667)--(-0.166667,-0.166667)--(0.333333,-0.666667)--(2,-0.666667)--cycle;
  \draw (0.666667,0.666667)--(0.666667,0.966667);
  \draw (0.333333,0.666667)--(0.333333,0.966667);
  \draw (-0.166667,0.166667)--(-0.466667,0.166667);
  \draw (-0.166667,-0.166667)--(-0.466667,-0.166667);
  \draw (0.333333,-0.666667)--(0.333333,-0.966667);
  \draw (2,-0.666667)--(2.6,-0.966667);
  \foreach \x/\y in {0.666667/0.666667,0.333333/0.666667,-0.166667/0.166667,-0.166667/-0.166667,0.333333/-0.666667,2/-0.666667}
    \fill (\x,\y) circle[radius=.65pt];
  \end{scope}
\end{tikzpicture}
\\[1pt]
  \hline
\raisebox{-.5\height}{\strut $S$} & {\raisebox{-.5\height}{\shortstack{\strut weak $dP_{3}$\\[-1pt]\strut $(A_{1})$}}} & {\raisebox{-.5\height}{\shortstack{\strut weak $dP_{3}$\\[-1pt]\strut $(A_{2}+A_{1})$}}} & {\raisebox{-.5\height}{\shortstack{\strut weak $dP_{4}$\\[-1pt]\strut $(2A_{1})$}}} & {\raisebox{-.5\height}{\shortstack{\strut weak $dP_{4}$\\[-1pt]\strut $(A_{2}+A_{1})$}}} & {\raisebox{-.5\height}{\shortstack{\strut weak $dP_{5}$\\[-1pt]\strut $(4A_{1})$}}} & {\raisebox{-.5\height}{\shortstack{\strut weak $dP_{5}$\\[-1pt]\strut $(A_{2}+2A_{1})$}}} & {\raisebox{-.5\height}{\shortstack{\strut weak $dP_{5}$\\[-1pt]\strut $(A_{3}+2A_{1})$}}} & {\raisebox{-.5\height}{\shortstack{\strut weak $dP_{6}$\\[-1pt]\strut $(3A_{2})$}}} \\[2pt]
  \hline
polygon &
\begin{tikzpicture}[reflexive atlas]
  \path[use as bounding box] (-1.7,-2.1) rectangle (1.7,2.1);
  \begin{scope}[shift={(-0,-0)},scale=1]
  \foreach \x in {-1,0,1}
    \foreach \y in {-1,0,1}
      \fill[black!30] (\x,\y) circle[radius=.3pt];
  \foreach \x/\y in {1/0,1/1,0/1,-1/1,-1/0,0/-1}
    \draw[dashed] (0,0)--(\x,\y);
  \draw (1,0)--(1,1)--(0,1)--(-1,1)--(-1,0)--(0,-1)--cycle;
  \foreach \x/\y in {1/0,1/1,0/1,-1/1,-1/0,0/-1}
    \fill (\x,\y) circle[radius=.85pt];
  \fill (0,0) circle[radius=1.1pt];
  \end{scope}
\end{tikzpicture}
&
\begin{tikzpicture}[reflexive atlas]
  \path[use as bounding box] (-1.7,-2.1) rectangle (1.7,2.1);
  \begin{scope}[shift={(0.5,-0)},scale=1]
  \foreach \x in {-2,-1,0,1}
    \foreach \y in {-1,0,1}
      \fill[black!30] (\x,\y) circle[radius=.3pt];
  \foreach \x/\y in {1/1,0/1,-1/1,-2/1,-1/0,0/-1}
    \draw[dashed] (0,0)--(\x,\y);
  \draw (1,1)--(0,1)--(-1,1)--(-2,1)--(-1,0)--(0,-1)--cycle;
  \foreach \x/\y in {1/1,0/1,-1/1,-2/1,-1/0,0/-1}
    \fill (\x,\y) circle[radius=.85pt];
  \fill (0,0) circle[radius=1.1pt];
  \end{scope}
\end{tikzpicture}
&
\begin{tikzpicture}[reflexive atlas]
  \path[use as bounding box] (-1.7,-2.1) rectangle (1.7,2.1);
  \begin{scope}[shift={(-0,-0)},scale=1]
  \foreach \x in {-1,0,1}
    \foreach \y in {-1,0,1}
      \fill[black!30] (\x,\y) circle[radius=.3pt];
  \foreach \x/\y in {1/0,1/1,0/1,-1/1,-1/0,-1/-1,0/-1}
    \draw[dashed] (0,0)--(\x,\y);
  \draw (1,0)--(1,1)--(0,1)--(-1,1)--(-1,0)--(-1,-1)--(0,-1)--cycle;
  \foreach \x/\y in {1/0,1/1,0/1,-1/1,-1/0,-1/-1,0/-1}
    \fill (\x,\y) circle[radius=.85pt];
  \fill (0,0) circle[radius=1.1pt];
  \end{scope}
\end{tikzpicture}
&
\begin{tikzpicture}[reflexive atlas]
  \path[use as bounding box] (-1.7,-2.1) rectangle (1.7,2.1);
  \begin{scope}[shift={(0.5,-0)},scale=1]
  \foreach \x in {-2,-1,0,1}
    \foreach \y in {-1,0,1}
      \fill[black!30] (\x,\y) circle[radius=.3pt];
  \foreach \x/\y in {1/0,1/1,0/1,-1/1,-2/1,-1/0,0/-1}
    \draw[dashed] (0,0)--(\x,\y);
  \draw (1,0)--(1,1)--(0,1)--(-1,1)--(-2,1)--(-1,0)--(0,-1)--cycle;
  \foreach \x/\y in {1/0,1/1,0/1,-1/1,-2/1,-1/0,0/-1}
    \fill (\x,\y) circle[radius=.85pt];
  \fill (0,0) circle[radius=1.1pt];
  \end{scope}
\end{tikzpicture}
&
\begin{tikzpicture}[reflexive atlas]
  \path[use as bounding box] (-1.7,-2.1) rectangle (1.7,2.1);
  \begin{scope}[shift={(-0,-0)},scale=1]
  \foreach \x in {-1,0,1}
    \foreach \y in {-1,0,1}
      \fill[black!30] (\x,\y) circle[radius=.3pt];
  \foreach \x/\y in {1/0,1/1,0/1,-1/1,-1/0,-1/-1,0/-1,1/-1}
    \draw[dashed] (0,0)--(\x,\y);
  \draw (1,0)--(1,1)--(0,1)--(-1,1)--(-1,0)--(-1,-1)--(0,-1)--(1,-1)--cycle;
  \foreach \x/\y in {1/0,1/1,0/1,-1/1,-1/0,-1/-1,0/-1,1/-1}
    \fill (\x,\y) circle[radius=.85pt];
  \fill (0,0) circle[radius=1.1pt];
  \end{scope}
\end{tikzpicture}
&
\begin{tikzpicture}[reflexive atlas]
  \path[use as bounding box] (-1.7,-2.1) rectangle (1.7,2.1);
  \begin{scope}[shift={(-0.5,-0)},scale=1]
  \foreach \x in {-1,0,1,2}
    \foreach \y in {-1,0,1}
      \fill[black!30] (\x,\y) circle[radius=.3pt];
  \foreach \x/\y in {1/0,2/1,1/1,0/1,-1/1,-1/0,-1/-1,0/-1}
    \draw[dashed] (0,0)--(\x,\y);
  \draw (1,0)--(2,1)--(1,1)--(0,1)--(-1,1)--(-1,0)--(-1,-1)--(0,-1)--cycle;
  \foreach \x/\y in {1/0,2/1,1/1,0/1,-1/1,-1/0,-1/-1,0/-1}
    \fill (\x,\y) circle[radius=.85pt];
  \fill (0,0) circle[radius=1.1pt];
  \end{scope}
\end{tikzpicture}
&
\begin{tikzpicture}[reflexive atlas]
  \path[use as bounding box] (-1.7,-2.1) rectangle (1.7,2.1);
  \begin{scope}[shift={(-0,-0)},scale=0.75]
  \foreach \x in {-2,-1,0,1,2}
    \foreach \y in {-2,-1,0,1,2}
      \fill[black!30] (\x,\y) circle[radius=.3pt];
  \foreach \x/\y in {1/0,2/1,1/1,0/1,-1/1,-2/1,-1/0,0/-1}
    \draw[dashed] (0,0)--(\x,\y);
  \draw (1,0)--(2,1)--(1,1)--(0,1)--(-1,1)--(-2,1)--(-1,0)--(0,-1)--cycle;
  \foreach \x/\y in {1/0,2/1,1/1,0/1,-1/1,-2/1,-1/0,0/-1}
    \fill (\x,\y) circle[radius=.85pt];
  \fill (0,0) circle[radius=1.1pt];
  \end{scope}
\end{tikzpicture}
&
\begin{tikzpicture}[reflexive atlas]
  \path[use as bounding box] (-1.7,-2.1) rectangle (1.7,2.1);
  \begin{scope}[shift={(-0.5,-0.5)},scale=1]
  \foreach \x in {-1,0,1,2}
    \foreach \y in {-1,0,1,2}
      \fill[black!30] (\x,\y) circle[radius=.3pt];
  \foreach \x/\y in {1/0,0/1,-1/2,-1/1,-1/0,-1/-1,0/-1,1/-1,2/-1}
    \draw[dashed] (0,0)--(\x,\y);
  \draw (1,0)--(0,1)--(-1,2)--(-1,1)--(-1,0)--(-1,-1)--(0,-1)--(1,-1)--(2,-1)--cycle;
  \foreach \x/\y in {1/0,0/1,-1/2,-1/1,-1/0,-1/-1,0/-1,1/-1,2/-1}
    \fill (\x,\y) circle[radius=.85pt];
  \fill (0,0) circle[radius=1.1pt];
  \end{scope}
\end{tikzpicture}
\\[1pt]
  \hdashline[2pt/2pt]
\shortstack{dual\\graph} &
\begin{tikzpicture}[reflexive atlas]
  \path[use as bounding box] (-1.7,-2.4) rectangle (1.7,2.4);
  \begin{scope}[shift={(-0.5,0)},scale=1]
  \draw (1.166667,0.166667)--(0.666667,0.666667)--(0.333333,0.666667)--(-0.166667,0.166667)--(-0.166667,-0.666667)--(1.166667,-0.666667)--cycle;
  \draw (1.166667,0.166667)--(1.466667,0.166667);
  \draw (0.666667,0.666667)--(0.666667,0.966667);
  \draw (0.333333,0.666667)--(0.333333,0.966667);
  \draw (-0.166667,0.166667)--(-0.466667,0.166667);
  \draw (-0.166667,-0.666667)--(-0.466667,-0.966667);
  \draw (1.166667,-0.666667)--(1.466667,-0.966667);
  \foreach \x/\y in {1.166667/0.166667,0.666667/0.666667,0.333333/0.666667,-0.166667/0.166667,-0.166667/-0.666667,1.166667/-0.666667}
    \fill (\x,\y) circle[radius=.65pt];
  \end{scope}
\end{tikzpicture}
&
\begin{tikzpicture}[reflexive atlas]
  \path[use as bounding box] (-1.7,-2.4) rectangle (1.7,2.4);
  \begin{scope}[shift={(-1.043478,0)},scale=0.978261]
  \draw (0.666667,0.666667)--(0.333333,0.666667)--(0,0.333333)--(-0.166667,0)--(-0.166667,-0.666667)--(2,-0.666667)--cycle;
  \draw (0.666667,0.666667)--(0.666667,0.966667);
  \draw (0.333333,0.666667)--(0.333333,0.966667);
  \draw (0,0.333333)--(0,0.633333);
  \draw (-0.166667,0)--(-0.466667,-0.3);
  \draw (-0.166667,-0.666667)--(-0.466667,-0.966667);
  \draw (2,-0.666667)--(2.6,-0.966667);
  \foreach \x/\y in {0.666667/0.666667,0.333333/0.666667,0/0.333333,-0.166667/0,-0.166667/-0.666667,2/-0.666667}
    \fill (\x,\y) circle[radius=.65pt];
  \end{scope}
\end{tikzpicture}
&
\begin{tikzpicture}[reflexive atlas]
  \path[use as bounding box] (-1.7,-2.4) rectangle (1.7,2.4);
  \begin{scope}[shift={(-0.5,0)},scale=1]
  \draw (1.166667,0.166667)--(0.666667,0.666667)--(0.333333,0.666667)--(-0.166667,0.166667)--(-0.166667,-0.166667)--(0.333333,-0.666667)--(1.166667,-0.666667)--cycle;
  \draw (1.166667,0.166667)--(1.466667,0.166667);
  \draw (0.666667,0.666667)--(0.666667,0.966667);
  \draw (0.333333,0.666667)--(0.333333,0.966667);
  \draw (-0.166667,0.166667)--(-0.466667,0.166667);
  \draw (-0.166667,-0.166667)--(-0.466667,-0.166667);
  \draw (0.333333,-0.666667)--(0.333333,-0.966667);
  \draw (1.166667,-0.666667)--(1.466667,-0.966667);
  \foreach \x/\y in {1.166667/0.166667,0.666667/0.666667,0.333333/0.666667,-0.166667/0.166667,-0.166667/-0.166667,0.333333/-0.666667,1.166667/-0.666667}
    \fill (\x,\y) circle[radius=.65pt];
  \end{scope}
\end{tikzpicture}
&
\begin{tikzpicture}[reflexive atlas]
  \path[use as bounding box] (-1.7,-2.4) rectangle (1.7,2.4);
  \begin{scope}[shift={(-0.5,0)},scale=1]
  \draw (1.166667,0.166667)--(0.666667,0.666667)--(0.333333,0.666667)--(0,0.333333)--(-0.166667,0)--(-0.166667,-0.666667)--(1.166667,-0.666667)--cycle;
  \draw (1.166667,0.166667)--(1.466667,0.166667);
  \draw (0.666667,0.666667)--(0.666667,0.966667);
  \draw (0.333333,0.666667)--(0.333333,0.966667);
  \draw (0,0.333333)--(0,0.633333);
  \draw (-0.166667,0)--(-0.466667,-0.3);
  \draw (-0.166667,-0.666667)--(-0.466667,-0.966667);
  \draw (1.166667,-0.666667)--(1.466667,-0.966667);
  \foreach \x/\y in {1.166667/0.166667,0.666667/0.666667,0.333333/0.666667,0/0.333333,-0.166667/0,-0.166667/-0.666667,1.166667/-0.666667}
    \fill (\x,\y) circle[radius=.65pt];
  \end{scope}
\end{tikzpicture}
&
\begin{tikzpicture}[reflexive atlas]
  \path[use as bounding box] (-1.7,-2.4) rectangle (1.7,2.4);
  \begin{scope}[shift={(-0.5,0)},scale=1]
  \draw (1.166667,0.166667)--(0.666667,0.666667)--(0.333333,0.666667)--(-0.166667,0.166667)--(-0.166667,-0.166667)--(0.333333,-0.666667)--(0.666667,-0.666667)--(1.166667,-0.166667)--cycle;
  \draw (1.166667,0.166667)--(1.466667,0.166667);
  \draw (0.666667,0.666667)--(0.666667,0.966667);
  \draw (0.333333,0.666667)--(0.333333,0.966667);
  \draw (-0.166667,0.166667)--(-0.466667,0.166667);
  \draw (-0.166667,-0.166667)--(-0.466667,-0.166667);
  \draw (0.333333,-0.666667)--(0.333333,-0.966667);
  \draw (0.666667,-0.666667)--(0.666667,-0.966667);
  \draw (1.166667,-0.166667)--(1.466667,-0.166667);
  \foreach \x/\y in {1.166667/0.166667,0.666667/0.666667,0.333333/0.666667,-0.166667/0.166667,-0.166667/-0.166667,0.333333/-0.666667,0.666667/-0.666667,1.166667/-0.166667}
    \fill (\x,\y) circle[radius=.65pt];
  \end{scope}
\end{tikzpicture}
&
\begin{tikzpicture}[reflexive atlas]
  \path[use as bounding box] (-1.7,-2.4) rectangle (1.7,2.4);
  \begin{scope}[shift={(-0.5,0)},scale=1]
  \draw (1.166667,0)--(1,0.333333)--(0.666667,0.666667)--(0.333333,0.666667)--(-0.166667,0.166667)--(-0.166667,-0.166667)--(0.333333,-0.666667)--(1.166667,-0.666667)--cycle;
  \draw (1.166667,0)--(1.466667,-0.3);
  \draw (1,0.333333)--(1,0.633333);
  \draw (0.666667,0.666667)--(0.666667,0.966667);
  \draw (0.333333,0.666667)--(0.333333,0.966667);
  \draw (-0.166667,0.166667)--(-0.466667,0.166667);
  \draw (-0.166667,-0.166667)--(-0.466667,-0.166667);
  \draw (0.333333,-0.666667)--(0.333333,-0.966667);
  \draw (1.166667,-0.666667)--(1.466667,-0.966667);
  \foreach \x/\y in {1.166667/0,1/0.333333,0.666667/0.666667,0.333333/0.666667,-0.166667/0.166667,-0.166667/-0.166667,0.333333/-0.666667,1.166667/-0.666667}
    \fill (\x,\y) circle[radius=.65pt];
  \end{scope}
\end{tikzpicture}
&
\begin{tikzpicture}[reflexive atlas]
  \path[use as bounding box] (-1.7,-2.4) rectangle (1.7,2.4);
  \begin{scope}[shift={(-0.5,0)},scale=1]
  \draw (1.166667,0)--(1,0.333333)--(0.666667,0.666667)--(0.333333,0.666667)--(0,0.333333)--(-0.166667,0)--(-0.166667,-0.666667)--(1.166667,-0.666667)--cycle;
  \draw (1.166667,0)--(1.466667,-0.3);
  \draw (1,0.333333)--(1,0.633333);
  \draw (0.666667,0.666667)--(0.666667,0.966667);
  \draw (0.333333,0.666667)--(0.333333,0.966667);
  \draw (0,0.333333)--(0,0.633333);
  \draw (-0.166667,0)--(-0.466667,-0.3);
  \draw (-0.166667,-0.666667)--(-0.466667,-0.966667);
  \draw (1.166667,-0.666667)--(1.466667,-0.966667);
  \foreach \x/\y in {1.166667/0,1/0.333333,0.666667/0.666667,0.333333/0.666667,0/0.333333,-0.166667/0,-0.166667/-0.666667,1.166667/-0.666667}
    \fill (\x,\y) circle[radius=.65pt];
  \end{scope}
\end{tikzpicture}
&
\begin{tikzpicture}[reflexive atlas]
  \path[use as bounding box] (-1.7,-2.4) rectangle (1.7,2.4);
  \begin{scope}[shift={(-0.5,0)},scale=1]
  \draw (1.166667,0.666667)--(0.5,0.666667)--(0.166667,0.5)--(-0.166667,0.166667)--(-0.166667,-0.166667)--(0.333333,-0.666667)--(0.666667,-0.666667)--(1,-0.333333)--(1.166667,0)--cycle;
  \draw (1.166667,0.666667)--(1.466667,0.966667);
  \draw (0.5,0.666667)--(0.8,0.966667);
  \draw (0.166667,0.5)--(-0.133333,0.5);
  \draw (-0.166667,0.166667)--(-0.466667,0.166667);
  \draw (-0.166667,-0.166667)--(-0.466667,-0.166667);
  \draw (0.333333,-0.666667)--(0.333333,-0.966667);
  \draw (0.666667,-0.666667)--(0.666667,-0.966667);
  \draw (1,-0.333333)--(1,-0.633333);
  \draw (1.166667,0)--(1.466667,0.3);
  \foreach \x/\y in {1.166667/0.666667,0.5/0.666667,0.166667/0.5,-0.166667/0.166667,-0.166667/-0.166667,0.333333/-0.666667,0.666667/-0.666667,1/-0.333333,1.166667/0}
    \fill (\x,\y) circle[radius=.65pt];
  \end{scope}
\end{tikzpicture}
\\[1pt]
  \hline
  \end{tabular}
  }
  \endgroup
  {
  \caption{The sixteen smooth toric weak del Pezzo surfaces and their polygons
  and dual graphs. Here $A_r$ denotes a chain of $r$ $(-2)$-curves;
  sums and multiples denote disjoint unions.}
  \label{fig:one-loop-examples}}
\end{figure}
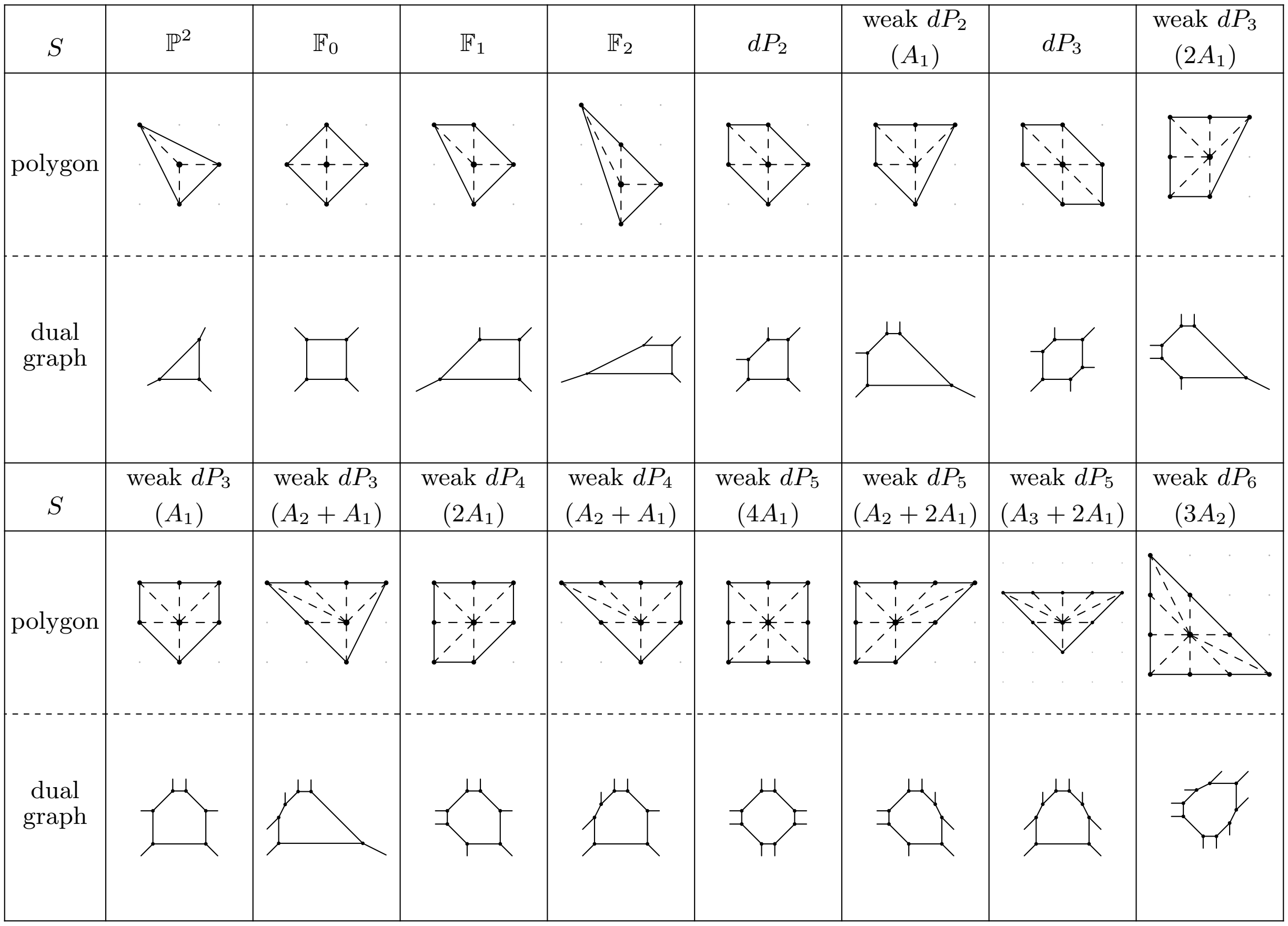

Figure~\ref{fig:one-loop-examples} displays all $16$ reflexive polygons, labelled by their associated smooth toric surfaces.
Here $\mathbb F_n=\mathbb P_{\mathbb P^1}(\mathcal O\oplus\mathcal O(n))$ is the $n$-th Hirzebruch surface, 
and (weak) $dP_n$ denotes a (weak) del Pezzo surface obtained by blowing up $\mathbb P^2$ at $n$ points in (almost) general position.
The labels below ``weak $dP_n$'' indicate the Dynkin types of their $(-2)$-curve configurations (see, e.g.,~\cite{DL10}).

Let $D_e$ denote the class of the torus-invariant curve corresponding to the ray $\mathbb R_{\geq0}v_e$, and let $t\in H^2(S;\mathbb C)$.
We introduce
\[
Q_e:=e^{-\int_{D_e}t}.
\]
Denote by $(v_{e,1},v_{e,2})$ the coordinate of $v_e$ in $\mathbb R^2$, then one has the principal divisor relations~\cite{CLS11}:
$\sum_{e=1}^{\mathfrak m} v_{e,1}D_e=\sum_{e=1}^{\mathfrak m} v_{e,2}D_e=0$.
This gives, for every $v\in\mathbb Z^2$,
\[
\prod_{e=1}^{\mathfrak m} Q_e^{\det(v,v_e)}=1.
\]
Let $f_e$ be the framing integer of the internal edge dual to $[0,v_e]$ as in~\eqref{eqn:framed-gluing-pair}.
With the chosen ordering of the legs, the framing rule of~\cite{AKMV} gives
$v_{e-1}+v_{e+1}=(1-f_{e})v_{e}$.
For a curve class $\beta=\sum_ed_eD_e$, it is known~\cite{CLS11} that 
\begin{equation*}
	D_{e}\cdot\beta=d_{e-1}+(f_{e}-1)d_{e}+d_{e+1},
	\qquad 1\leq e\leq {\mathfrak m}.
\end{equation*}
Moreover, let $E=-K_S=\sum_e D_e$ be the anticanonical class of $S$, then we have 
$$
E\cdot\beta=\sum_{e=1}^{\mathfrak m}(f_e+1)d_e,\qquad
E^2=\sum_{e=1}^{\mathfrak m}(f_e+1)=12-\mathfrak m.
$$

Place the vertex state $|C_i\rangle$ at vertex $i$, with its legs
ordered as external, outgoing, and incoming.
Let $\langle P_i|$ be the pairing~\eqref{eqn:framed-gluing-pair}
with parameters $(Q_i,f_i)$ on the internal edge dual to $[0,v_i]$.
The glued state is then given by~\eqref{eqn:glued-state}.
Write its coefficients as $V_{\lambda_1\cdots\lambda_{\mathfrak m}}$,
where $\lambda_i$ labels the external leg at vertex $i$.
Explicitly, the gluing rule~\cite{AKMV,LLLZ09,MOOP11,DZ12} gives
\begin{equation}\label{eqn:one-loop-gluing}
 V_{\lambda_1\cdots\lambda_{\mathfrak m}}(\mathbf Q;\sfq)
 =\sum_{\eta_1,\ldots,\eta_{\mathfrak m}}\prod_{i=1}^{\mathfrak m}
 \bigl((-1)^{f_{i+1}+1}Q_{i+1}\bigr)^{|\eta_i|}
 \sfq^{f_{i+1}\kappa_{\eta_i}/2}C_{\lambda_i,\eta_i,\eta_{i-1}^t}(\sfq).
\end{equation}
Here $\eta_0=\eta_{\mathfrak m}$, and $Q_{\mathfrak m+1}=Q_1$, $f_{\mathfrak m+1}=f_1$.

For an outer leg $\ell$, let $L$ be the corresponding outer Aganagic--Vafa brane with framing zero.
We use the geometric invariants of~\cite[\S 3.13]{FLZ20} and the correlators defined in~\eqref{eqn:corr-GW}.
Set $V_\lambda=V_{\varnothing\cdots\lambda\cdots\varnothing}$, with $\lambda$ in the $\ell$-th position, and $|V^{(\ell)}\rangle=\sum_\lambda V_\lambda|\lambda\rangle$.
The one-leg partition function is
\begin{equation}\label{eqn:one-loop-Z-TV}
 Z^{\rm TV}(\mathbf Q,\mathbf p;\sfq)
 =\langle0|e^{\sum_{k\geq1}p_k\alpha_k/k}|V^{(\ell)}\rangle
 =\sum_\lambda V_\lambda(\mathbf Q;\sfq)s_\lambda(\mathbf p).
\end{equation}
With $\sfq=e^{-\hbar}$ and the conventions of Section~\ref{sec:localp2}, the vertex/GW comparison~\cite{LLLZ09,MOOP11,FLZ20} reads
\begin{equation}\label{eqn:one-loop-TV-geometry}
 \langle\alpha_{\mu_1},\ldots,\alpha_{\mu_n}\rangle^{\mathcal X,L}_{g,n}
 =(-1)^{g-1+n+|\mu|}[\hbar^{2g-2+n}]
\prod_{i=1}^n(\mu_i\partial_{p_{\mu_i}})\log Z^{\rm TV}\bigg|_{\mathbf p=0},\qquad n\geq1,
\end{equation}
where $|\mu|=\mu_1+\cdots+\mu_n$.
This includes the disk and annulus.
The coefficient $V_\varnothing$ is the closed vertex partition function, normalized by $V_\varnothing(\mathbf0;\sfq)=1$; classical and constant-map contributions are omitted.

\subsection{Mirror map and Mirror curve}
\label{sec:one-loop-TR}
On the B-side, we denote the corresponding variable by ${\cq}$, which satisfies $\cq^{\beta+\beta'}=\cq^\beta\cq^{\beta'}$ and $\cq^0=1$.
Let $\cq_e=\cq^{D_e}$, then for every $v\in\mathbb Z^2$,
\(
 \prod_{e=1}^{\mathfrak m}\cq_e^{\det(v,v_e)}=1.
\)
Following~\cite{FLZ20}, we define
\begin{equation}\label{eqn:one-loop-interior-mirror-series}
 \I(\cq)
 =\sum_{\substack{E\cdot\beta>0;\,  D_i\cdot\beta\geq0}}
 \frac{(-1)^{E\cdot\beta-1}\bigl(E\cdot\beta-1\bigr)!}
      {\prod_{i=1}^{\mathfrak m}(D_i\cdot\beta)!}\,\cq^\beta,
\end{equation}
and
\begin{equation}\label{eqn:one-loop-mirror-series}
A_e(\cq)
=\sum_{\substack{E\cdot\beta=0;\,  D_{e}\cdot \beta <0 \\ D_i\cdot\beta\geq0,\, (i\ne e)}}
\frac{(-1)^{-D_e\cdot\beta-1}(-D_e\cdot\beta-1)!}
{\prod_{\substack{1\leq i\leq {\mathfrak m}\\i\ne e}}(D_i\cdot\beta)!}
\,\cq^\beta.
\end{equation}
Then the closed mirror map is
\begin{equation}\label{eqn:one-loop-mirror-map}
 \log Q_e=\log\cq_e-(f_e+1)\I(\cq)
 +\sum_{i=1}^{\mathfrak m}(D_e\cdot D_i)A_i(\cq),
 \qquad 1\leq e\leq {\mathfrak m}.
\end{equation}

The polygon $P$ is also the Newton polygon of the mirror curve, whose smooth compactification has genus
\(
\#\bigl(\operatorname{Int}(P)\cap\mathbb Z^2\bigr)=1.
\)
See~\cite[\S4.2]{FLZ20}.
To describe the mirror curve, fix an outer leg $\ell$ (dual to segment $[v_\ell,v_{\ell+1}]$).
In the lattice basis $(-v_{\ell+1},v_\ell-v_{\ell+1})$, we write
\[
 v_k=n_kv_\ell-(m_k+n_k)v_{\ell+1}.
\]
Clearly, $n_k=\det(v_k,v_{\ell+1})$, $m_k=-\det(v_k,v_{\ell+1})-\det(v_\ell,v_k)$, and $m_k\geq-1$ by convexity.
Let
$\gamma=\gamma(\cq)=\left(\prod_{i=1}^{\mathfrak m}\cq_i\right)^{-{1}/{(12-{\mathfrak m})}}$,
and
$$
 r_k=r_k(\cq)
 =\gamma(\cq)^{m_k+1}e^{n_k(A_{\ell+1}(\cq)-A_\ell(\cq))}
 \prod_{j=k}^{\ell}\cq_j^{\det(v_k,v_j)},
 \qquad 1\leq k\leq {\mathfrak m}.
$$
Here, the product follows the cyclic order from $k$ to $\ell$. 
The mirror curve~\cite{FLZ20} is given by
\begin{equation}\label{eqn:one-loop-mirror-curve}
	H(X,Y;\cq)
	=\gamma(\cq)+\sum_{k=1}^{\mathfrak m}
	r_k(\cq)X^{m_k}Y^{n_k}=0.
\end{equation}
For this mirror curve, we have a boundary point $(X,Y)=(0,-1)$.

\begin{remark}
By summing~\eqref{eqn:one-loop-mirror-map}, we have
$\gamma(\cq)e^{\I(\cq)}
=\left(\prod_{i=1}^{\mathfrak m} Q_i\right)^{-1/{(12-{\mathfrak m})}}$.
And we have by the relation $v_{i-1}+v_{i+1}=(1-f_{i})v_{i}$ that
\begin{equation*}\textstyle
 \cq_i
 =\gamma(\cq)^{-(f_i+1)}\prod_{j=1}^{\mathfrak m} r_j(\cq)^{D_i\cdot D_j},
 \qquad 1\leq i\leq {\mathfrak m}.
\end{equation*}
For local $\mathbb P^2$, one has ${\mathfrak m}=3$,
$\gamma=\cq^{-1/3}=-3\phi$, and $r_1=r_2=r_3=1$.
\end{remark}

Let $\Sigma$ be the smooth compactification of~\eqref{eqn:one-loop-mirror-curve}.
We take $x=-\log X$, $y=-\log Y$, and the $A$-normalized Bergman
kernel compatible with the mirror theorem~\cite{FLZ20}.
Let $b$ be the boundary point $(X,Y)=(0,-1)$.
These data define the TR partition functions
$Z(\epsilon{,}\mathbf p;\hbar)$ and
$Z^{\rm NP}_{\mu,\nu}(\epsilon{,}\mathbf p;\hbar;w)$
as in Section~\ref{sec:quantum_curve}.

We take the filling fraction variation at fixed boundary coefficients
$r_1,\ldots,r_{\mathfrak m}$.
Varying the coefficient $\gamma$ of the unique interior monomial gives
a holomorphic variation of $y\,dx$~\cite[\S 5.3]{FLZ20}, and hence
the filling fraction direction~\cite[\S 5.3]{EO07}.
We use the large-radius period $\epsilon=\frac{1}{2\pi i}\oint_{\cyA}y\,dx$,
with $\gamma\partial_\gamma\epsilon=1+O(\cq)$.
In this normalization, the flat-coordinate relations
\cite{GKMR15} identify the filling fraction
direction with the anticanonical direction:
\begin{equation}\label{eqn:one-loop-TR-Novikov-direction}
 \partial_\epsilon t=c_1(S),\qquad
 \partial_\epsilon Q^\beta=-(E\cdot\beta)Q^\beta.
\end{equation}
Comparing this with~\cite{FGZblowup}, we see $\pd_{\epsilon}$ equals to the operator $-D$ in~\cite{FGZblowup}.
Similarly as the local $\mathbb P^2$ case, we have
${\partial_\epsilon^mF_g=(-1)^m\mathcal T_{g,m}((-1)^E\mathbf Q)}$ and
$\mathcal R_{g,1}((-1)^E\mathbf Q)=\partial_\epsilon F_g^{S/E}(\mathbf Q)$
for $m\geq1$ and $2g-2+m>0$.
Therefore, we have the following formula of the canonical vanishing point 
\begin{equation}\label{eqn:one-loop-TR-NS-point}
 w_*^+=\frac{1}{2\pi i}\sum_{g\geq1}\hbar^{2g-1}
 \partial_\epsilon F_g^{S/E}(\mathbf Q).
\end{equation}
By~\cite[Corollary~4.2]{BS24}, the relative potential in positive anticanonical degree is the geometric NS limit of the corresponding refined GW theory.

\subsection{Coefficient relation}
\label{sec:one-loop-TR-Schur}
We establish the coefficient relation for local toric weak del Pezzo surfaces, generalizing that for the local $\mathbb P^2$ case and preparing for the proof of the canonical quantum curve conjecture.
Similar as local $\mathbb P^2$, we consider the Schur expansion:
$$
 Z(\epsilon,\mathbf p;\hbar)=\sum_\lambda Z_\lambda(\epsilon;\hbar)s_\lambda(\mathbf p).
$$
\begin{proposition}\label{prop:one-loop-TR-row-hook}
For $n\geq1$, we have
\begin{align}
 \sum_{k=1}^{\mathfrak m}(-1)^{n_k}r_k \sfq^{-n_k(n-\frac{m_k+1}{2})}Z_{(n-m_k)}
 &+\bigg(\sum_{m_k=-1}(-1)^{n_k}r_k\sfq^{n_k}\bigg)Z_{(n,1)}\nonumber \\
 &-\bigg(\sum_{m_k=0}(-1)^{n_k}r_k\sfq^{\frac{n_k}{2}}\bigg)Z_{(n)}=0.\label{eqn:one-loop-TR-row-hook}
\end{align}
\end{proposition}
\begin{proof}
The proof is parallel to the local $\mathbb P^2$ case, so we just show the sketch and the necessary formulas and omit the details.

For $(a,b)\in\mathbb Z^2\setminus\{(0,0)\}$, define $a_i$ and $b_i$ by
\[
 b_iv_i+(a_i-b_i)v_{i+1}=bv_1+(a-b)v_2,\qquad 1\leq i\leq\mathfrak m.
\]
Applying~\eqref{eqn:localp2-vertex-symmetry} at every vertex and
transferring the internal currents through the pairings,
we have (similar as~\eqref{eqn:localp2-three-leg-general-symmetry})
\begin{equation}\label{eqn:one-loop-TR-multileg-Ward}
\sum_{i=1}^{\mathfrak m}(-1)^{b_i}
\bigg(\prod_{j=1}^{i-1}\bigl((-1)^{f_{j+1}+1}Q_{j+1}\bigr)^{b_j}\bigg)
\mathcal E_{a_i,b_i}^{(i)}|V\rangle=0.
\end{equation}

For each boundary point $v_k$, apply this identity with the cyclic order
starting at $k$ and initial pair $(a_k,b_k)=(1,1)$.
We multiply these specialized identities by $e^{-A_k}$ and sum over $k$.
As in the proof of~\eqref{eqn:localp2-one-leg-symmetry},
projecting all legs except $\ell$ to the vacuum gives the
following one-leg identity
\begin{equation}\label{eqn:one-loop-TR-vertex-one-leg-Ward}
 \bigg[\sum_{m_k\ne0}\widetilde r_k\mathcal E_{-m_k,n_k}
 +\sum_{m_k=0}\widetilde r_k
 \bigg(\mathcal E_{0,n_k}-\frac{1}{\sfq^{-n_k/2}-\sfq^{n_k/2}}\bigg)
 \bigg]|V^{(\ell)}\rangle=0,
\end{equation}
where
$\widetilde r_k=(-1)^{m_k+n_k+1}e^{-A_k}\prod_{j=k}^{\ell}Q_j^{\det(v_k,v_j)}$.

By the open mirror map and the all-genus mirror theorem~\cite[(7.2), Theorem~7.5]{FLZ20}, together with the vertex/GW comparison~\eqref{eqn:one-loop-TV-geometry}, we have 
\begin{equation}\label{eqn:one-loop-TR-Schur-comparison}
 \frac{Z_\lambda(\epsilon;\hbar)}{Z_\varnothing(\epsilon;\hbar)}
 =\left(-\gamma e^{\I-A_{\ell+1}}\right)^{|\lambda|}
 \frac{V_\lambda(\mathbf Q;\sfq)}{V_\varnothing(\mathbf Q;\sfq)}.
\end{equation}
Since $\widetilde r_k\left(-\gamma e^{\I-A_{\ell+1}}\right)^{m_k+1}=(-1)^{n_k}e^{-A_{\ell+1}}r_k$
and $\mathcal E_{-m_k,n_k}$ shifts partition size by $m_k$,
these identities transform~\eqref{eqn:one-loop-TR-vertex-one-leg-Ward}
into
\begin{equation}\label{eqn:one-loop-TR-Ward}
 \bigg[\sum_{m_k\ne0}(-1)^{n_k}r_k\mathcal E_{-m_k,n_k}
 +\sum_{m_k=0}(-1)^{n_k}r_k
 \bigg(\mathcal E_{0,n_k}-\frac{1}{\sfq^{-n_k/2}-\sfq^{n_k/2}}\bigg)
 \bigg]\sum_\lambda Z_\lambda|\lambda\rangle=0.
\end{equation}
Taking the row matrix elements as in the proof of~\eqref{eqn:localp2-row-hook}, we obtain~\eqref{eqn:one-loop-TR-row-hook}.
\end{proof}

\subsection{The canonical quantum curve}
\label{sec:one-loop-TR-quantum-curve}
Now we are ready to prove the canonical quantum curve conjecture.
Similarly as local $\mathbb P^2$ case, we have $Z^{\rm NP}_{\lambda,*}$, $\Psi^X_*(X)$, and $\Psi_*(X)$.

\begin{theorem}\label{thm:one-loop-TR-quantum-curve}
The canonical wave function $\Psi_*$ satisfies
\begin{equation}\label{eqn:one-loop-TR-quantum-curve}
 \left(\sum_{k=1}^{\mathfrak m}
 r_k\sfq^{-m_kn_k/2}X^{m_k}\widehat Y^{\,n_k}
 +\gamma+\cW^+\right)\Psi_*
 =0,
\end{equation}
where $\cW^+=\cW(\mathbf Q;\hbar)-\cW(\mathbf Q;0)$ and
\begin{equation}\label{eqn:one-loop-TR-Wilson-coefficient}
 \cW=
 -\left(\sum_{m_k=-1}(-1)^{n_k}r_k\sfq^{-n_k}\right)
 \frac{Z^{\rm NP}_{(2),*}}{Z^{\rm NP}_{(1),*}}
 -\sum_{m_k=0}(-1)^{n_k}r_k\sfq^{-n_k/2}.
\end{equation}
In particular, $\cW(\mathbf Q;0)=\gamma$.
\end{theorem}
\begin{proof}
Since $\partial_\epsilon r_k=0$, equation~\eqref{eqn:one-loop-TR-row-hook} remains valid with $Z_\lambda$ replaced by $Z^{\rm NP}_{\lambda,*}$.
As in the proof of Theorem~\ref{thm:quantum-localp2}, the homogeneous
Pl\"ucker relation~\eqref{eqn:localp2-TR-rank-two} and the closed
vanishing equation give
\[
 Z^{\rm NP}_{(n,1),*}
 =\frac{Z^{\rm NP}_{(1,1),*}}{Z^{\rm NP}_{(1),*}}
 Z^{\rm NP}_{(n),*},\qquad n\geq1.
\]
The case $n=1$ of~\eqref{eqn:one-loop-TR-row-hook} at this specialization
gives
\begin{equation}\label{eqn:wilsonloopdelpezzo}
 \cW=-\sum_{m_k=0}(-1)^{n_k}r_k\sfq^{n_k/2}
 +\left(\sum_{m_k=-1}(-1)^{n_k}r_k\sfq^{n_k}\right)
 \frac{Z^{\rm NP}_{(1,1),*}}{Z^{\rm NP}_{(1),*}}.
\end{equation}
Substituting these identities into the NP coefficient relation, we see
\begin{equation}\label{eqn:one-loop-TR-closed-row}
 \sum_{k=1}^{\mathfrak m}(-1)^{n_k}r_k
 \sfq^{-n_k(n-(m_k+1)/2)}Z^{\rm NP}_{(n-m_k),*}
 +\cW Z^{\rm NP}_{(n),*}=0,\qquad n\geq1.
\end{equation}
The constant coefficient vanishes since $\left.XH(X,Y;\cq)\right|_{X=0,Y=-1}=0$.
Since $\omega_{0,1}^{\mathrm{sing},b}=\pi i\,dX/X$ and $dX/dx=-X$,
the coefficient comparison in the proof of
Theorem~\ref{thm:quantum-localp2} identifies this relation with
\[
 \left(\sum_{k=1}^{\mathfrak m}
 r_k\sfq^{-m_kn_k/2}X^{m_k}\widehat Y^{\,n_k}
 +\cW\right)\Psi_*=0.
\]
The disk contribution gives
$\widehat Y^{\,n_k}\Psi_*/\Psi_*=Y^{n_k}+O(\hbar)$.
Taking the classical limit and
using~\eqref{eqn:one-loop-mirror-curve}, we obtain
\[
 \cW(\mathbf Q;0)
 =-\sum_{k=1}^{\mathfrak m}r_kX^{m_k}Y^{n_k}
 =\gamma.
\]
This proves~\eqref{eqn:one-loop-TR-quantum-curve}.
\end{proof}

\begin{remark}
The identification of the scalar $\cW$ given by~\eqref{eqn:one-loop-TR-Wilson-coefficient} with a geometrically defined NS Wilson loop remains conjectural for the general cases considered here.
\end{remark}

\addtocontents{toc}{\protect\setcounter{tocdepth}{0}}

\end{document}